\documentclass{article}

\usepackage[english]{babel}
\usepackage{booktabs, longtable, array, multirow}
\usepackage[justification=centering, singlelinecheck=false]{caption}

\usepackage[letterpaper,top=2cm,bottom=2cm,left=3cm,right=3cm,marginparwidth=1.75cm]{geometry}

\usepackage{setspace}
\usepackage{parskip}
\usepackage{amsmath}
\usepackage{graphicx}
\usepackage[colorlinks=true, allcolors=blue]{hyperref}

\usepackage{xr-hyper}
\title{FlexBSS: A flexible multi-objective framework for bike-sharing station optimization}
\author{Jordi Grau-Escolano\textsuperscript{1,2}, David Duran-Rodas\textsuperscript{3}, Julian Vicens\textsuperscript{1}\\
\small
\textsuperscript{1} Eurecat, Centre Tecnològic de Catalunya, Barcelona, Spain\\
\small
\textsuperscript{2} Universitat Politècnica de Catalunya, Barcelona, Spain\\
\small
\textsuperscript{3} Technical University of Munich, Munich, Germany}

\begin{document}
\maketitle

\begin{abstract}
Bike-sharing systems (BSS) are key components of urban mobility, promoting active travel and complementing public transport. This paper presents a flexible, data-driven framework for optimizing BSS station placement. Existing methods usually focus on a single planning objective, such as maximizing demand, or on a fixed set of two or three objectives, such as combining demand and social equity. This restricts their adaptability to evolving priorities and diverse contexts. Moreover, they often aggregate candidate sites into coarse spatial units and neglect features such as topography or directed cycling networks.
The framework addresses these limitations through four components: (i) a multi-criteria decision-making model that evaluates candidate sites using configurable spatial factors and adjustable weights, thereby enabling planners to represent any combination of planning objectives rather than being limited to a fixed set; (ii) network-level metrics that explicitly balance station proximity and system-wide accessibility on directed bicycle networks; (iii) slope-adjusted distances that account for uphill effort and downhill facilitation; and (iv) a genetic algorithm that generates feasible station sets while respecting minimum-distance constraints.
A case study in Barcelona demonstrates its application using open data for scenarios focused on demand, multimodal integration, and social equity. Results show that network-level metrics reshape the spatial arrangement of stations while keeping the same planning objectives, slope-adjusted distances improve station availability in hilly districts, and an expansion scenario in l’Eixample district produces coherent layouts complementing existing stations. By integrating spatial factors and network representation, the framework provides a versatile decision-support tool for BSS planning in complex urban environments.
 
\end{abstract}

\section{Introduction}
Bike-sharing systems (BSS) are becoming a key component of urban mobility offering a sustainable and active mode of transport~\cite{Fishman2016, Meddin2025}. As their popularity has grown, the strategic placement of stations has become a decisive factor of operational success, influencing ridership levels~\cite{NACTO2016, Kabra2019} and cost efficiency of the system~\cite{Fu2022}. Historically, BSS station planning has relied heavily on heuristics based on expert judgment~\cite{Ferrando2007, Yanocha2018, CROW2006}. Planners commonly identified broad zones of high population density or proximity to major trip generators, such as public transport stations, employment districts, or shopping areas, and then refined locations iteratively based on budget and qualitative assessments of suitability. Similarly, fleet sizes were set according to rough demand estimates or available budgets. While this approach laid the groundwork for many early systems, it often relied more on intuition and practical experience than on formal analytical methods, and was primarily oriented toward utilitarian needs. Consequently, systematically evaluating competing goals and adapting to changing urban conditions was challenging.

The proliferation of high-resolution spatial data and advanced computational methods has enabled a shift toward data-driven decision-making in BSS planning. Various methods, including location– allocation models~\cite{Frade2015, Mete2018}, multicriteria decision-making (MCDM)~\cite{Kabak2018, Bahadori2022}, and exact and metaheuristic optimization techniques~\cite{Cintrano2018, Qian2022, Caggiani2020} have been applied to optimize station locations. Yet, regardless of the methodology used, the design and implementation of most docked systems still follow an iterative workflow in which the first and most critical step is the clear definition of planning objectives~\cite{Duran-Rodas2021, Ferrando2007, Yanocha2018}. These may include expanding active mobility~\cite{tfgmActiveTravel2025}, improving first- and last-kilometer connectivity~\cite{MTC2013BayAreaBSS}, reducing greenhouse gas emissions~\cite{ctcn_velib}, or promoting social inclusion~\cite{tembiciBogota2023}. Such objectives can differ across regions within the same system and often evolve over time; for instance, the expansion of an existing BSS may prioritize goals that are different from those guiding its original implementation.

Despite advances, existing data-driven methods exhibit limitations: they are often applied to pursue a single dominant planning objective, making it difficult to explicitly model trade-offs among genuinely competing goals, such as maximizing demand and improving social inclusion. For instance, while MCDM approaches can incorporate many data sources, they are typically used to optimize a single dominant objective, often corresponding to maximizing coverage~\cite{Fazio2021} or demand~\cite{Kabak2018, Bahadori2022}. Even approaches explicitly designed to handle multiple planning objectives usually limit the analysis to a small, predefined set of goals, most often only two, such as demand and social equity~\cite{Conrow2018, Qian2022, Caggiani2020, Duran-Rodas2021} or social equity and first–last-mile connectivity~\cite{Fan2024}. Moreover, in one identified case~\cite{Nikiforiadis2021} three planning objectives were considered, namely demand, social equity, and cost. Relying on such predefined sets of planning objectives constrains the ability to explore a wider range of scenarios without using multiple models or redesigning the optimization framework.

In addition, the spatial resolution of candidate station locations remains a persistent challenge. Many studies aggregate these into coarse administrative zones or regular grids to reduce computational complexity~\cite{Frade2015, Fazio2021, Veillette2018, Nikiforiadis2021}. Although this simplification can ease model formulation and reduce the risk of overfitting when modelling, it can have its drawbacks. First, it prevents planners from working directly with the street layout, which is the level of detail needed for deciding the final station locations. Second, it can hide important local factors, such as the exact location of trip generators (e.g. workplaces, schools, or public facilities) or nearby cycling infrastructure.

Beyond the spatial resolution of candidate locations, how the transport network is represented becomes equally important, as it shapes the estimation of how users travel between stations and, ultimately, how they use the system. Few studies have gone beyond simply mapping potential sites onto a street-level cycling network to actually integrating the network as a graph into the optimization process, where station placement is evaluated based on a more realistic travel paths and connectivity~\cite{Cintrano2018, Xin2013}. In these cases, the network is represented as an undirected graph, meaning that all links are treated as bidirectional regardless of actual circulation rules. This simplification is already a step toward more realistic modelling compared to not acknowledging the distance between stations or considering purely Euclidean distances, but it can still misrepresent actual cycling conditions. In practice, bikes should follow directed networks that respect one-way links, turn restrictions, and other routing constraints. Incorporating these constraints would enhance realism of how users use the bikes and the evaluation of key metrics, such as station-to-station travel distance.

Furthermore, slope has been found to be a key determinant of cycling behaviour and effort in the literature~\cite{Grau-Escolano2024, MateoBabiano2016}, with different effects on mechanical (m-bike) and electric bikes (e-bike)~\cite{Grau-Escolano2024}. Despite this, most BSS placement studies commonly omit topography when optimizing locations~\cite{Fazio2021, Kabak2018, Ebrahimi2022, Tera2023}. This limitation is particularly relevant in cities with pronounced elevation changes, where neglecting slope can distort estimates of station accessibility and user suitability. In this regard, one exception is the work of García-Palomares et al.~\cite{GarciaPalomares2012}, who incorporated slope as a uniform penalty on effective travel time. While this is a relevant step in incorporating topography effects, their approach did not distinguish between uphill and downhill gradients, thereby overlooking the facilitating effect of moderate declines and the safety-related penalties of steep descents. By neglecting these effects, the accessibility and suitability of certain locations can be misrepresented.

Finally, the ability to expand and adapt BSS networks over time is essential for ensuring long-term system efficiency and alignment with changing mobility patterns. However, many BSS planning methods are designed for single-phase deployments, in which the entire network is planned at once for the initial roll-out, and thus lack mechanisms to support incremental expansion or reconfiguration while considering existing BSS stations~\cite{Kabak2018, Bahadori2022, Duran-Rodas2021}.

Although these models can be applied to extend an existing BSS, they are typically not designed to consider how existing stations are positioned relative to each other and their urban context. As a result, optimizing new stations as part of a cohesive system is often challenging, and they risk being treated as isolated additions disconnected from the existing network.

Taken together, these insights highlight the opportunity to develop planning tools that integrate multiple objectives, fine-grained spatial detail, and realistic modeling of the operational environment. To address these gaps, this paper introduces a flexible, data-driven framework for optimizing the placement of BSS stations. Its main contributions are as follows: 

\begin{enumerate}
    \item  \textbf{A general MCDM scoring framework:} a modular scoring approach that quantifies the suitability of each candidate station location by aggregating spatial attributes within adjustable buffers that capture the surrounding conditions of potential sites. 
    
    Unlike models limited to a predefined set of planning objectives, this framework is designed to allow planners to integrate any spatial factor based on the specific planning objectives of the BSS, and to freely choose their relative weights to reflect any combination of priorities. This flexibility supports the creation of alternative planning scenarios, enabling the design of station networks aligned with multiple objectives without requiring separate models.

    \item \textbf{Integrated network metrics on directed graphs:} the framework incorporates network-level measures of station dispersion and accessibility, enabling planners to balance trade-offs between station clustering and system accessibility explicitly. These metrics are computed on a directed network that respects one-way streets, turn restrictions, and other routing constraints, and consider all intersections of the bicycle network as potential station locations. This provides a realistic basis for evaluating and comparing configurations.

    \item \textbf{Topography-adjusted distance modeling:} to improve the accuracy of proximity and accessibility assessments, the model optionally integrates elevation-adjusted distance calculations that reflect slope-induced variations in cycling effort. This feature is particularly important in cities with significant topographic variation and supports planning for both m-bikes and e-bikes.

    \item \textbf{Support for incremental expansion and reconfiguration:} the framework can be applied not only to the design of new BSS networks but also to the incremental expansion or reconfiguration of existing systems. By explicitly accounting for the existing station layout and its network relationships, the model facilitates planning that strengthens network cohesion.

    \item \textbf{Metaheuristic optimization approach:} a genetic algorithm (GA) is used to search for high-quality station configurations while satisfying operational constraints such as the number of stations and the minimum inter-station distance. This approach enables the exploration of diverse scenarios and the generation of near-optimal solutions for large problem instances within reasonable computational time.

\end{enumerate}

To demonstrate the applicability of this framework, we conduct a case study for the city of Barcelona, Spain, using its extensive open data resources. The results illustrate how our approach may help planners systematically explore alternative configurations, assess trade-offs among multiple planning objectives, and design station networks that are tailored to the city's socioeconomic context, infrastructure and topography.

The remainder of the paper is structured as follows. Section 2 reviews the related literature. Section 3 details the methodology, including the scoring framework, optimization procedure, and implementation. Section 4 reports the results of the Barcelona case study. Finally, Section 5 discusses the results and outlines directions for future research.

\section{Literature review}

The siting of BSS stations has been addressed through a wide range of methodological perspectives. Contributions in the literature vary in terms of the data they use, the analytical tools they apply, and the planning objectives they pursue. This section reviews these contributions to clarify the main ways the problem has been addressed and to highlight insights relevant to this study.

\subsection{From heuristics to digital decision-support tools in BSS planning}

Docked BSS are typically planned through an iterative process, in which goals are first defined and station networks are progressively refined across successive deployment phases~\cite{Duran-Rodas2021, Ferrando2007, Yanocha2018}. Rather than fixing all the station locations at the start, planning guidelines recommend setting clear objectives such as maximizing ridership, promoting social equity, or enhancing first–last kilometer access, and using them to guide preliminary site selection, field validation, and ongoing adjustment. These revisions should be informed by potential demand patterns, operational performance, and stakeholder input~\cite{Ferrando2007, Yanocha2018}. Importantly, such planning goals are not necessarily fixed over time or consistent across a single BSS. For instance, a BSS expansion may pursue different priorities than those of the initial implementation.

In practice, early deployments relied on a combination of expert judgment and heuristics. Expert knowledge highlighted key factors shaping cycling demand, such as topography, land use, and connectivity, while heuristics translated these insights into simplified rules that guided site selection. Planning guides reflected this process by recommending station locations in areas of high expected demand, such as near public transportation, or commercial hubs, and encouraged the use of origin-destination surveys and stakeholder consultations to support site selection~\cite{Ferrando2007, Yanocha2018}. These documents emphasized intermodal integration and land-use alignment as core planning principles, often suggesting station placement in visible, accessible, and well-connected locations~\cite{Ferrando2007, Yanocha2018, CROW2006}. Furthermore, bicycle infrastructure manuals underscored the importance of coherent and continuous cycling networks~\cite{AASHTO1999bicycle, CROW2006}. While these guidelines provided valuable practical recommendations, they lacked analytical tools or optimization frameworks capable of systematically evaluating locations or adapting the BSS to diverse urban conditions. Nonetheless, heuristics and optimization tools can complement each other: practical rules can inform modelling choices, while optimization results can in turn be used to refine or validate heuristic guidelines.

Beyond expert judgment and heuristics, subsequent studies began adopting more systematic methods. As richer spatial data and computational tools became more available in the early 2010s, researchers began to incorporate more formal methods. Geographic Information Systems (GIS) were increasingly used to integrate and visualise multiple spatial layers, supporting more data-informed site selection (see Section~\ref{sec:integrating_gis_to_station_placement}). In parallel, spatial optimization models, particularly location–allocation formulations, were introduced to frame station siting as a mathematical problem with explicit objectives and constraints (see Section~\ref{sec:location_allocation_models}). Together, these developments expanded the range of tools available for BSS planning, combining practical heuristics with data-driven and model-based decision support.

\subsection{Location-allocation models} \label{sec:location_allocation_models}

Location-allocation models are spatial optimization tools that decide where to place a fixed number of service facilities and how to allocate demand to them, with the aim of optimizing one specific performance metric~\cite{Daskin1995}. In these models, demand is typically defined over predefined zones or grid cells, while facilities correspond to candidate sites. In practical applications, these are usually framed as discrete-location problems~\cite{Teixeira2008}, although continuous formulations also exist~\cite{Yeh1996}. These formalizations include the \textit{p}-median problem, which minimizes the total demand-weighted distance to assigned facilities~\cite{Hakimi1965}, and the Maximal Covering Location Problem (MCLP), which seeks to maximize demand served within a defined service radius~\cite{Church1974} (i.e., the maximum distance or travel time from a facility within which users are considered effectively covered). Another influential formulation is the \textit{p}-center problem, which minimizes the maximum distance between any user and their nearest facility~\cite{Hakimi1964}. Depending on the formulation, they can be broadly classified as either efficiency-oriented (e.g., \textit{p}-median), which aim to minimize overall travel effort or system-wide costs, or equity-oriented (e.g., \textit{p}-center),  which aim to ensure fair access by minimizing the longest distance any user must travel to reach a facility~\cite{Murray2010}.

Several studies have applied these classical formulations to BSS station placement~\cite{Frade2015, Cintrano2018, Lin2011, Mete2018}. In these cases, they typically rely on single-objective optimization frameworks, most often aiming to either maximize spatial coverage or minimize user distance. To ensure a context-sensitive BSS deployment, models usually incorporate practical constraints such as a fixed number of stations, budget limits, or minimum inter-station distances. For instance, Frade and Ribeiro~\cite{Frade2015} used a variant of the MCLP to maximize population coverage under budget constraints. Their model optimizes the number, capacity, and location of stations and fleet size, while also considering bike relocations and operational costs. Similarly, Cintrano et al.~\cite{Cintrano2018} studied the \textit{p}-median problem to identify optimal station locations. Lin and Yang~\cite{Lin2011} extended this paradigm by proposing a mixed-integer nonlinear programming model (MINLP) to minimize the total system costs, accounting for user travel distance, station and lane installation, bike inventory, and unmet demand penalties, while ensuring coverage and service availability. Moreover, Mete et al.~\cite{Mete2018} applied several coverage-based formulations (\textit{p}-center, \textit{p}-median, set covering) to locate stations across a university campus, using GIS to identify and evaluate candidate sites based on proximity to key facilities.

These models have provided reproducible alternatives to heuristic siting but remain focused on classical objectives such as spatial coverage and distance. Broader priorities like social equity or multimodal integration are largely absent, though some aspects have been indirectly considered. For instance, Frade and Ribeiro~\cite{Frade2015} accounted for demand heterogeneity by weighting zones according to estimated trip generation and attraction. This ensured that covering high-demand zones contributed more to the optimization objective than covering low-demand regions, thereby embedding potential usage indirectly rather than treating demand maximization as a distinct objective.

\subsection{Integrating GIS to station placement} \label{sec:integrating_gis_to_station_placement}

Many studies have also incorporated GIS as a key feature to enrich spatial context and support more informed decision-making~\cite{Fazio2021, Kabak2018, GarciaPalomares2012, Chou2019, Tera2023, Ebrahimi2022}. By integrating and visualizing diverse data layers (e.g., land use, population, slope, and transport networks), GIS allows planners to evaluate site suitability with greater precision.

One purely GIS-focused approach is that of Fazio et al.~\cite{Fazio2021}, who proposed a GIS-based multi-criteria framework to prioritize cycle parking locations in Catania without relying on formal optimization algorithms. Working with a 100×100 meter grid, they integrated spatial data on population, employment, the distribution of POIs, and public transport accessibility to calculate thematic indexes for each cell, which were then aggregated into a composite score used to rank and identify the most suitable areas. Similarly, Kabak et al.~\cite{Kabak2018} developed a GIS-based MCDM method combining Analytic Hierarchy Process (AHP) and Multi-Objective Optimization by Ratio Analysis (MOORA) to evaluate and rank potential station locations based on twelve spatial criteria, also avoiding formal mathematical optimization but producing explicit suitability maps and priority rankings.

Other studies have combined GIS with location–allocation models to decide BSS station locations. García-Palomares et al.~\cite{GarciaPalomares2012} developed one of the earliest GIS-supported methodologies for siting stations in central Madrid. They estimated potential demand by integrating building-level residential and employment data with zonal trip generation and attraction rates. Using a slope-aware street network to compute bicycle travel times, they applied $p$-median and MCLP formulations to identify optimal station locations under different network density scenarios.

More recently, Chou et al.~\cite{Chou2019} developed an optimization approach that integrates GIS-derived spatial data on public transport networks as key model inputs. Their framework incorporates proportional flow constraints to estimate how bicycles circulate between stations and to assess station capacity requirements. While primarily focused on optimizing fleet allocation and redistribution, their methodology also leverages spatial information to identify candidate station locations and better align supply with latent demand for short-distance trips. Similarly, Tera et al.~\cite{Tera2023} applied a GIS-based MULTIMOORA model to recommend new station locations in Tartu, Estonia. By integrating spatial data on population density, observed station demand, and proximity to schools and shopping centers, they systematically ranked and compared candidate sites, demonstrating how multi-criteria evaluation can complement network design decisions.

While previous approaches focus on placing new stations, Ebrahimi et al.~\cite{Ebrahimi2022} demonstrated how GIS can also support the optimization of existing networks. They applied the MCLP and Target Market Share models to evaluate how well current stations served residential and transit demand, identifying gaps in accessibility and highlighting areas where additional capacity could be prioritized. Although their analysis aimed at optimizing subsets of existing facilities rather than selecting entirely new sites, it illustrates how spatial optimization can inform both operational improvements and broader strategic planning.

\subsection{Multi-criteria and multi-objective station placement}

While many BSS station placement models have focused on optimizing a single planning goal such as maximizing demand~\cite{GarciaPalomares2012}, maximizing spatial coverage~\cite{Frade2015}, or minimizing user distance~\cite{Cintrano2018}, researchers have increasingly incorporated multiple criteria to better capture the complex factors influencing system performance~\cite{Kabak2018, Guler2021a, Guler2021b, Bahadori2022, Wang2021, Veillette2018, Fazio2021, Tera2023, Hafezalkotob2019}. In most cases, these \textit{multi-criteria approaches} are used to combine diverse indicators into a single objective function, where all factors are assumed to contribute to the same planning goal. For example, variables such as population density, proximity to public transport, and cycling infrastructure may be weighted and aggregated to identify sites that best support overall demand. More recently, however, some studies have extended multi-criteria methods to explicitly define multiple planning objectives~\cite{Duran-Rodas2021, Caggiani2020, Fan2024, Qian2022, Nikiforiadis2021}. In these \textit{multi-objective approaches}, different factors are treated as representing distinct goals. This distinction allows planners to explore trade-offs among competing objectives instead of aggregating all criteria into a single composite measure. The following discussion reviews both in detail.

\textit{Multi-criteria approaches} rely on combining diverse spatial and contextual factors into a unified metric. Several strategies have been adopted to determine the importance of each factor, such as the AHP, Fuzzy AHP, Best–Worst Method (BWM), Analytic Network Process (ANP), or Technique for Order of Preference by Similarity to Ideal Solution (TOPSIS) to prioritize criteria according to planners' or stakeholders' judgments~\cite{Kabak2018, Guler2021a, Guler2021b, Bahadori2022}. Other studies have used data-driven weighting, where factor importance is derived from statistical models (e.g. regression coefficients or variable importance scores in machine learning) based on observed BSS usage~\cite{Wang2021}. A third approach involves flexible weighting, where authors suggest predefined weights to reflect the assumed importance of each factor. For example, Veillette et al.~\cite{Veillette2018} propose a weighting scheme based on existing and potential cyclist demand and proximity to public transport. However, they explicitly recommend adapting the scheme to match the specific priorities and planning goals of different regions. Similarly, Fazio et al.~\cite{Fazio2021} showcase their methodology using equal weights across all criteria, but emphasise that the weights should ultimately be calibrated in collaboration with local stakeholders and aligned with the intended planning objectives. On the other hand, Tera et al.~\cite{Tera2023} applied the MULTIMOORA model with equal weights across criteria in their evaluation of candidate station locations in Tartu, Estonia. Although the authors did not specify custom weighting in their application, the framework inherently allows planners to adjust weights to reflect specific planning priorities~\cite{Hafezalkotob2019}. 

Multi-criteria approaches typically account for spatial factors such as population and employment density, land-use mix, proximity to POIs, cycling infrastructure, and access to public transport. A systematic review by Bahadori et al.~\cite{Bahadori2021} provides an overview of these factors, classifying them into four categories: bike network (e.g., station density, infrastructure), operator (e.g., budget, maintenance costs), user (e.g., walking distance, safety), and city infrastructure (e.g., public transport connectivity, POIs, population).

In contrast, \textit{multi-objective approaches} treat factors that pertain to different planning goals separately. A fundamental property of multi-objective optimization problems is that no single solution optimizes all objectives simultaneously; instead, they yield a set of Pareto-optimal solutions that capture the balance between competing objectives. This separation makes the contribution of each objective explicit and enables systematic exploration of alternatives, rather than concealing them within a single aggregated metric~\cite{Rahimi2023, honegumi}. For example, Conrow et al.~\cite{Conrow2018} developed a bi-objective model that maximizes coverage of both residential population and bicycle network infrastructure. 

Several multi-objective studies have incorporated social equity more explicitly. Duran-Rodas et al.~\cite{Duran-Rodas2021} proposed the Demand And/or Equity (DARE) approach, which generates alternative station plans under different emphasis scenarios to illustrate how prioritizing demand, social equity, or a blend of both reshapes spatial allocation. Similarly, Caggiani et al.~\cite{Caggiani2020} integrated multimodal accessibility and social equality into station placement. Their approach minimizes disparities in accessibility among different population groups, measured using a Theil index of multimodal accessibility combining bike-sharing and public transport travel times. This model balances equality objectives with minimum requirements for system coverage and average accessibility. Moreover, Fan and Harper~\cite{Fan2024} developed a bi-objective integer programming model to jointly maximize coverage of potential demand and improve public transport for underserved communities. Their approach explicitly weights equity among social groups, showing that even a modest emphasis on equity can yield significant gains in public transport accessibility for disadvantaged populations with minimal loss in system-wide service quality.

Other multi-objective contributions have placed greater emphasis on explicitly exploring trade-offs along the Pareto frontier. Qian et al.~\cite{Qian2022} focused on the reallocation of existing stations to improve both revenue generation and equitable access to jobs and essential services, particularly for disadvantaged communities. Their framework estimates trip patterns using socioeconomic and built environment data, distributes trips via a gravity model, and employs a GA to identify station configurations that balance profitability and social equity objectives within a fixed overall system size. 

Finally, Nikiforiadis et al.~\cite{Nikiforiadis2021} used a tri-objective optimization through a two-phase framework. In the first phase, each objective (maximizing demand coverage, maximizing area coverage, and minimizing rebalancing needs) is optimized independently to establish target benchmark values. In the second phase, these objectives are combined by minimizing the weighted percentage deviations from the targets, enabling planners to adjust the importance of each goal and systematically explore trade-offs. Their approach integrates spatial analysis of BSS rental patterns and the built environment with practical constraints such as budget, site availability, and minimum station spacing.

Overall, multi-criteria and multi-objective literature on BSS station placement has substantially advanced the field. Multi-criteria studies have clarified how diverse spatial factors can be assembled into a composite suitability score aligned with a single overarching goal, typically from a predefined set of factors~\cite{Kabak2018, Guler2021a, Guler2021b, Bahadori2022}. Multi-objective formulations build on this by making trade-offs between compiting planning goals explicit and allowing the emphasis across objectives to be adjusted when selecting among the Pareto set~\cite{Duran-Rodas2021, Caggiani2020, Fan2024, Qian2022, Nikiforiadis2021}. Yet in both approaches, sets of spatial factors and planning objectives are commonly fixed \textit{a priori}; adding new dimensions (e.g., safety or environmental criteria) would require reformulating the model. This can constrain the transparent exploration of wider policy scenarios. To adress this, we introduce a modular, adaptable MCDM framework that allows new criteria to be incorporated without altering the overall structure. Moreover, planners can flexibly assign weights to these criteria to reflect local priorities, ensuring that different planning objectives can be emphasised as needed.

\subsection{Spatial granularity and network representation}

Studies vary in the spatial scope they consider for planning new BSS stations, ranging from entire cities to small regions. Many analyses cover a full urban area such as an entire city. For instance, Conrow et al.~\cite{Conrow2018} applied their model to the city of Phoenix and Duran-Rodas et al.~\cite{Duran-Rodas2021} to Munich. In other cases, the focus is on a limited study area. Examples of smaller-scale planning include García-Palomares et al.~\cite{GarciaPalomares2012}, who optimized stations in central Madrid, and Mete et al.~\cite{Mete2018}, who planned stations on a university campus with only 20 candidate sites. 

Equally important is the spatial granularity of potential station locations. Early and some recent optimization models define candidate sites at the zonal level, either by aggregating demand within predefined administrative or planning zones or by applying a uniform grid over the study area. For example, Frade and Ribeiro~\cite{Frade2015} aggregated demand to find potential station locations into predefined neighbourhoods-level regions and Mete et al.~\cite{Mete2018} did similarly using campus sub-areas. Fazio et al.~\cite{Fazio2021} applied a uniform 100×100\,m grid to define candidate cells across the city of Catania, Tera et al.~\cite{Tera2023} and Veillete et al.~\cite{Veillette2018} generated evenly spaced 300\,m grid points covering Tartu and Québec to identify candidate locations, whereas Nikiforiadis et al.~\cite{Nikiforiadis2021} combined grid cells with predefined zones and selected nodes of the transport network. Conrow et al.~\cite{Conrow2018} used census block groups to represent small neighbourhood-scale regions. Duran-Rodas et al.~\cite{Duran-Rodas2021} computed demand and social equity indicators over road network-based catchment zones created by placing regularly spaced virtual station points every 300\,m and generating non-overlapping 300\,m service areas around them using Voronoi tessellation to fully cover the study area. Although these regional approaches simplify the problem and improve tractability, they sacrifice spatial precision, as the exact placement within each cell or zone remains unspecified.

Moving beyond aggregated zones, a second group of studies relies on network-based approaches, selecting discrete candidate points strategically located along major corridors, at intersections of principal streets, or near key points of interest (POIs) such as public transport stops. This approach offers more detail than coarse zoning but avoids the computational burden of enumerating every possible location. For instance, Conrow et al.~\cite{Conrow2018}, sampled several hundred potential sites distributed along the bicycle network in their bi-objective model. Similarly, Caggiani et al.~\cite{Caggiani2020} defined their candidate locations as a curated set of around 200 unserved public transport adjacent nodes that primarily corresponded to bus and metro stops with no existing BSS stations. Moreover, Fan and Harper~\cite{Fan2024} adopt a similar strategy, generating over 1,200 candidate stations distributed along existing and proposed bike infrastructure, with points placed systematically every 400 metres to ensure coverage of both high-demand and disadvantaged areas. 

To our knowledge, only a few studies pursue high-resolution approaches by using street-network-derived candidate sets for station locations. For example, Cintrano et al.~\cite{Cintrano2018} enumerate an exceptionally large set of 33,550 street segments across Málaga as potential sites, while Xin’s Vancouver analysis~\cite{Xin2013} evaluates 1,489 street intersections within the downtown area. Beyond site definition, these high-resolution approaches and most of the literature rely on network representations in the optimization process to evaluate realistic travel paths rather than simple Euclidean distances or coarse zoning. In these cases, the network is represented as undirected, treating all links as bidirectional with identical distances. Although this simplification does not capture features such as one-way streets and asymmetric accessibility between origins and destinations, these studies show how network-based optimizations can better represent cycling conditions and provide a basis for future refinements.

Additionally, slope is a critical factor in BSS cycling behaviour, with fewer trips observed on uphill trips and more on downhill ones~\cite{Morency2015, Grau-Escolano2024, MateoBabiano2016}. While Mix et al.~\cite{Mix2022} found that elevation was not a significant predictor in Santiago de Chile, slope has been shown to strongly influence cycling patterns in other cities with heterogeneous topography, such as Barcelona~\cite{Grau-Escolano2024} and Brisbane~\cite{MateoBabiano2016}. In terms of modelling, García-Palomares et al.~\cite{GarciaPalomares2012} took an important step by incorporating slope as a penalty factor uniformly increasing effective travel time. Their approach, however, did not differentiate between uphill and downhill gradients, thereby overlooking the potentially facilitating effect of moderate declines and the safety-related penalties of steep descents.

Our work builds directly on previous high-resolution approaches that derive candidate sites from the street network, such as Cintrano et al.~\cite{Cintrano2018} in Málaga and Xin~\cite{Xin2013} in Vancouver, as well as on the early recognition by García-Palomares et al.~\cite{GarciaPalomares2012} of the need to incorporate slope to achieve a more realistic estimation of user access distances. We extend this line of research by addressing two key limitations. First, unlike prior work that relied on undirected networks, we employ a directed street-level representation that captures circulation constraints such as one-way streets and asymmetric accessibility. This refinement is important because shortest routes in opposite directions can differ substantially in length due to one-way restrictions, with empirical evidence showing asymmetries of more than 10\% in some urban areas~\cite{Melo2022}. Second, we replace the uniform slope penalty of García-Palomares et al. with an effort-based distance measure that differentiates between uphill and downhill gradients. Empirical studies have shown that uphill slopes strongly reduce speed in a near-linear fashion, whereas downhill effects are non-linear: cyclists accelerate only up to a certain gradient and often brake on steeper descents due to safety concerns~\cite{Ryeng2016, Flugel2019}. By incorporating these asymmetries, our approach reflects the increased effort of climbs, the facilitating effect of moderate declines, and the risks associated with steep descents.

\subsection{Spatial arrangement of stations}

In BSS station planning, a critical design consideration is the spatial arrangement of stations, meaning the overall pattern of how stations are distributed across the service area. Classical network-based formulations such as the $p$-median problem or the $p$-center problem~\cite{Hakimi1964, Hakimi1965} improve accessibility, understood here as reducing the distance between users and their nearest station, by minimizing either average or worst-case user distance. However, in these models the spatial arrangement of facilities remains an emergent outcome rather than an explicit design objective that the planner can control.

The same pattern is evident in BSS-specific applications of these classical formulations. Although they differ in detail, most focus on improving accessibility: Frade and Ribeiro~\cite{Frade2015} maximized population coverage within a service threshold, Cintrano et al.~\cite{Cintrano2018} minimized average distance, Lin and Yang~\cite{Lin2011} incorporated distance-related costs, and Mete et al.~\cite{Mete2018} explored coverage-based models to balance worst-case distance and facility coverage. While these contributions demonstrate the importance of accessibility in station planning, the spatial arrangement of stations is still treated as a secondary outcome of the optimization rather than a planning objective in its own right.

A complementary dimension of spatial arrangement is proximity, reflecting how close stations are to one another. Accessibility and proximity are distinct dimensions: a tightly clustered set of stations at one edge of the service area may have high inter-station proximity but low accessibility for most users, whereas relocating the same cluster to a central area leaves proximity unchanged while improving accessibility. Considering proximity can be valuable for enabling short trips and facilitating bike rebalancing, yet most models don't explicitly optimize proximity but only impose minimum inter-station distances to avoid excessive clustering of stations~\cite{Nikiforiadis2021, Kabak2018}. In this line, concepts from network science such as farness~\cite{Sabidussi1966}, defined as the sum of shortest-path distances from one node to all others, provide a way to quantify spatial separation among stations within the cycling network. Adapting such measures offers a principled approach to evaluating inter-station spacing, complementing accessibility in capturing the overall structure of the station network.

Building on earlier studies that primarily optimized accessibility~\cite{Frade2015, Cintrano2018, Lin2011, Mete2018} or imposed minimum spacing rules to avoid redundant stations~\cite{Nikiforiadis2021, Kabak2018}, our framework evaluates both accessibility and proximity as complementary dimensions of spatial arrangement. Accessibility is measured by adapting the $p$-median formulation to compute the distance from each node to its nearest station, while proximity is captured by summing pairwise distances among selected stations. These metrics are then combined into a composite score with adjustable weights, making spatial arrangement trade-offs explicit. Moreover, this score works alongside the MCDM framework, where additional spatial factors can be added with chosen weights, so that results reflect broader planning goals while balancing accessibility and proximity.

\subsection{Algorithmic approaches}

GIS-based methods have been used to systematically integrate diverse spatial data~\cite{GarciaPalomares2012, Ebrahimi2022}. Combined with MCDM techniques such as AHP and MOORA, they allow factors to be weighted and aggregated into composite suitability scores or ranked lists of candidate sites~\cite{Kabak2018, Bahadori2022, Guler2021a, Guler2021b}. Applied on their own, however, GIS methods mainly support the visualisation and ranking of potential sites rather than computing a single optimal solution.

In contrast, optimization models frame station placement as a mathematical problem to maximize or minimize a defined objective under constraints. As described earlier, common formulations include the $p$-median, maximal covering, and $p$-center models. These problems are typically set up as integer programming models with decision variables indicating whether a station is placed at each candidate location~\cite{Frade2015, Lin2011, Mete2018, Cintrano2018}. Constraints often enforce budget limits (or a fixed number of stations), and additional constraints such as minimum spacing between stations and the coverage area of each station are sometimes incorporated in the optimizations~\cite{Conrow2018, Nikiforiadis2021, Caggiani2020}. These models are then solved for an optimal solution using optimization solvers such as CPLEX~\cite{Mete2018, Chou2019}, Gurobi~\cite{Mix2022},  LINGO~\cite{Lin2011}, or FICO Xpress-IVE~\cite{Conrow2018}. While exact solvers guarantee optimality, they can struggle as the candidate pool grows large or the objective function becomes more complex. To address this computational burden, some researchers have used the location-allocation tools in GIS software such as ArcGIS~\cite{GarciaPalomares2012, Ebrahimi2022}, which implement heuristics for $p$-median or covering problems and can quickly generate near-optimal solutions.

Given the computational burden of exact optimization methods, recent research has increasingly turned to metaheuristic algorithms. Many studies in diverse fields have adopted these methods to search for high-quality solutions within practical computational times~\cite{Tomar2023}. Widely used approaches include GA, Simulated Annealing (SA), Particle Swarm Optimization (PSO), and Ant Colony Optimization (ACO), all of which have demonstrated strong performance in large-scale combinatorial optimization. These methods have been successfully applied in various domains such as facility location~\cite{Chadry2003, Chalupa2019}, transport planning~\cite{Holliday2023, Sachan2024}, and network design~\cite{Gallo2010, Madadi2024}. Although these metaheuristic approaches were originally designed for single-objective problems, they have often been adapted to multi-objective settings by reformulating the problem into a series of single-objective ones~\cite{Rahimi2023}. Common strategies include weighted aggregation, where objectives are combined into a single fitness function with varying weights, and the $\varepsilon$-constraint method, where one objective is optimized while the others are imposed as constraints with different bounds. Such reformulations, however, usually require multiple runs to approximate the Pareto front and may fail to capture non-convex or discontinuous trade-off surfaces.

To overcome these limitations, dedicated multi-objective metaheuristics have been developed. These algorithms are designed to approximate the Pareto front in a single run by maintaining a diverse set of nondominated solutions, rather than collapsing multiple objectives into one. Well-known examples include NSGA-II (Nondominated Sorting Genetic Algorithm II)~\cite{Deb2002}, SPEA2 (Strength Pareto Evolutionary Algorithm 2)~\cite{Zitzler2001} and its extension SPEA2+\cite{Kim2004}, and swarm-based extensions such as MOPSO (Multi-Objective Particle Swarm Optimisation)~\cite{Coello2002} and MONACO (Multi-Objective Network Ant Colony Optimisation)~\cite{Cardoso2010}.

In the context of BSS, several studies have leveraged metaheuristic methods to optimize station placement. Cintrano et al.~\cite{Cintrano2018} evaluated SA, PSO, Iterated Local Search, and Variable Neighbourhood Search on a $p$-median formulation, finding that the GA consistently achieved the best solution quality. Qian et al.~\cite{Qian2022} applied a GA in Chicago to balance revenue and social equity objectives, demonstrating its flexibility for planning under competing priorities. Liu et al.~\cite{Liu2015} combined demand prediction using neural networks with a GA-based optimization model to select station sites that maximize usage and minimize rebalancing needs in New York City. Caggiani et al.~\cite{Caggiani2020} also implemented a GA, highlighting its suitability for equity-focused planning objectives. Finally, Askari et al.~\cite{Askari2017} proposed a hybrid GA–PSO approach to address a capacitated location–allocation problem under demand uncertainty, while Chen et al.~\cite{Chen2020} developed a hybrid PSO that incorporates GA operators to solve a mixed-integer programming formulation of e-bike station deployment in Ningbo, China, improving population coverage and accessibility.

Despite the variety of metaheuristic approaches, GAs have been especially prominent in BSS research~\cite{Cintrano2018, Liu2015, Caggiani2020, Qian2022}, where their population-based design and flexibility in operator definition make them well suited to the spatial encoding of station locations while also being able to handle multiple planning objectives and constraints. Comparative experiments further suggest that GAs can outperform some metaheuristic alternatives: Cintrano et al.~\cite{Cintrano2018}, for instance, reported that GA produced higher-quality solutions than SA, PSO, Iterated Local Search, and Variable Neighbourhood Search on a $p$-median formulation. On this basis, a GA algorithm was adopted in our framework due to: (i) its use in prior BSS optimization studies; (ii) its ability to incorporate custom, constraint-aware operators; (iii) its suitability for encoding solutions at the street-network level; and (iv) its stronger performance in preliminary tests.

In optimization literature, metaheuristic algorithms are typically validated by evaluating computational time, memory requirements, and the time needed to obtain a reasonably good solution, although the primary focus is often on the solution quality or an error measure~\cite{Osaba2021}. Moreover, because metaheuristics are stochastic, it is standard practice to perform multiple independent runs and report aggregate performance statistics (e.g., best, mean, worst solutions and standard deviations) to assess reliability and variability~\cite{Osaba2021}. The robustness of results across these runs and diverse problem instances is judged by the consistency, ensuring that the algorithm performs well under varying conditions. Furthermore, rigorous comparative studies frequently include statistical significance tests to confirm whether performance differences are meaningful~\cite{Derrac2011, Danka2013}.

Researchers commonly evaluate solution quality by comparing a metaheuristic’s best-found objective value against known optima or exact methods when available. Where problem size permits, benchmarking against exact solvers (e.g., MILP or IP) is a standard approach to compute optimality gaps ~\cite{Roshanaei2010}. For instance, in their review of the Travelling Salesman Problem, Alkhalifa et al.~\cite{Alkhalifa2025} describe this gap in terms of how far a solution is from the best possible one, typically reported as a percentage. For larger or more complex instances where exact solutions are impractical, comparisons are conducted against simpler heuristics (e.g., greedy or rule-based methods) or previously published metaheuristic results. In the case where the global optimum is known, several studies report that well-tuned metaheuristics can achieve near-optimal solutions within a few percent of the optimal value~\cite{Helsgaun2000, JohnsonMcGeoch2003, Schwinn2018}.

In contrast to most previous BSS optimization studies, which rarely benchmark metaheuristic solutions against exact methods or systematically compare them with alternative metaheuristics, our framework evaluates performance by quantifying optimality gaps relative to exact MILP solutions under simplified objectives. This would provide evidence that the proposed GA can deliver competitive solutions in these scenarios.

\section{Methodology}

This section begins with the presentation of the Barcelona case study and the description of the datasets employed in the analysis, which serve to illustrate the proposed optimization framework. The framework is then detailed, encompassing the mathematical formulation of the problem through a network-based representation with spatial constraints, followed by a two-step evaluation procedure composed of an MCDM-based scoring stage and a network-level adjustment that balances proximity among stations and overall system accessibility. Next, the topography module is introduced, which further refines distance calculations by accounting for slope effects, and the section concludes with the configuration of the GA used for the optimization.

    \subsection{Case study}
    
Barcelona, the capital of Catalonia in northeastern Spain, occupies approximately 101 km² and was home to roughly 1.7 million residents in 2024, making it one of Europe’s most densely populated cities. Its municipal boundary is delineated by two important geographic features: the Mediterranean Sea to the southeast and the Collserola mountain range to the northwest. These natural limits not only constrain urban expansion but also shape travel patterns, as coastal flatlands yield to the steep slopes of Collserola rising to nearly 450 m (Figure~\ref{fig:Barcelona_use_case}C). Administratively, Barcelona is divided into 10 districts and 73 neighborhoods (Figure~\ref{fig:Barcelona_use_case}A), which display marked socioeconomic contrasts (Figure~\ref{fig:Barcelona_use_case}B and D). Districts such as Sarrià–Sant Gervasi and Les Corts stand out with the highest average incomes, while Nou Barris and parts of Ciutat Vella record much lower income levels. The Eixample district, although more socioeconomically mixed, concentrates the largest number of amenities and points of interest, making it a major hub of urban activity. 

Over the past two decades, Barcelona has actively pursued policies to promote sustainable and active transport. Initiatives such as the superblock program, low-emission zones, and the expansion of the cycling network reflect this agenda. Today, the city offers more than 260 km of dedicated bike lanes and nearly 2,000 km of bike-friendly streets (Figure~\ref{fig:Barcelona_use_case}C), complemented by eight metro lines, six tram lines, and over 100 bus routes with more than 8,000 stops. Beyond the city limits, rail networks (FGC and Rodalies) and interurban buses extend connectivity to other municipalities.

How residents make use of this infrastructure is reflected in the 2024 survey \textit{Enquesta de Mobilitat en Dia Feiner} (EMEF). According to this survey~\cite{EMEF2024}, Barcelona residents made 5.75 million weekday trips in 2024, excluding intensive work-related travel, which equals an average of 3.9 trips per person per day. Overall, active mobility is the leading mode, followed by public transport and private vehicles (Table~\ref{tab:modal_split}). However, patterns differ depending on the type of trip. For internal trips within Barcelona, active mobility dominates, whereas for trips connecting Barcelona with surrounding municipalities, public transport is the main mode. Within active mobility, the vast majority of trips are made on foot, while cycling and personal mobility vehicles (PMV) (e.g., e-scooters, segways, or similar devices) together account for less than 4\% of trips.

\begin{table}[hpbt!]
    \centering
    \caption{\textbf{Modal split of trips in Barcelona based on EMEF 2024 survey.}}
    \label{tab:modal_split}
    \begin{tabular}{lccc}
        \hline
        & \textbf{Total (\%)} & \textbf{Internal (\%)} & \textbf{Connections (\%)} \\
        \hline
        Active mobility & 54.5 & 60 & 5 \\
        \quad Walking & 51.3 & -- & -- \\
        \quad Bicycle & 2.4 & -- & -- \\
        \quad PMV (e-scooters, Segways, others) & 0.8 & -- & -- \\
        Public transport & 28.8 & 28 & 53 \\
        Private vehicles & 16.7 & 12 & 42 \\
        \hline
    \end{tabular}
\end{table}

Within this landscape, \textit{Bicing}--the city's public BSS--plays a complementary role. Launched in 2007, it rapidly expanded to more than 400 stations in 2008. A major renewal in 2019 introduced e-bikes and raised the fleet to about 6,700 units, with e-bikes representing 15\% of the total. Moreover, the number of stations was once more increased to over 500 (Figure~\ref{fig:Barcelona_use_case}C). Since then, around 2,000 m-bikes have been converted to e-bikes and additional units have been added, so that by December 2022 e-bikes made up to 47\% of a fleet of nearly 7,000 bicycles. Building on this trend, the system is now undergoing a further expansion, with 30 new stations added between 2024 and 2025, bringing the total to 557. The fleet has also grown to nearly 8,000 bicycles, of which about 5,000 are electric. This steady growth in supply has been accompanied by growing demand: by the end of 2023, the system had approximately 147,000 subscribers and recorded 18 million trips, according to Barcelona's City Council~\cite{AjuntamentBarcelona2024_Bicing}. Empirical studies further reveal distinct usage patterns~\cite{Grau-Escolano2024}: e-bikes are more intensively used in hilly districts near Collserola, whereas m-bikes dominate in the city’s flatter coastal areas.

Taken together, these features make Barcelona a compelling case study. The city combines compactness and high population density with marked topographic contrasts between flat coastal areas and hilly districts, as well as pronounced socioeconomic diversity. Although active mobility is widespread, cycling remains marginal at only 2.4\% of trips, highlighting clear potential for growth. This aligns with the city’s Urban Mobility Plan 2025–2030, which sets the goal of increasing the combined share of cycling and personal mobility vehicles by nearly 50\% by 2030~\cite{PMU2025}. Coupled with the extensive open data resources, these conditions position Barcelona as a highly suitable real-world environment for evaluating the proposed framework.

    \subsection{Data}
\begin{figure}[!htbp]
    \centering
    \includegraphics[width=0.65\linewidth]{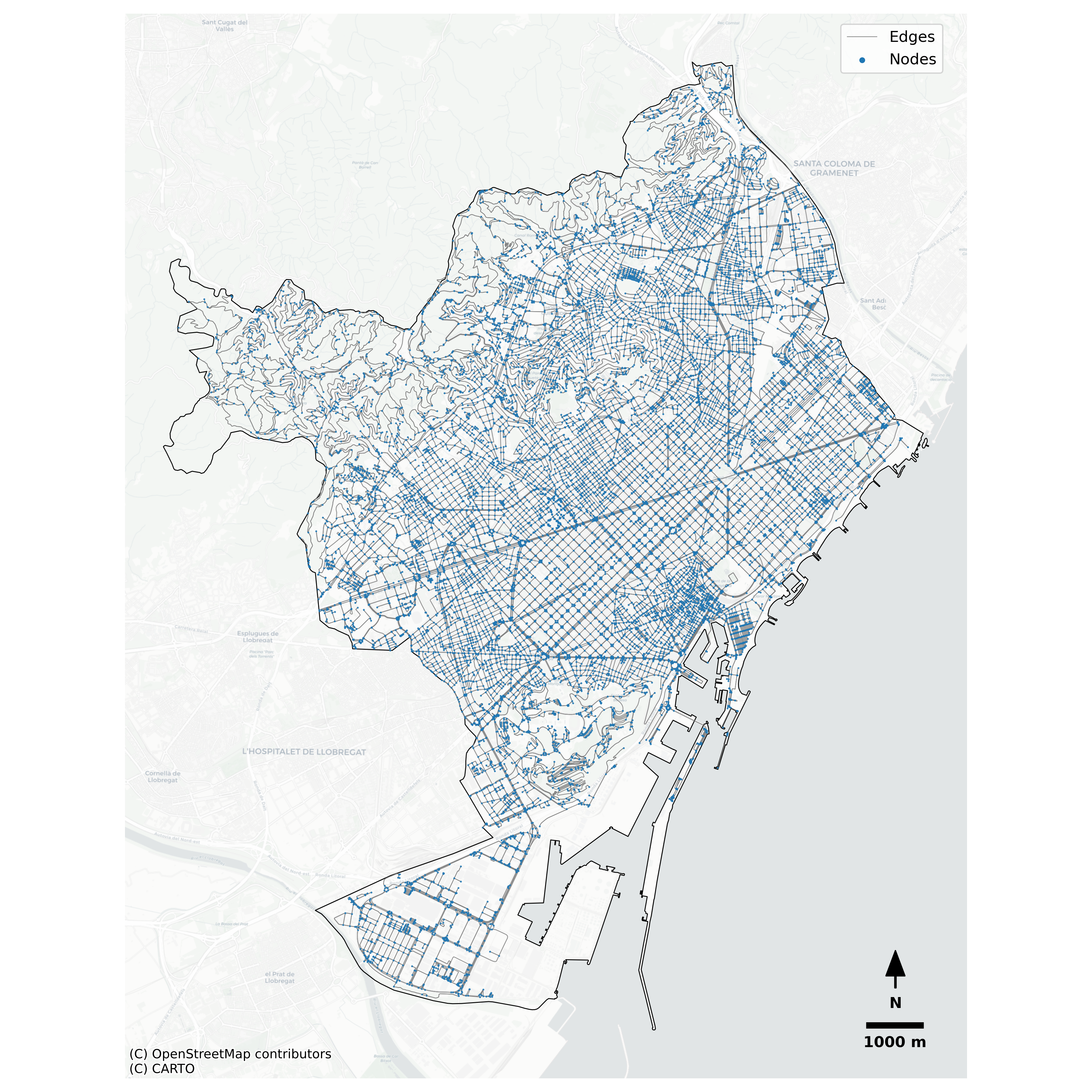}
    \caption{\textbf{Barcelona’s OSM bike network after cleaning}. See Appendix~\ref{app:graph_cleaning} for details on the cleaning procedure.}
    \label{fig:osm_network}
\end{figure}

    This study draws on multiple open data sources available for Barcelona, including the Instituto Nacional de Estadística (INE)~\cite{INE}, OpenStreetMap (OSM)~\cite{OpenStreetMap_web}, Open Data Barcelona (ODB)~\cite{ODB}, and OpenTopoData (OTD)~\cite{OTD}, which provide the following datasets:

\begin{itemize}

\item \textbf{Bicycle‐accessible street network:} the bicycle‐accessible street network of Barcelona was obtained by querying OSM via OSMnx~\cite{OSMNx2025} with the network type ‘bike’, yielding a directed graph that comprises approximately 18,700 nodes and 38,000 edges. The graph is formally defined as:
    
    $$
        G = (V, E)
    $$
    
    where $V$ is the set of nodes, with each node $v \in V$ representing street intersections or endpoints; and $E$ is the set of directed edges, where each edge $e \in E$ corresponds to a street link connecting two nodes (Table~\ref{table:notation}). The procedure used to clean and unify this directed graph, removing isolated components and ensuring full bidirectional reachability, is described in Appendix~\ref{app:graph_cleaning}. The resulting, fully connected bicycle network is shown in Figure~\ref{fig:osm_network}.

    \item \textbf{Locally relevant variables:} alongside the bicycle network, a set of locally relevant variables was collected to characterize each node \(i\in V\). These variables come from the open‐data providers and fall into four categories: socioeconomic (population and income from INE, 2022; education, nationality, household size, vehicle ownership, and unemployment from ODB, 2022–2024), public transport (metro, tram, and bus stops from OSM, accessed 2024), built environment (bike lanes and POIs from OSM, accessed 2024), and topographic (elevation values from OTD, accessed 2024). Table~\ref{table:local_factors_nodes} summarizes each variable’s preprocessing, and full source details appear in Appendix~\ref{app:data_sources}.

\end{itemize}

\begin{table}[!htbp]
  \centering
  \caption{\textbf{Notation for the BSS optimization model.}}
  \label{table:notation}
  \footnotesize
  \begin{tabular}{p{0.15\linewidth}p{0.20\linewidth}p{0.60\linewidth}}
    \toprule
    \textbf{Category} & \textbf{Symbol} & \textbf{Definition} \\
    \midrule
    \multirow{3}{*}[0.5em]{\textbf{Graph}} 
      & $G=(V,E)$ & Directed bicycle network with a set of nodes $V$ and a set of edges $E$ \\
      & $d(i,j)$ & Shortest-path distance between nodes $i,j$ \\
    \midrule
    \multirow{5}{*}[0.5em]{\textbf{Stations}} 
      & $S \subset V$ & Set of selected station locations \\
      & $k = |S|$ & Total number of stations \\
      & $x_i \in \{0,1\}$ & Binary decision variable: 1 if node $i$ selected as a station, 0 otherwise \\
      & $X$ & Minimum inter-station distance constraint \\
      & $Y$ & Service coverage radius of the stations \\
    \midrule
    \multirow{4}{*}[0.5em]{\textbf{Local utility}} 
      & $F$ & Total number of local factors \\
      & $V_{\mathrm{norm},f}^i$ & Normalized value of factor $f$ at node $i \in [0,1]$ \\
      & $W_f$ & Weight assigned to factor $f$, with $\sum_{f=1}^F W_f = 1$ \\
      & $U_i$ & Node utility:  \\
    \midrule
    \multirow{4}{*}[0.5em]{\textbf{Network metrics}} 
      & $S_{\mathrm{prox}}$ and $S_{\mathrm{acc}}$ & Inverted normalized proximity and accessibility scores, respectively \\
      & $\alpha \in [0,1]$ & Trade-off parameter balancing proximity and accessibility \\
      & $CS$ & Combined network score: $CS = \alpha \cdot S_{\mathrm{prox}} + (1-\alpha) \cdot S_{\mathrm{acc}}$ \\
    \midrule
    \textbf{Objective} & $U^*_{\mathrm{BSS}}$ & Total BSS utility: $U^*_{\mathrm{BSS}} = \left(\sum_{i \in S} U_i\right) \times CS$ \\
    \bottomrule
  \end{tabular}
\end{table}

    \subsection{Optimization problem}
    Building on the data described above, this study formulates BSS planning as an optimization problem in which station locations are chosen to maximize the system’s total utility (Eq.~\ref{eq:full_formula}). The framework is flexible in that it can incorporate a wide range of spatial factors, adjust weighting schemes, and prioritize proximity among stations or accessibility of the system to reflect different planning priorities.

The aim of the optimization is to select a set $ S \subset V $ of $ k $ nodes from the OSM bike graph to serve as BSS stations. To do this, each node $i \in V$ is associated with a binary decision variable $x_i$, where $x_i = 1$ if node $i$ is selected as station and $x_i = 0$ otherwise. The only constraint in this optimization is spatial (Eq.~\ref{eq:spatial_constraint}): any two selected stations $ i $ and $ j $ must be separated by at least $X$ meters apart considering the paths of the directed bike network: 

\begin{equation}\label{eq:spatial_constraint}
    d(i,j) \geq X, \quad \text{if } x_i = 1 \text{ and } x_j = 1
\end{equation}

where $ d(i,j) $ is the shortest‐path distance. Minimum inter-station separation distances commonly adopted in the BSS literature range from 250\,m to 500\,m \cite{Kabak2018, Mateo_Babiano2016, Croci2014}; in this study, a fixed value of  $X$ = 300\,m is adopted to illustrate the presented framework, which can accommodate any alternative value of \(X\) as required. Furthermore, the basic path‐length metric \(d(i,j)\) can be substituted with an elevation‐adjusted “equivalent flat distance”  $d_{\mathrm{eq}}(i,j)$, thereby accounting for uphill and downhill trip effects (see Section~\ref{sec:slopes}). 

Subject to the spatial‐separation constraint above, this work proposes to optimize station locations to maximize overall system utility, defined by two complementary components: the suitability of individual candidate nodes and the structural performance of the network in terms of system compactness and accessibility. These components can sometimes conflict: a candidate location that scores highly on local factors (e.g. locations with high population density or public transport connectivity) may not improve network compactness or accessibility, while a location that contributes to a more compact or accessible network may have lower local suitability. To reconcile these considerations, the framework integrates both perspectives through a two-step approach:

\begin{enumerate}
  \item \textbf{Station‐level scoring:} assign each node \(i\) a utility \(U_i\) via a weighted sum of normalized local factors chosen based on the system's planning objectives (e.g. population, number of POIs, income, etc.). Each factor is measured within a configurable circular buffer of radius $Y$ around the node, so that the score reflects the surrounding conditions of the potential station location (see Section~\ref{sec:section_nodes_utility}).

  \item \textbf{Network‐level adjustment:} combine the sum of selected node utilities with a composite network metric that balances station proximity and system accessibility according to a user-defined parameter $\alpha$ (see Section~\ref{sec:section_network_utility}).
\end{enumerate}

This two‐step procedure yields a utility score for any station set $S$, but does not itself solve the placement problem. Given the combinatorial nature of the placement problem, a GA is employed to identify high-utility BSS station sets (Section \ref{sec:GA}).

    \subsection{Scoring framework}
    \subsubsection{Nodes utility} \label{sec:section_nodes_utility}

The utility score \(U_i\) for each candidate node \(i\) is computed as a spatial weighted sum over \(F\) factors:

\begin{equation} \label{eq:node_utility}
    U_i \;=\;\sum_{f=1}^F W_f\,V_{\mathrm{norm},f}^i,
\end{equation}

where \(V_{\mathrm{norm},f}^i\) represents the normalized value of factor \(f\) within a buffer of radius \(Y\) meters around node \(i\), and \(W_f\) is the user-defined weight for that factor. By adjusting the weights \(W_f\), planners can shift emphasis among competing objectives, while the buffer radius \(Y\) determines the spatial extent of each node’s local context.

Table~\ref{table:local_factors_nodes} enumerates the local factors considered in this study. These are meant as examples of factors commonly considered in BSS planning rather than a fixed set: any spatial attribute such as land‐use mix, crime rates, ridership, or parking availability can be incorporated by first measuring its value within a specified buffer around each candidate node and then assigning a weight to each spatial attribute. In this study, absolute totals are used (e.g., the number of residents or the count of amenities within the buffer), so that higher underlying values translate directly into higher node scores. Relative indicators, such as per-capita or per-area measures, could also be incorporated, but they express values in relation to a denominator rather than absolute magnitude, which can lead to different interpretations across planning scenarios. Regardless of the choice, all variables are normalized to ensure comparability. Details on the normalization procedures and before/after distributions for these variables are provided in Appendix~\ref{app:normalization}. Once the normalized scores have been computed, the BSS utility for a given set of stations is determined by the following formula:

\begin{equation}\label{eq:BSS_utility_no_metrics}
        U_{BSS} = \max \sum_{i\in V} x_i \cdot U_i
\end{equation}

where $V$ is the set of candidate nodes, $x_i$ is the binary decision variable for node $i$ (1 if selected as a station, 0 otherwise), and $U_i$ is the utility score of node $i$.

\begin{table}[!htbp]
    \centering
    \caption{\textbf{Local factors considered for BSS station placement}. Data sources are detailed in Appendix~\ref{app:data_sources}.}
    \label{table:local_factors_nodes}
    \scriptsize
    \renewcommand{\arraystretch}{1.2}
    \begin{tabular}{p{0.15\linewidth} p{0.12\linewidth} p{0.73\linewidth}}
        \toprule
        \textbf{Category} & \textbf{Variable} & \textbf{Preprocessing and aggregation within radius \(Y\)} \\
        \midrule
         Socioeconomics & Population 
            & Section-level counts apportioned to residential building footprints (area-share, see Appendix~\ref{app:pop_distribution}); building-level counts summed within \(Y\), with partial overlaps weighted by intersecting area.\\
          & Sex 
            & Male and female counts per section allocated to buildings and aggregated to nodes using the same method as population. \\
          & Age 
            & Counts in 10-year bins (10–19, …, 70+) per section allocated to buildings and aggregated to nodes using the same method as population. \\
          & Immigration 
            & Non-Spanish citizen counts per section redistributed to buildings and aggregated to nodes using the same method as population. \\
          & Education 
            & Counts by attainment level (primary, secondary, tertiary) per section redistributed to buildings and aggregated to nodes using the same method as population. \\
          & Unemployment 
            & Neighborhood-level unemployment rate applied to section population aged 19–69; resulting counts redistributed to buildings and aggregated to nodes as population. \\
          & Car ownership 
            & Motorization index (vehicles per 1 000) converted to absolute counts (index × section population / 1 000), redistributed to buildings and aggregated as population. \\
          & Household size 
            & Average dwelling size (m²) per section aggregated to each node via area-weighted mean over intersecting sections. \\
          & Income 
            & Average net income per capita per section aggregated to each node via area-weighted mean over intersecting sections. \\
        \midrule
        
        Public transport  & Metro lines 
            & Number of distinct metro lines whose entrances lie within \(Y\) of each node. \\
          & Tram lines 
            & Number of distinct tram lines within \(Y\). \\
          & Bus stops 
            & Number of bus stops (by line) within \(Y\). \\
        \midrule
        
        Built environment & Bike lanes 
            & Total length of OSM “cycleway” segments within \(Y\). \\
          & POI count 
            & Number of POIs within \(Y\) considering the categories: healthcare, culture, tourism, recreation, sport, economic and retail, industrial, green, civic, worship, and education, based on a 15-minute city review~\cite{PAPADOPOULOS2023104875}. Additionally, an overall count of all POIs is also considered.\\
          & POI diversity 
            & Shannon entropy of POI categories within \(Y\) (zero when no POIs present). \\
        \bottomrule
    \end{tabular}
\end{table}

\subsubsection{Network utility} \label{sec:section_network_utility}

After computing the node's local utility for a candidate set $S$ (Eq.~\ref{eq:BSS_utility_no_metrics}), the BSS score is complemented by an evaluation of the BSS network structure. This evaluation must account for both the proximity among the chosen stations and the coverage of the BSS service area. These two aspects capture different characteristics of the station placement, as a configuration that maximizes station dispersion does not necessarily ensure optimal coverage, and vice versa. A composite metric is constructed by combining the two individual metrics to provide a unified measure of overall network spatial arrangement:

\begin{itemize}
  \item \textbf{Proximity metric} measures how close the stations in $S$ are, building on the concept of farness~\cite{Sabidussi1966}, which sums the shortest-path distances from one node to all others, and is thus closely related to the reciprocal of the unnormalized closeness centrality. Farness quantifies how far a node is from all other nodes in a network. Here, this idea is adapted to evaluate the spatial distribution of BSS stations by computing the total pairwise farness within the selected stations:

    \begin{equation}
        Pro = \sum_{{s,t} \subset S} d(s,t)
    \end{equation}

  where $d(s,t)$ represents the shortest-path distance between stations $s$ and $t$, computed using the edges of the directed graph $G$. A larger $Pro$ value indicates that the selected stations are more dispersed across the network.

  \item \textbf{Accessibility metric} measures how well the stations provide accessibility across the entire network. It is based on the $p$-median problem~\cite{Daskin2015, Revelle2008}, a well-known facility location model that seeks to minimize the sum of distances between demand points and their nearest facility. The metric is computed as the total shortest-path distance from each node $v \in V$ to its nearest station $s \in S$:

      \begin{equation}
          d(v, S) = \min_{s \in S} d(v,s),
      \end{equation}

  Then, the accessibility metric is defined as:
    
    \begin{equation}
          Acc = \sum_{v \in V} d(v, S)
      \end{equation}

  Lower values of $Acc$ indicate higher accessibility, meaning that, on average, stations are better positioned to minimize travel distance from any node of $G$.

\end{itemize}

Because both $Pro$ and $Acc$ depend on network size and station count, they are first min–max scaled to $[0,1]$ for comparability, and then inverted so that higher values denote higher proximity and higher access, as shown in Eq.~\ref{eq:normalization}. The resulting normalized inverted metrics are denoted as $S_{prox}$ and $S_{acc}$.

\begin{equation} \label{eq:normalization}
        M_{\mathrm{inv}} = 1 - \frac{M - M_{\min}}{M_{\max}-M_{\min}}.
\end{equation}

Since finding the truly minimum or maximum values of the proximity metric is NP-hard~\cite{Hassin1997, Ravi1994} (it requires choosing a subset of nodes to optimize the sum of pairwise shortest‐path distances), its normalization bounds are approximated using heuristics. Similarly, for the accessibility metric, which is based on the \textit{p}-median problem, determining its optimal values is also NP-hard~\cite{Kariv2006, Hakimi1965}, necessitating the use of heuristic approximations for its bounds. The approximate minimum and maximum values are computed as follows (Figure~\ref{fig:graph_metric_bounds}):

\begin{figure}[!htbp]
    \centering
    \includegraphics[width=0.8\linewidth]{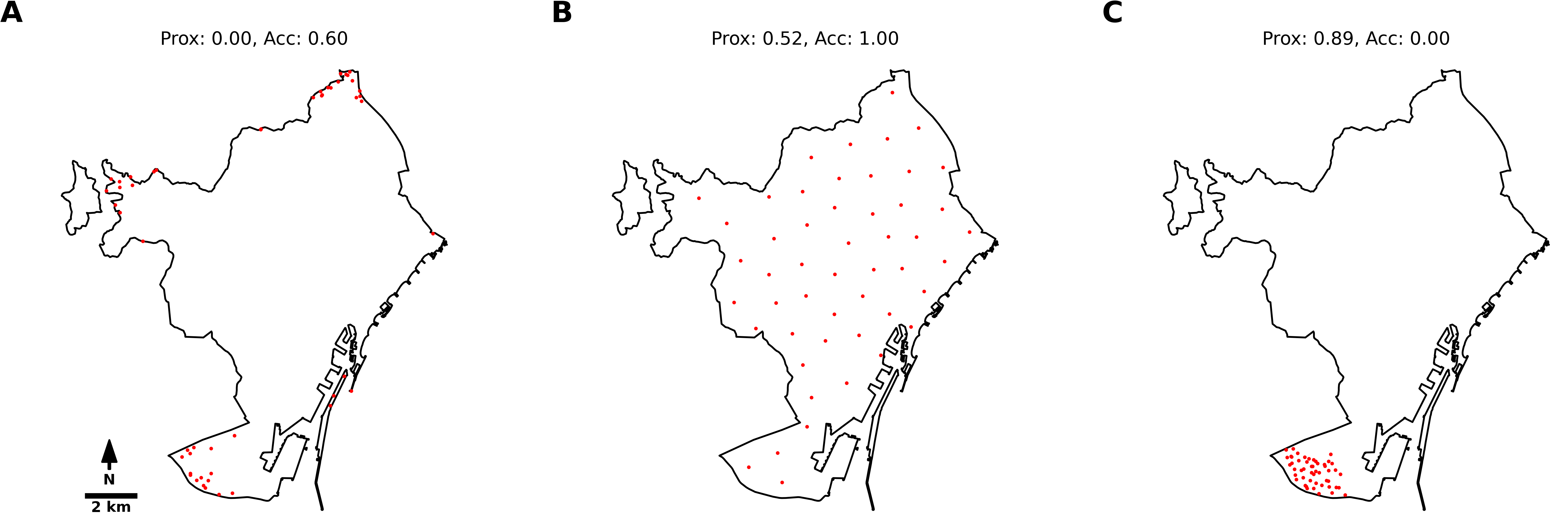}
    \caption{\textbf{Normalization bounds for proximity and accessibility metrics on Barcelona’s bike network with \(k=50\) stations}. Stations are shown in red, and the city boundary is outlined in grey.
    \textbf{(A)} Maximum proximity (\(\mathrm{Pro}_{\max}\)), representing the worst‐case proximity of stations.  
\textbf{(B)} Minimum accessibility (\(\mathrm{Acc}_{\min}\)), representing the best‐case coverage of all nodes.  
\textbf{(C)} Maximum accessibility (\(\mathrm{Acc}_{\max}\)), representing the worst‐case coverage.  
   Above each subplot, the inverse‐normalized value of the corresponding metric is shown. Note that the proximity lower bound (Eq.~\ref{eq:pro_min}) is agnostic to the city graph and hence omitted.}
    \label{fig:graph_metric_bounds}
\end{figure}

\begin{itemize}
  \item \textbf{Minimum proximity distance}: this is the most compact configuration, where each pair of $k$ stations is exactly $X$\,m apart:
      
  \begin{equation}
      \mathrm{Pro}_{min} =  \binom{k}{2}X.
        \label{eq:pro_min}
      \end{equation}

  \item \textbf{Maximum proximity distance}: this corresponds to the most dispersed BSS, approximated via a \textit{farthest‐first} heuristic. First, the node with the greatest eccentricity is  chosen as the initial station,
    
    \begin{equation}
      s_{1} = \arg\max_{v\in V}\;\max_{u\in V} d(v,u),
      \label{eq:seed}
    \end{equation}
    ensuring a peripheral starting point. Then, subsequent stations are added one by one by selecting the node whose minimum distance to the existing set is largest:
      \begin{equation}
        s_i = \arg\max_{v \notin \{s_1,\dots,s_{i-1}\}} \min_{s \in \{s_1,\dots,s_{i-1}\}} d(v,s).
        \label{eq:farthest}
      \end{equation}
      The resulting total pairwise distance gives the upper bound for proximity:
      \begin{equation}
        \mathrm{Pro}_{max} = \sum_{\{s,t\} \subset S} d(s,t).
        \label{eq:pro_max}
      \end{equation}

  \item \textbf{Minimum accessibility distance:} this corresponds to  the scenario in which all nodes in \(V\) are as close as possible to a station. It is approximated via \(k\)-means clustering on node coordinates. A station is placed at the node closest to each cluster centroid forming the station set $S_{\mathrm{acc,min}}$. The bound is:

    \begin{equation}
        \mathrm{Acc}_{min} \approx \sum_{v\in V} \min_{s \in S_{\mathrm{acc,min}}} d(v,s).
        \label{eq:acc_min}
      \end{equation}

  \item \textbf{Maximum accessibility distance:} this corresponds to the scenario in which all nodes are as far as possible from their nearest station. Using the same definition in Eq.~\ref{eq:acc_min}, the bound is evaluated with $S_{\mathrm{acc,max}}$ instead of $S_{\mathrm{acc,min}}$. In this case, stations are concentrated in a peripheral area of the network. Starting from a peripheral seed node (Eq.~\ref{eq:seed}), subsequent stations are iteratively placed at the closest available nodes while respecting the minimum separation constraint.
\end{itemize}

To assess both network proximity and accessibility at the same time, the normalized inverted scores of both are combined into a single score defined as:
\begin{equation}
    CS(\alpha) = \alpha S_{\text{pro}} + (1 - \alpha) S_{\text{acc}}
\end{equation}
where $\alpha \in [0,1]$ controls the weighting between proximity and accessibility, $S_{\text{pro}}$ is the inverted normalized $Pro$, and $S_{\text{acc}}$ the inverted normalized $Acc$. A value of $\alpha = 0$ prioritizes accessibility, while $\alpha = 1$ focuses solely on station proximity, and intermediate values allow planners to balance these two objectives according to their priorities.

Based on this, the overall BSS utility is obtained by extending Eq.~\ref{eq:BSS_utility_no_metrics} to combine individual station scores with network performance: 
\begin{equation} \label{eq:full_formula}
  U_{\text{BSS}}^* = U_{\text{BSS}} \times \left( \alpha \cdot S_{\text{pro}} + (1 - \alpha) \cdot S_{\text{acc}} \right)
  \end{equation}
where $U_{\text{BSS}}$ is the sum of individual station scores, ensuring that the final utility reflects both geographical coverage and network proximity, as balanced by $\alpha$.

    \subsection{Distance adjustment based on topography}
    \label{sec:slopes}
    Beyond the baseline formulation, it is important to consider how topography influences cycling effort. Since shortest-path distances $d(i,j)$ play a central role in the optimization, they should capture not only geometric length but also the effort associated with elevation changes. Uphill segments demand greater exertion, while downhill sections may ease travel but, in the case of steep declines, can also force cyclists to brake to avoid accidents. To represent these effects, the framework allows the use of an elevation-adjusted ‘equivalent flat distance’ $d_{\mathrm{eq}}(i,j)$. In this formulation, uphill segments are penalised and downhill segments rewarded, yielding distances that more closely approximate the perceived effort of cycling, particularly for m-bikes.

The equivalent flat distance for a single edge \(\ell\) of horizontal length \(d_{\ell}\) and elevation change \(\Delta h_{\ell}\) is computed in the following steps. First, the slope of the edge \(\ell\) is computed and decomposed into uphill and downhill components in one step:
    \[
    g_{+,\ell} = \max\!\bigl(g_{\ell},0\bigr), 
    \quad
    g_{-,\ell} = \min\!\bigl(g_{\ell},0\bigr),
    \quad\text{where}\quad
    g_{\ell} = \frac{\Delta h_{\ell}}{d_{\ell}}.
    \]
To align with standard cycling guidelines, which recommend a maximum slope of 10\%, \(g_{\ell}\) is then clamped to the interval $[-0.10,0.10]$ \cite{AASHTO1999bicycle, CycleHighways2025, CROW2006}. Applying Parkin \& Rotheram’s linear slope–speed relation \cite{PARKIN2010335}, the segment speed becomes  
    \[
    v(g_{\ell})=v_{0}+b_{\mathrm{up}}\,g_{+,\ell}+b_{\mathrm{down}}\,g_{-,\ell},
    \]
with \(v_{0}=6.01\) m/s, \(b_{\mathrm{up}}=-40.02\), and \(b_{\mathrm{down}}=-23.79\) (m/s per unit slope). The resulting travel time \(t_{\ell}=d_{\ell}/v(g_{\ell})\) is then converted back into distance at flat‐ground speed via \(d_{\mathrm{eq},\ell}=t_{\ell}\,v_{0}\). Summing these adjusted lengths over the shortest‐path \(P_{ij}\) between nodes \(i\) and \(j\) yields  
    \[
    d_{\mathrm{eq}}(i,j)=\sum_{\ell\in P_{ij}}d_{\mathrm{eq},\ell},
    \]
which may replace the original \(d(i,j)\) in the spatial constraint whenever slope effects are deemed relevant. 

Thus, the procedure provides an adjusted distance matrix where each shortest path $d_{eq}(i,j)$ reflects slope effects. This measure can directly replace the original distances $d(i,j)$ in the spatial constraint, penalizing uphill segments and rewarding downhill ones.

    \subsection{Optimization modelling}
    \label{sec:GA}
    In this study, a GA is employed to optimize the placement of BSS stations. GAs are population-based search procedures inspired by natural selection and genetics, well suited to exploring large combinatorial spaces and handling multiple objectives and constraints without requiring differentiable or convex formulations. These properties make them appropriate for station location problems, where the search space is discrete, constraints such as minimum station spacing must be enforced, and solutions must adapt to different planning objectives while retaining the same underlying spatial structure. A GA evolves a population of fixed-length ‘chromosome’ candidates, each encoding all decision variables, through the operators of \textit{selection}, which favors higher-fitness individuals; \textit{crossover}, which splices gene segments from two parents to create offspring; and \textit{mutation}, which randomly alters individual genes to maintain diversity and avoid premature convergence. This evolutionary cycle continues until a stopping criterion such as a maximum number of generations or a convergence threshold is met. For a comprehensive overview of GA concepts and variants, the reader is referred to Katoch et al.~\cite{Katoch2021}.

Our algorithm follows a single-objective, non-adaptive GA with elitism. Each chromosome is a binary vector over all candidate sites (‘station’ versus ‘no station’), encoding a complete BSS deployment plan $S$ with $k$ stations. The initial population of $N$ candidate plans is generated by constrained random sampling: a first station is selected at random, and subsequent stations are added only if they lie at or beyond the prescribed minimum distance $X$ from all previously chosen sites, until $k$ stations are placed. The population then evolves through selection, crossover, and mutation, while elitism ensures that a fixed fraction of the best solutions is preserved across generations. 

This design was chosen for its robustness and compatibility with custom, constraint-aware operators. Although other variants such as steady-state, adaptive, or multi-objective GA variants exist, the chosen formulation proved sufficient to capture the spatial structure of the problem and to generalize across diverse planning scenarios. As demonstrated in Section~\ref{GA_vs_MILP}, this consistency allows a single hyper-parameter optimization process to be reused across all experimental configurations.

\subsubsection{Hyper-parameter tuning}

To ensure adaptability across diverse planning objectives, a comprehensive hyper-parameter search was conducted on a representative scenario (\(k=60\) with a randomly generated local-factor weight vector), hypothesizing that these hyper-parameter settings would generalize to other station counts and weight combinations (see Section~\ref{GA_vs_MILP}). Each hyper-parameter configuration was assessed in terms of solution quality and population diversity over time. The main components of the algorithm, along with the alternative configurations explored, are detailed below and in Table~\ref{table:ga_grid}. 

{%
  \setlength{\parskip}{0.45\parskip}%
\begin{itemize}
\item \textbf{Population size.} The algorithm starts with $N$ solutions, each comprising $k$ station locations that satisfy $X$. A larger $N$ enhances diversity, giving the algorithm a broader view of the solution space and increasing the chances of escaping local optima, but it also increases computational time.

\item \textbf{Selection strategy.} Determines how parent solutions are selected for reproduction:
\begin{itemize}
    \item \textit{Roulette-wheel:} individuals are selected with a probability proportional to their fitness score. High-performing solutions are more likely to be chosen, but lower-quality ones retain a non-zero chance.
    \item \textit{Tournament:} a small subset of individuals is randomly drawn from the population, and the one with the highest fitness within this group is selected.  This approach introduces selection pressure while still allowing less-fit individuals a chance to be chosen.

\end{itemize}

\item \textbf{Crossover strategy.} Controls how offspring are generated from parent solutions: 
\begin{itemize}
    \item \textit{Greedy:} candidate nodes from both parents are ranked by score, and the best ones are iteratively added while respecting spatial constraints. 
    \item \textit{Top-first:} the top 25\% of nodes by score are shuffled and selected first, with remaining slots filled by random picks from the other 75\%.
    \item \textit{Weighted-random:} candidates are sampled with probabilities proportional to their scores, balancing exploration and exploitation. 
\end{itemize}

\item \textbf{Mutation strategy.} each node of non-elite offspring has a probability of mutation, capped at 20\% of nodes. Node replacements must satisfy the distance constraint \(X\).

\item \textbf{Elitism.} A fixed proportion of the best-performing solutions is directly transferred to the next generation. This guarantees that high-quality individuals are preserved and not lost through random variation.

\item \textbf{Early stopping.} To avoid unnecessary computation, the algorithm is terminated if no improvement is observed in either the best or average fitness score after 300 consecutive generations.
\end{itemize}
}%

\begin{table}[!htbp]
  \centering
  \caption{\textbf{GA grid search hyper-parameter values}. Combinations of hyperparameters that were not feasible due to insufficient numbers in the different GA operators were skipped.}
  \label{table:ga_grid}
  \footnotesize
    \begin{tabular}{ll}
    \toprule
    \textbf{Hyperparameter} & \textbf{Values tested} \\
    \midrule
    Population size & 25, 50, 100, 200 \\
    Mutation rate & 0.01, 0.02, 0.05, 0.1, 0.2 \\
    Elite fraction & 0.01, 0.02, 0.05, 0.1, 0.2 \\
    Selection strategy & Tournament, Roulette-wheel \\
    Crossover strategy & Greedy, Weighted-random, Top-first \\
    Max generations & 10,000 (with early stopping)\\
    \bottomrule
    \end{tabular}
\end{table}

The best‐performing set comprised tournament selection, greedy crossover, \(N=50\), mutation rate 0.05, and elite fraction 0.05.

\subsubsection{GA accuracy bench-marking}
\label{GA_vs_MILP}

Although the GA hyper-parameters were tuned on a single scenario, they are expected to generalize across planning scenarios, as the network topology remains fixed. To validate this hypothesis, GA solution utilities have been compared against the exact optima obtained from a Mixed Integer Linear Programming (MILP) model. GAs are meta-heuristic and cannot guarantee global optimality, whereas MILP solvers exhaustively search the binary decision space for linear objectives and certify the optimal solution, making them ideal benchmarks.

Because the full BSS utility combines local-factor and network-structure metrics in a non-linear objective, it cannot be solved exactly by MILP. Instead, the MILP benchmark optimizes only the linear component
\[
f_{\mathrm{MILP}} = \max_{x}\;\sum_{i\in V} x_i\,U_i
\]
subject to  
\[
\sum_{i\in V} x_i = k,\quad
x_i + x_j \le 1\;\;\forall\,(i,j)\in\mathcal{E},\quad
x_i\in\{0,1\},
\]
where \(V\) is the set of candidate nodes, \(U_i\) the local-factor utility of node \(i\), \(k\) the prescribed number of stations, and \(\mathcal{E}=\{(i,j)\mid d(i,j)<X\}\) enforces the minimum inter-station distance \(X\). This linear formulation preserves the same stations' deployment space while substantially reducing computational complexity and avoiding approximation errors required to linearize non-linear terms. It should be emphasized, however, that this MILP benchmark is based on a reduced formulation of the problem that excludes the non-linear component (i.e., incorporating the network-structure metrics into $U_{\mathrm{BSS}}$). As such, it provides only a partial measure of GA optimality. To enable a fair comparison, GA runs were also applied to obtain $U_{\mathrm{BSS}}$. GA robustness was assessed using two experiments: 

\begin{enumerate}
  \item \textbf{Weight sensitivity} (fixed $k=60$).
    A total of 118 weight vectors for the local factors were generated by Monte Carlo sampling. A sample size of 118 was chosen based on a preliminary power analysis with 20 samples to ensure that a one-sided test at the 0.5\% gap threshold would achieve at least 90\% power with \(\alpha=0.05\). For each weight vector, the MILP solved the linear model to optimality and the GA was executed with the tuned hyper-parameters, enabling paired comparisons of \(U_{\mathrm{BSS,GA} }\) and \(U_{\mathrm{BSS,MILP}}\).

  \item \textbf{Station-count sensitivity (fixed weights, 24 vectors).}  
    Four station counts (\(k=30,60,90,120\)) were each tested against 96 randomly generated weight vectors. The choice of 96 weight combinations per \(k\) is guided by the same power analysis.

\end{enumerate}

For every test instance, the optimality gap  

\[
\mathrm{Gap} = 100\times\frac{f_{\mathrm{BSS,MILP}} - f_{\mathrm{BSS,GA}}}{f_{\mathrm{BSS,MILP}}}
\]

was computed as in Alkhalifa et al.~\cite{Alkhalifa2025} and runs with $\mathrm{Gap}\leq1\%$ were classified as near-optimal, following the convention in the heuristic optimization literature \cite{Lagos2023, Madadi2024, Pisinger2007}. Results are summarized in Table~\ref{table:MILP_combined}. In all instances the GA achieved gaps under 1\% of the MILP optimum, confirming that neither weight variation nor station count degrades GA accuracy under the simplified linear formulation. To statistically validate this equivalence, paired percentage gaps were tested for normality (Shapiro--Wilk) and, depending on the result, subjected to either one-sample $t$-tests or Wilcoxon signed-rank tests against a $0.5\%$ threshold. In every scenario the one-tailed $p$-value was below $0.001$, and observed mean gaps ranged from $0.23\%$ to $0.35\%$. 

These findings demonstrate that the GA can approximate exact solutions very closely when the problem is linear. Since the MILP benchmark excludes the network metric of the full BSS utility $U^*_{\mathrm{BSS}}$, the results should be interpreted as an indicative rather than exhaustive validation. On the other hand, although GA runtimes (10–66 min) exceed MILP times (6 min), only the GA approach, among the two, can optimize the full $U^*_{\mathrm{BSS}}$, and its near-optimal performance on the linear proxy gives confidence in its ability to tackle the complete problem.

\begin{table}[!htbp]
  \centering
  \caption{\textbf{GA vs.\ MILP benchmark results.} Results for both the weight and station count sensitivity analysis. GA scores correspond to the best solution of the population.}
  \label{table:MILP_combined}
  \footnotesize
  \begin{tabular}{r r r cc cc}
  \toprule
   & & & \multicolumn{2}{c}{\textbf{MILP}} & \multicolumn{2}{c}{\textbf{GA}} \\
  \cmidrule(lr){4-5} \cmidrule(lr){6-7}
  \textbf{Scenario} & \textbf{\(k\)} & \textbf{Weights} 
    & \textbf{Score (Avg\,$\pm$\,Std)} & \textbf{Time (Avg\,$\pm$\,Std)} 
    & \textbf{Score (Avg\,$\pm$\,Std)} & \textbf{Time (Avg\,$\pm$\,Std)} \\
  \midrule
  Weights & 60 & 118 & 53.48 $\pm$ 0.51 & 5.92 $\pm$ 0.07 & 53.29 $\pm$ 0.54 & 33.60 $\pm$ 10.25 \\
  \midrule
  Stations & 30 & 24  & 27.73 $\pm$ 0.10 & 5.92 $\pm$ 0.08 & 27.67 $\pm$ 0.13 & 10.51 $\pm$ 2.76 \\
                     & 60 & 24  & 53.44 $\pm$ 0.43 & 5.93 $\pm$ 0.11 & 53.26 $\pm$ 0.46 & 21.45 $\pm$ 4.69 \\
                     & 90 & 24  & 77.77 $\pm$ 0.77 & 5.94 $\pm$ 0.06 & 77.50 $\pm$ 0.76 & 38.01 $\pm$ 11.86 \\
                     & 120& 24  & 100.89 $\pm$ 1.03 & 5.96 $\pm$ 0.10 & 100.60 $\pm$ 1.03 & 65.66 $\pm$ 20.23 \\
  \bottomrule
  \end{tabular}
\end{table}

\section{Results}
The results are organized to evaluate the proposed optimization framework under varying planning objectives. The analysis begins with deployments based solely on node utilities, illustrating how alternative weight combinations reflect different goals such as demand coverage, multimodal integration, or social equity. The framework is then extended by incorporating network-level metrics, demonstrating how trade-offs between proximity and accessibility reshape the resulting station layouts. The role of slope is subsequently examined through altitude-adjusted distances, which better represent the effort of cycling in hilly areas. Finally, the framework’s applicability to system growth is illustrated through a case study of Barcelona’s l'Eixample district, where new stations are optimized while accounting for the existing network of \textit{Bicing}.

\subsection{BSS deployments based solely on node utilities}

To evaluate the behavior of the node‐utility function \(U_{i}\) (Equation \ref{eq:BSS_utility_no_metrics}) under different planning objectives, three representative weight–combination scenarios have been selected as illustrative examples. A buffer radius \(Y = 300\,\mathrm{m}\), minimum inter‐station separation \(X = 300\,\mathrm{m}\), and total station count \(k = 250\) are held constant; only the weight vectors vary. Table~\ref{tab:scenario-defs} defines the weights for each scenario and Appendix \ref{app:scenarios} displays the spatial distributions of the variables considered in each scenario. The first scenario (S1) intends to maximize the overall ridership by placing stations in areas of high residential density coupled with a high total count of POIs, thus capturing both trip origins and attractors. The second scenario (S2) aims to promote multimodal integration by favoring sites within easy walking distance of existing transit stops (bus, metro, tram) and directly connected by bike lanes to locations with abundant POIs. Finally, the third scenario (S3) promotes social equity by targeting neighborhoods with lower incomes, higher unemployment, and lower educational attainment.

\begin{table}[!htpb]
    \caption{\textbf{Definition of scenarios (S1–S3).} Each weight vector $W^f$ sums to 1.}
    \footnotesize
    \centering
    \begin{tabular}{p{0.20\linewidth} p{0.75\linewidth}}
      \toprule
      \textbf{Scenario}  &  \textbf{$W_f$}\\
      \midrule
      Maximum demand coverage (S1)& \(\{\,w^{\mathrm{population}}:0.4,\;w^{\mathrm{pois\_count}}:0.35,\;w^{\mathrm{pois\_diversity}}:0.25\}\)\\
        First–Last kilometer (S2)& \(\{\,w^{\mathrm{bus}}:0.2,\;w^{\mathrm{metro}}:0.2,\;w^{\mathrm{tram}}:0.1,\;w^{\mathrm{bike\_lane}}:0.4,\;w^{\mathrm{pois\_count}}:0.1\}\)\\
        Social equity (S3)& \(\{\,w^{\mathrm{income}}:0.4,\;w^{\mathrm{unemp}}:0.3,\;w^{\mathrm{edu\_primary}}:0.3\}\)\\
      \bottomrule
    \end{tabular}
  
    \label{tab:scenario-defs}
\end{table}

\begin{figure}[!htbp]
    \centering
    \includegraphics[width=0.8\linewidth]{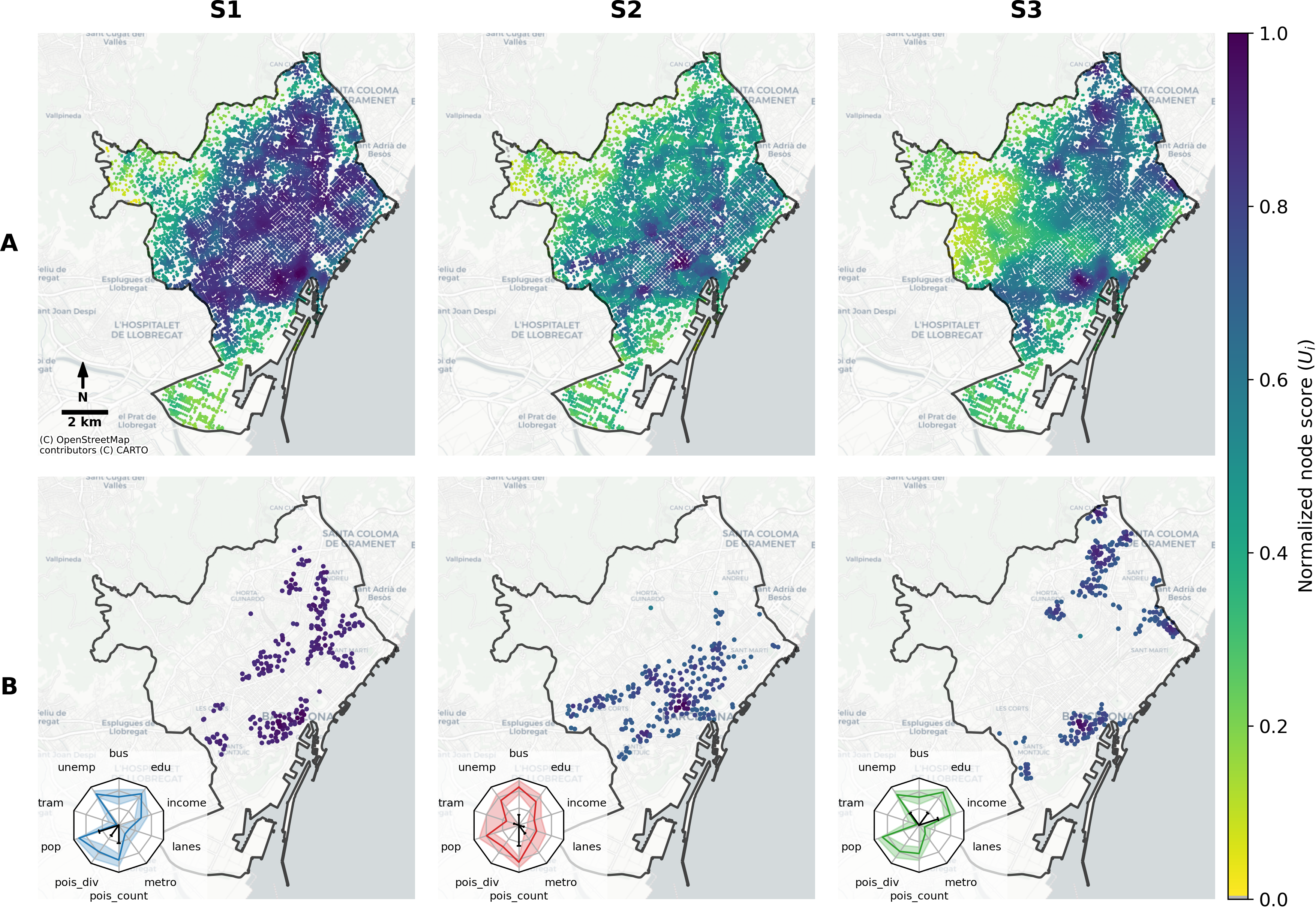}
    \caption{
        \textbf{Station-level utility and BSS optimization for 250 stations across scenarios S1-S3.} \textbf{(A)} Maps of normalized utility scores $U_i$ for all candidate nodes under three scenarios (S1, S2, S3), each defined by a different vector of factor weights. Node utilities have been min–max scaled to $[0,1]$ after weighting to facilitate direct comparison across scenarios. \textbf{(B)} Station locations selected by optimizing solely on $U_i$ (no network‐based objective). Each map includes a radar‐chart inset in which the black line traces the input factor weights, the colored line shows the mean normalized factor values of the chosen nodes, and the shaded band represents $\pm1$ standard deviation of those values. The concentric grid lines on each radar chart correspond to values of 0, 0.33, 0.66, and 1. Variables correspond to the following local factors: bus stops (bus), metro lines (metro), and tram lines (tram); people with only primary education (edu); average income (income); bike lanes (lanes); POI diversity (pois\_div); total count of POIs (pois\_count); population (pop); and unemployment rate (unemp). More details on the variables are provided in Table~\ref{table:local_factors_nodes}.}
    \label{fig:BSS_no_graph_metric}
\end{figure}

Figure \ref{fig:BSS_no_graph_metric} presents the station‐placement results under the three scenarios without incorporating network‐ level metrics $S_{pro}$ or $S_{acc}$. In all cases, the optimization behaves as expected, successfully prioritizing nodes aligned with the goals of each scenario. In the \textit{maximum demand coverage} scenario (S1), stations concentrate in locations with very high residential density and abundant POIs. This is reflected in S1’s radar chart, which shows that on average, selected nodes exhibit a population score of $0.81 \pm 0.06$, POI count of $0.68 \pm 0.12$, and POI diversity of $0.64 \pm 0.16$, the highest among all scenarios. In contrast, the \textit{first–last kilometre} scenario (S2) shifts station selection toward locations near bus, metro, and tram stops, as well as dense bike lanes, resulting in higher average public transport and bike‐lane scores. Average scores increase to $0.74 \pm 0.13$ for bus, $0.40 \pm 0.21$ for metro, and $0.36 \pm 0.19$ for bike lanes, all notably higher than in S1, while also maintaining a high POI count of $0.72 \pm 0.10$. However, this comes at the cost of lower population $0.66 \pm 0.15$ and POI diversity $0.52 \pm 0.11$ values relative to the first scenario.  

Finally, the \textit{social equity} scenario (S3) focuses on disadvantaged regions, which is reflected in the high values of unemployment $0.75 \pm 0.09$, primary‐education share $0.82 \pm 0.08$, and lower income $0.65 \pm 0.13$. This confirms that its weights have shifted the optimization towards areas of greater social need. Yet, when compared to S1, the socioeconomic values of S3 improve only marginally. For instance, unemployment and education already reach $0.74 \pm 0.08$ and $0.75 \pm 0.10$ in S1, despite the absence of social equity weights. This occurs because regions with many unemployed residents and low education or low income tend to coincide geographically with areas of high population, with correlations of 0.96 and 0.93 respectively. In other words, by emphasizing population in S1, many of the same disadvantaged regions that S3 targets are already being captured.

Another important effect comes from the 300 m minimum‐distance constraint. In scenarios S2 and S3, where there are small regions with a lot of high score nodes, the minimum inter-station distance constraint prevents the GA from placing multiple stations within these areas. Instead, stations are spread over larger regions where normalized node scores, while still high, are closer to citywide averages. This means that very localized, high‐score regions could be relatively under-served if they fall within the minimum separation distance. 

Finally, the tight clustering of selected stations under the three scenarios demonstrates that relying solely on local factor scores can produce an overly compact network. If the planning goal demands a more evenly distributed and accessible layout, it is essential to incorporate network‐level considerations alongside the node‐level utility.

\subsection{BSS deployments with integrated node and network utilities}

 \begin{figure}[!htbp]
     \centering
     \includegraphics[width=1\linewidth]{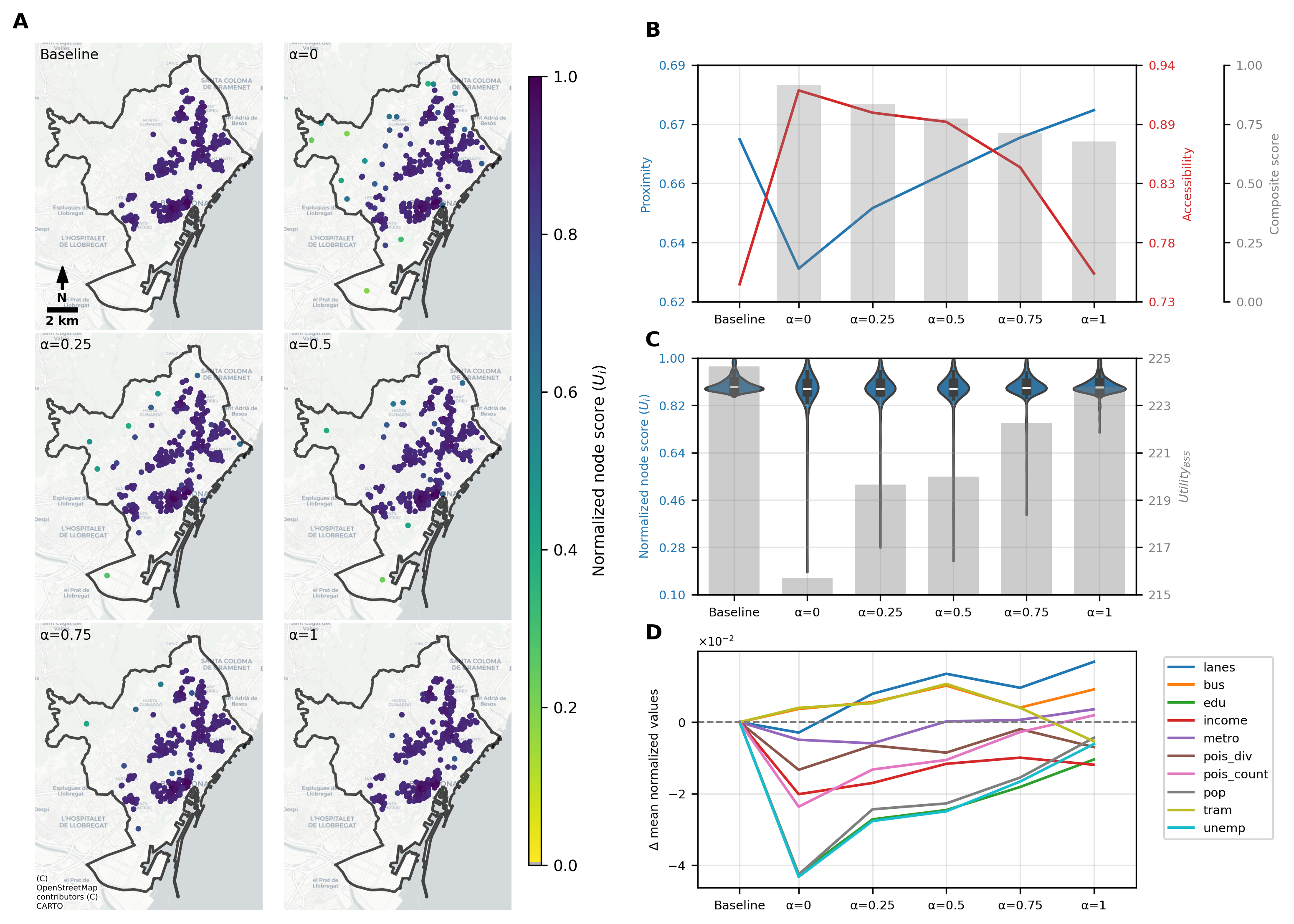}
     \caption{\textbf{Impact of network‐metrics trade-off ($\alpha$) on scenario S1.}  All panels use the same S1 input weights to illustrate how modifying $\alpha$ shifts the BSS network and the node utilities $U_i$. \textbf{(A)} Maps of selected stations under six conditions: a baseline optimization without network metrics, and optimizations with $\alpha$ = 0, 0.25, 0.50, 0.75, and 1. \textbf{(B)} displays proximity score $S_{pro}$ and accessibility score $S_{acc}$ versus $\alpha$. \textbf{(C)} BSS utility $U_{BSS}$ alongside box-plots of individual station utilities for each $\alpha$. \textbf{(D)} Change in mean local factor values of the selected stations relative to the no network metrics baseline.}
     \label{fig:alpha_screening_s1}
 \end{figure}

When network-level metrics are incorporated into the optimisation  (Equation \ref{eq:full_formula}) the resulting station sets can exhibit markedly different spatial structures. This effect is illustrated in Figure~\ref{fig:alpha_screening_s1}, which shows how the S1 scenario is reshaped under varying values of~$\alpha$. Corresponding results for S2 and S3 are also provided in Appendix~\ref{app:network_scenarios}. As seen in Figure~\ref{fig:alpha_screening_s1}A, increasing~$\alpha$ towards~1 leads the algorithm to prioritize proximity between stations, whereas decreasing it towards~0 shifts the focus towards system-wide accessibility. However, setting $\alpha$ to its extreme values (0 or 1) does not guarantee a purely proximity‐ or accessibility‐optimal solution, because the overall objective $U_{BSS}^{*}$ always retains a local‐suitability component. In other words, even when $\alpha= 1$ (nominally maximum station proximity), each candidate node must still meet a minimum value of local factors (e.g. population and POI density). If one truly wanted a pure proximity or accessibility solution, the local‐suitability term could be neutralized by assigning every node the same nonzero normalized utility; in that case, $\alpha=1$ or $0$ would indeed optimize only the chosen network metric.

Under scenario S1, stations selected without any network‐level adjustment are already tightly clustered, as all high‐utility nodes lie in close proximity. Consequently, using $\alpha = 1$ produces only marginal gains in stations' proximity (Figure~\ref{fig:alpha_screening_s1}B). By contrast, setting $\alpha$ = 0 substantially boosts $S_{acc}$ from 0.75 to 0.91. As a result, nodes with lower utility scores are chosen (Figure~\ref{fig:alpha_screening_s1}C). In this case, prioritizing accessibility comes at the expense of population and POI count, however, the POIs diversity is less affected (Figure~\ref{fig:alpha_screening_s1}D). Moreover, as previously noted, some factors are spatially correlated, so optimizing for accessibility can indirectly influence variables not explicitly weighted in the objective function, most notably, unemployment and educational attainment. Conversely, when prioritizing station proximity for this scenario, the algorithm tends to select locations with better cycling infrastructure, resulting in an increase in bike lane coverage among the selected stations.

\subsection{Altitude adjustment}

\begin{figure}[htbp!]
    \centering
    \includegraphics[width=1\linewidth]{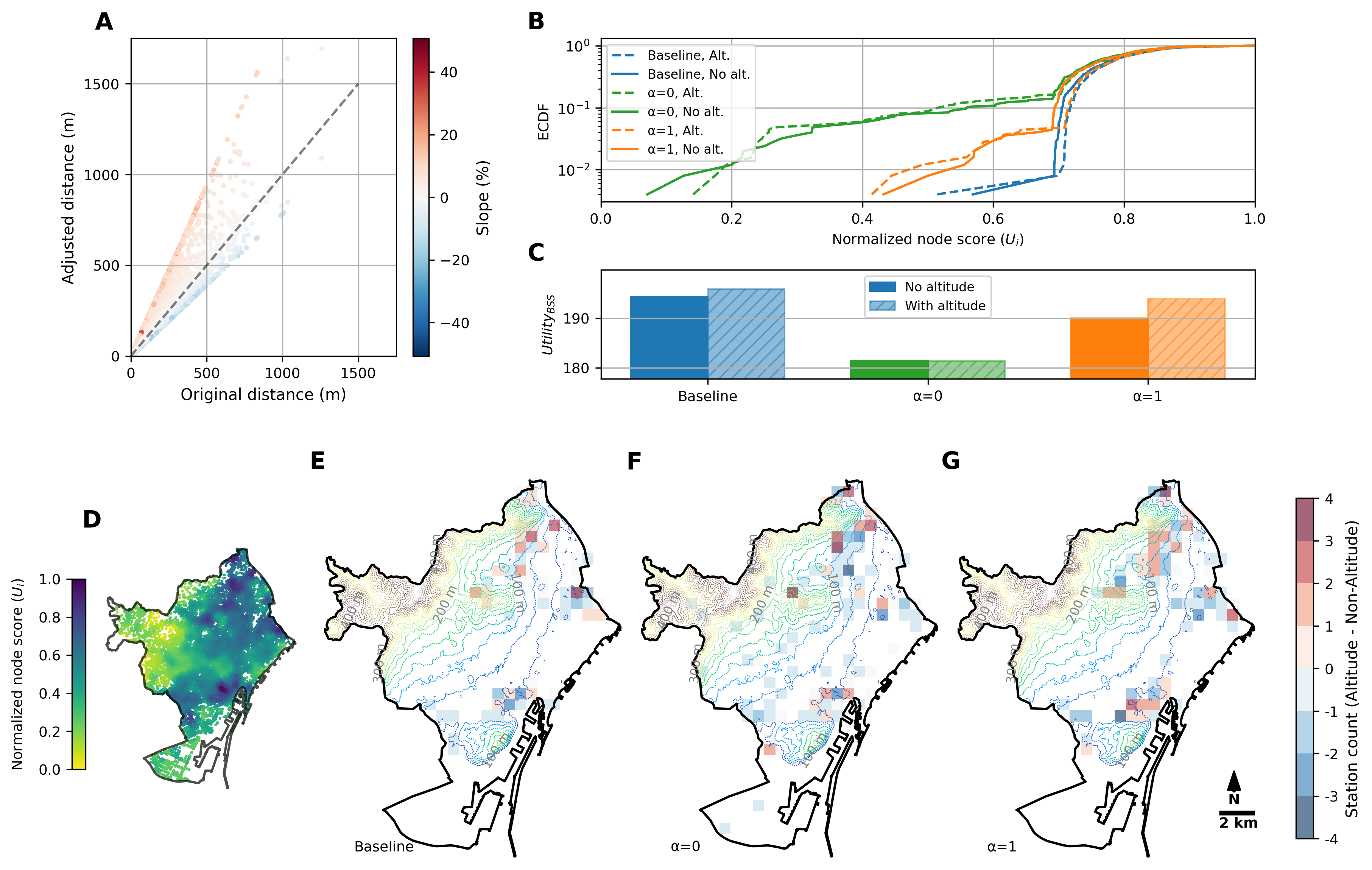}    \caption{
\textbf{Impact of altitude-adjusted distances on station selection.} \textbf{(A)} Original edge distances vs. altitude-adjusted distances, coloured by slope of the between origin and destination. \textbf{(B)} ECDF of normalized node scores for different $\alpha$ values under the S3 scenario, with and without altitude adjustments; the baseline corresponds to the optimization without network metrics for S3. \textbf{(C)} Composite utility $U^*_{BSS}$ for the resulting station sets, showing a slight trade-off when elevation is considered. \textbf{(D)} Spatial distribution of normalized node scores in the study area. \textbf{(E–G)} Differences in selected station locations (grid cell counts) between configurations with and without altitude adjustment under the \textbf{(E)} S3 scenario, and its network-aware versions using \textbf{(F)}~$\alpha=0$ and \textbf{(G)}~$\alpha=1$. Grids cells are 500 m by 500 m. Positive values indicate more stations were selected in a given cell when altitude was considered. Background contour lines represent elevation levels.}
    \label{fig:altitude_s3}
\end{figure}

To assess how altitude-based distance adjustments influence the station selection process, Figure~\ref{fig:altitude_s3} presents results for scenario~S3 under three different optimization strategies (Appendix \ref{app:results_altitude} shows the corresponding figures for S1 and S2). This scenario is particularly illustrative, as it features distinct clusters of high-scoring nodes situated in steep terrain. Figure~\ref{fig:altitude_s3}A shows how original edge distances $d(i,j)$ are transformed into equivalent flat distances \(d_{\mathrm{eq}}(i,j)\) by penalizing uphill segments and rewarding downhill ones. These adjustments, clamped to the interval \([-0.10, 0.10]\) to avoid extreme effects, reshape the shortest-path distance matrix and alter the resulting station configurations.

Across all strategies, incorporating altitude tends to increase distances in steep areas, which relaxes the minimum separation constraint locally and enables more stations to be placed within high-slope, high-score clusters. This generally results in a rightward shift in the node score distribution (Figures~\ref{fig:altitude_s3}B and~\ref{fig:altitude_s3}C), as more top-performing nodes can be included. These shifts are particularly evident in steep districts such as Horta-Guinardó, Nou Barris, Sant Martí, and the Montjuïc area (Figures~\ref{fig:altitude_s3}E–G), where station counts rise in the altitude-aware solutions.

The effects that the altitude adjustment generates on the station placement varies depending on the optimization objective. In proximity-focused configurations ($\alpha=1$), the objective favors compact clusters. Altitude adjustments slightly expand the spacing within steep regions, allowing a few additional high-score nodes to be included without violating the minimum separation constraint. This results in marginal gains in node scores while preserving overall spatial compactness.

In accessibility-focused configurations ($\alpha=0$), slope-adjusted distances increase the perceived remoteness of steep areas. When these regions also contain high-score nodes, (as in S1 and S3) the algorithm assigns more stations there to improve network-wide accessibility, often at the expense of flatter, low-score areas. This trade-off is visible in the spatial maps and is evident in the left tail of the ECDF, where altitude-aware solutions include fewer low-score nodes. In contrast, when high scores are located in flatter areas (as in S2), the altitude adjustment has a more modest impact, leading to less pronounced redistribution. This demonstrates that while the underlying mechanism is consistent, its effect depends strongly on the spatial correlation between slope and node utility.

Together, these results highlight a consistent effect: adjusting distances for slope enables better exploitation of steep, high-score areas that would otherwise be constrained by proximity limits or undervalued in flat-distance metrics. The nature and extent of this effect, however, depend on whether the optimization emphasizes station's proximity, system accessibility, or local suitability alone.

\subsection{BSS expansion}

\begin{figure}[!htbp]
    \centering
    \includegraphics[width=1\linewidth]{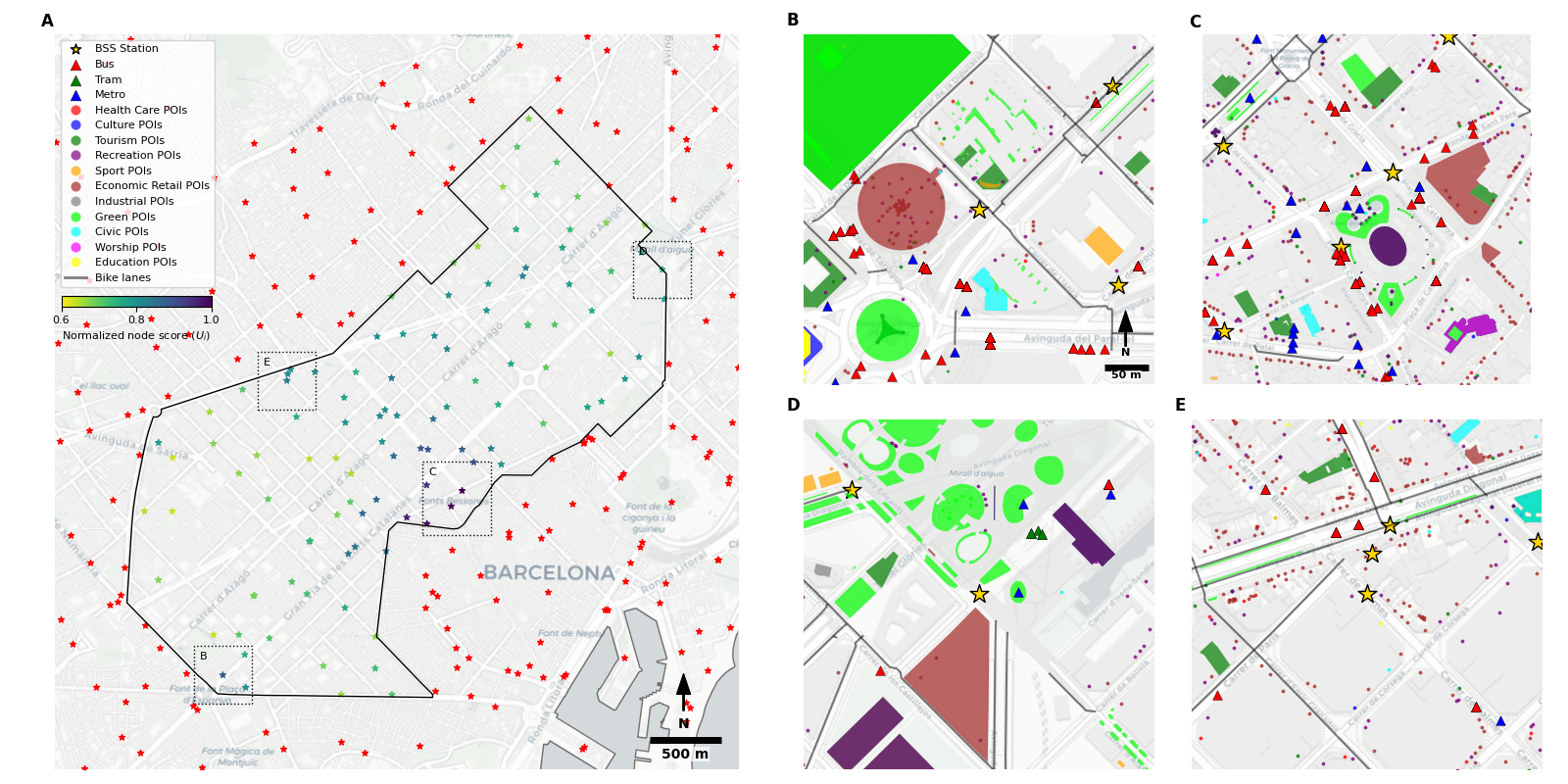}  \caption{
        \textbf{BSS expansion in Barcelona’s l'Eixample district under scenario S2 } \textbf{(A)} Selected locations for the \textit{Bicing} system expansion within l'Eixample district, whose boundaries are delineated by the black line. Existing \textit{Bicing} stations are shown as red stars. The optimization was performed with $\alpha$=0.5, balancing proximity and accessibility objectives. The colour of each selected station reflects its normalized local score based on contextual factors. Dashed rectangles indicate some illustrative areas. \textbf{(B-E)} Close-up views of the highlighted areas display the spatial context around selected stations, including public transport and various categories of POIs. Each subpanel corresponds to a specific rectangle marked in panel (A).}    
    \label{fig:results_eixample}
\end{figure}

\begin{table}[!htbp]
    \caption{\textbf{Normalized node scores $U_i$ and local factor values (mean $\pm$ standard deviation) for the expansion stations in four highlighted subareas of l'Eixample under scenario~S2.} The subareas correspond to the close-up views marked in Figure~\ref{fig:results_eixample}. "Plç." stands for Plaça, and "Av." stands for Avinguda.}
    \footnotesize
    \centering
    \begin{tabular}{lccccccc}
        \toprule
        Location & Stations & $U_i$ & Bus lines & Metro lines & Tram lines & POIs total & Bike lanes \\
        \midrule
        Plç. Espanya (Fig.~\ref{fig:results_eixample}B) & 3 & $0.81 \pm 0.07$ & $0.93 \pm 0.10$ & $0.46 \pm 0.19$ & $0.00 \pm 0.00$ & $0.73 \pm 0.01$ & $0.46 \pm 0.05$ \\
        Plç. Catalunya (Fig.~\ref{fig:results_eixample}C) & 5 & $0.97 \pm 0.04$ & $0.97 \pm 0.01$ & $0.88 \pm 0.14$ & $0.00 \pm 0.00$ & $0.82 \pm 0.04$ & $0.32 \pm 0.14$ \\
        Glòries (Fig.~\ref{fig:results_eixample}D) & 2 & $0.76 \pm 0.02$ & $0.57 \pm 0.04$ & $0.28 \pm 0.00$ & $1.00 \pm 0.00$ & $0.68 \pm 0.01$ & $0.40 \pm 0.07$ \\
        Av. Diagonal (Fig.~\ref{fig:results_eixample}E) & 4 & $0.81 \pm 0.01$ & $0.66 \pm 0.00$ & $0.50 \pm 0.07$ & $0.00 \pm 0.00$ & $0.85 \pm 0.01$ & $0.46 \pm 0.06$ \\
        \bottomrule
    \end{tabular}
    \label{tab:subarea_factors}
  \end{table}

The proposed method can easily be expanded to generate the expansion of an existing BSS. This functionality is illustrated in l'Eixample district of Barcelona under scenario~S2, which emphasizes first--last kilometer connectivity. To simulate a realistic expansion process, all 108 \textit{Bicing} stations within the district were temporarily removed and replaced with an equal number of newly selected stations, while maintaining the 410 existing stations outside the district. The optimization was carried out with $\alpha=0.5$, balacing proximity between stations and system-wide accessibility. L'Eixample district was selected due to its dense public transport infrastructure and high concentration of POIs (Figure~\ref{fig:pois_and_pt_eixample}), making it a suitable context for evaluating the methodology.

As shown in Figure~\ref{fig:BSS_no_graph_metric}A, l'Eixample district contains many high-scoring candidate nodes under scenario~S2. The optimized stations selected in this expansion achieve an average score of $0.76 \pm 0.08$ (range $0.63$–$1.0$), indicating that the algorithm consistently prioritizes top-scoring locations while maintaining a relatively narrow spread of values. The newly selected stations complement the spatial distribution of the existing \textit{Bicing} network while strictly respecting the minimum distance constraint, both among themselves and in relation to the fixed stations outside the district. Moreover, although the existing \textit{Bicing} stations that serve as base for the expand may include stations closer than the imposed minimum inter-station distance, the framework is designed to be capable of generating feasible solutions that comply with this constraint.  

Several of the optimized stations are located in highly significant areas of the district. Figures~\ref{fig:results_eixample}B--E present close-up views of a selection of these locations and detailed local factor values for the chosen stations of these subareas are summarized in Table~\ref{tab:subarea_factors}. Stations in Figure~\ref{fig:results_eixample}B are situated near Plaça d’Espanya, a major multimodal hub surrounded by green spaces and retail-oriented POIs such as the Arenas shopping mall. Here, the three optimized stations combine very high bus connectivity (normalized value of $0.93$), moderate metro accessibility ($0.46$), and strong POI density ($0.73$), with average node scores around $0.81$. Figure~\ref{fig:results_eixample}C shows Plaça de Catalunya, one of the central locations of the city, characterized by an exceptional concentration of POIs and public transport connections. The five selected stations in this area achieve the highest scores in the district, with an average node score of $0.97$, reflecting very high levels of bus and metro accessibility ($0.97$ and $0.88$, respectively) and very dense POI coverage ($0.82$), which capture the area’s role as a focal point of commerce and tourism. Figure~\ref{fig:results_eixample}D depicts Glòries, a redeveloped area that combines strong metro and tram connectivity with a mix of cultural and commercial POIs, including the Mercat dels Encants and the Catalan National Theater. The optimized stations here record a mean node score of $0.76$, supported by the maximum tram accessibility of the BSS ($1.0$) together with balanced values for bus ($0.57$), metro ($0.28$), POIs ($0.68$), and bike lanes ($0.40$). Finally, Figure~\ref{fig:results_eixample}E focuses on Avinguda Diagonal, a major avenue lined with economic and retail-related amenities. The selected stations achieve high POI coverage ($0.85$), good accessibility to bus ($0.66$) and metro ($0.50$), and moderate integration with the cycling network ($0.46$), yielding node scores around $0.81$. In some cases (such as in Figure~\ref{fig:results_eixample}E), selected stations may appear geographically too close to one another, even though they satisfy the minimum distance constraint imposed by the model. This is because distance calculations are based on the directed graph, which represents the actual navigable paths that bicycles must follow. As a result, stations that seem close in straight-line terms may still be sufficiently distant in graph-based terms, as traveling between them can require a longer detour to comply with traffic regulations and use designated cycling infrastructure. This highlights the importance of understanding how the underlying graph is constructed and how distances are defined when applying spatial constraints.

\begin{figure}[!htbp]
    \centering
    \includegraphics[width=\textwidth]{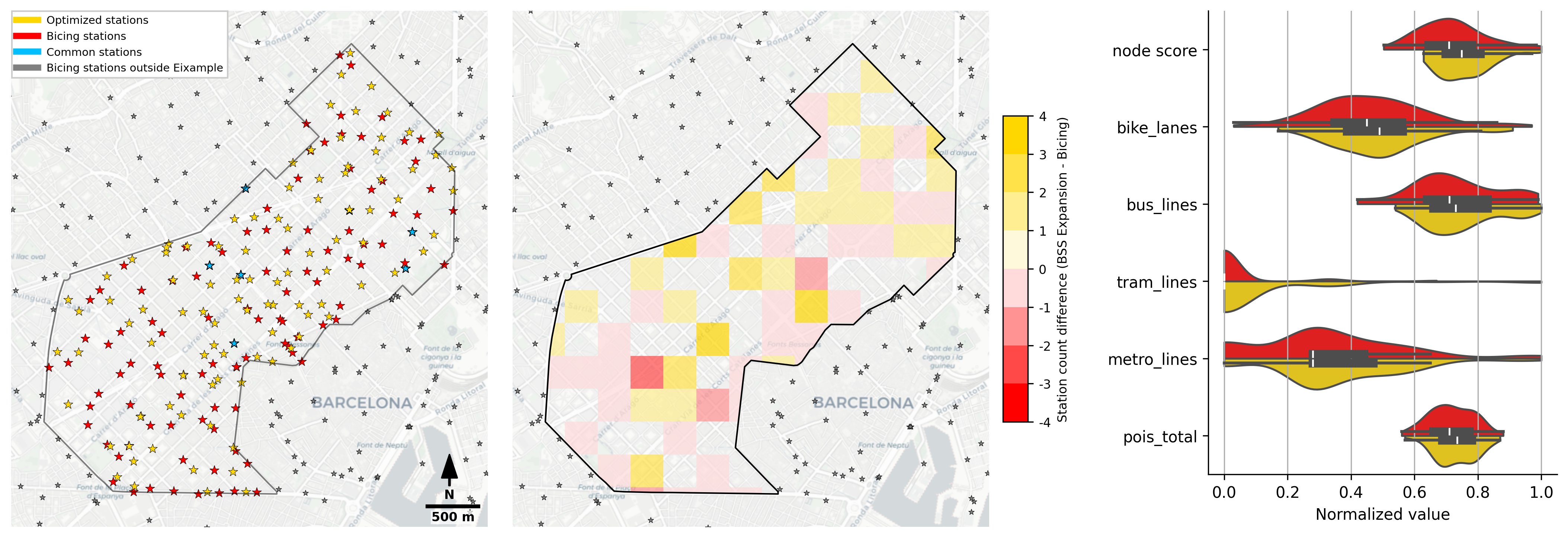}
    \caption{
    \textbf{Comparison between optimized and actual Bicing stations in l'Eixample.}
    \textbf{(A)} Overlay of optimized (yellow), actual (red), and common (blue) stations. 
    \textbf{(B)} Grid heatmap showing station count differences (optimized minus actual). 
    \textbf{(C)} Violin plots comparing normalized values of key factors across the two sets of stations.
    }
    \label{fig:results_eixample_comparison}
\end{figure}

\begin{table}[!htbp]
    \centering
    \footnotesize
    \caption{Comparison of normalized values of node utilities and their local factors values used for the optimization under scenario S2 considering the expansion stations and the actual Bicing stations in l'Eixample (Average $\pm$ standard deviation).}
    \begin{tabular}{lcccccc}
        \toprule
        & Bus lines & Metro lines & Tram lines & POIs total & Bike lanes & Node utility \\
        \midrule
        Actual stations & $0.73 \pm 0.14$ & $0.31 \pm 0.24$ & $0.03 \pm 0.12$ & $0.72 \pm 0.07$ & $0.46 \pm 0.17$ & $0.72 \pm 0.11$ \\
        Optimized stations & $0.76 \pm 0.12$ & $0.36 \pm 0.21$ & $0.09 \pm 0.23$ & $0.73 \pm 0.06$ & $0.50 \pm 0.16$ & $0.76 \pm 0.08$ \\
        \bottomrule
    \end{tabular}
    \label{tab:bicing_comparison}
\end{table}

To further evaluate the expansion results, the optimized stations were directly compared with the existing \textit{Bicing} network in l’Eixample. Although the current \textit{Bicing} stations were not placed through an explicit optimization under scenario~S2, their performance can still be assessed with the same node score. On this basis, \textit{Bicing} stations achieve an average of $0.72 \pm 0.11$ (range $0.50$–$1.0$), while the optimized set reaches $0.76 \pm 0.08$ (range $0.63$–$1.0$). This reflects a systematic shift toward higher-scoring nodes and a narrower spread of values. At the same time, the similarity in the score distributions suggests that the real-world \textit{Bicing} implementation implicitly incorporates many of the contextual factors considered in scenario~S2. Figure~\ref{fig:results_eixample_comparison} illustrates these points: Figure~\ref{fig:results_eixample_comparison}A shows that, despite only seven stations display exact overlaps, several more optimized and \textit{Bicing} stations are located on or near the same street intersection; Figure~\ref{fig:results_eixample_comparison}B reveals that most cells differ by no more than one or two stations, confirming broadly similar spatial patterns; and Figure~\ref{fig:results_eixample_comparison}C and Table~\ref{tab:bicing_comparison} indicate that the main differences in the scores lie in higher tram accessibility (+173\%), metro accessibility (+17\%), and bike-lane integration (+8\%) for the optimized stations, while bus accessibility and POI density remain comparable. Moreover, a closer look at the few cells with larger differences between optimized and \textit{Bicing} stations shows that the number of POIs is the most influential factor, as shown in Figure~\ref{fig:pois_and_pt_eixample}. Overall, the optimized network delivers slightly higher node scores while reproducing much of the structure of the existing system.

\section{Discussion and conclusions}

This study advances the literature on BSS planning by demonstrating how a flexible optimization framework can address multiple planning goals, explicitly manage the spatial arrangement of stations, and integrate topographic and expansion considerations within a single framework. Its versatility was illustrated using a MCDM approach applied to three scenarios targeting demand, first–last kilometer connectivity, and social equity. Building on these scenario-specific outcomes, the integration of station proximity and system accessibility metrics, derived from directed network distances and balanced through a tunable parameter, provided explicit control over the spatial arrangement of stations while still reflecting the underlying planning goals set by the local factor weights. Moreover, incorporating topography-adjusted distances further refined planning in steep districts such as Horta-Guinardó and Nou Barris, enabling a redistribution of stations toward areas that would otherwise face reduced accessibility. Finally, the l’Eixample district expansion experiment demonstrated the framework’s ability to guide incremental growth, ensuring that new stations integrate coherently with existing networks while respecting spatial constraints.

A first aspect concerns the definition of candidate sites. Many studies have defined potential stations at the level of zones or regular grids~\cite{Frade2015, Mete2018, Fazio2021, Veillette2018, Tera2023}, which offers tractable formulations but limits the precision needed for final deployment. Other contributions have moved toward higher resolution by considering all street intersections as candidate sites to better capture local conditions~\cite{Cintrano2018, Xin2013}. We follow this line of work, since planning decisions are ultimately made at the street level, and evaluating the full set of intersections ensures comprehensive coverage of the urban network. In practice, this assumption should be interpreted flexibly: if a node is identified as promising, the actual station would likely be placed in its surroundings where conditions are suitable, such as adjacent sidewalks or plazas. At the same time, this approach excludes mid-block locations and overlooks practical considerations such as the availability of suitable public space for station installation. Future extensions could enrich the computational framework by integrating contextual information—potentially derived from satellite or street-level imagery—to better approximate where stations could realistically be deployed in the urban layout. For example, recent studies have demonstrated how deep-learning models applied to street-view imagery and point clouds can automatically identify and classify street-level elements such as vegetation, roads, sidewalks and street furniture~\cite{Ning2022, Zhou2022}, offering a promising direction for detecting suitable public space for station installation.

Beyond the choice of candidate locations, another key dimension is how planning objectives are framed. Multi-criteria methods have proven valuable for assembling diverse spatial factors such as population and transport accessibility into composite suitability scores~\cite{Kabak2018, Fazio2021, Veillette2018, Tera2023}, but they usually reduce all factors into a single aggregated objective, most often demand coverage. Multi-objective formulations represent an important step forward because they separate competing priorities such as social equity, demand, or multimodal integration and make the trade-offs explicit~\cite{Conrow2018, Duran-Rodas2021, Caggiani2020, Qian2022, Fan2024, Nikiforiadis2021}. Yet, these models also typically rely on a fixed set of goals and spatial factors, so exploring alternative priorities often requires new frameworks or reformulations. Our contribution is not to replace these approaches but to highlight that the multi-criteria logic can be applied more openly: by flexibly deciding the weights and incorporating any data source relevant to the case at hand, planners can generate multi-objective scenarios with a wide range of planning goals within a single framework, avoiding the need to use multiple models. Still, the process of assigning weights remains partly subjective unless supported by stakeholder input or empirical evidence. To address this, decision-making techniques such as AHP, TOPSIS, or MOORA could be incorporated to make weighting more systematic and transparent, as in previous works~\cite{Kabak2018, Guler2021a, Bahadori2022}.

Complementing these aspects, a third contribution concerns the spatial arrangement of stations, a dimension often overlooked in earlier optimization studies. Previous optimization models have typically treated the spatial arrengement of stations as an incidental outcome of the optimization process~\cite{Frade2015, Cintrano2018, Mete2018, Lin2011}. Our framework addresses this by incorporating proximity and accessibility metrics that make spatial arrangement explicit, enabling planners to balance clustered versus dispersed layouts while still reflecting the local factor weights. It relies on a directed graph rather than the undirected representations common in earlier studies~\cite{Cintrano2018, Xin2013}. This is because circulation rules of directed networks such as one-way streets and turn restrictions create asymmetric accessibility, meaning that shortest-path distances can differ depending on travel direction~\cite{Melo2022}. Slope is also accounted for, as it strongly influences cycling behaviour, generally discouraging uphill trips while encouraging downhill ones~\cite{MateoBabiano2016, Grau-Escolano2024}. To our knowledge, García-Palomares et al.~\cite{GarciaPalomares2012} is the only optimization study to incorporate slope, applying a uniform penalty on travel times of climbs. While this was an important advance, the uniform treatment overlooked the different effects of climbs and descents. Our framework refines this with an equivalent flat distance formulation based on the slope–speed relation proposed by Parkin \& Rotheram~\cite{PARKIN2010335}, penalizing climbs while recognizing that moderate declines can facilitate cycling and steep descents may reduce effective speed. Incorporating directionality and topography together provides a more realistic basis for evaluating inter-station proximity and system accessibility, and thus for designing station networks that better reflect actual cycling conditions.

In addition to modelling choices, the algorithmic approach also deserves reflection. Most BSS optimization studies that use metaheuristics rely on genetic algorithms, with only a few exploring alternatives such as simulated annealing or particle swarm optimization~\cite{Cintrano2018, Qian2022, Liu2015, Caggiani2020}. While these methods are effective for handling complex objectives, they are typically applied without formal validation against exact formulations, leaving open the question of how close the solutions are to the true optimum. In our case, the GA was adapted to the specific structure of the problem and benchmarked against a linear MILP proxy. The algorithm’s hyperparameters were tuned on a single optimization scenario, yet the comparison performed across multiple settings—combining different spatial factor weights and station counts—showed that the GA consistently achieved very small accuracy gaps relative to the MILP solutions. These results demonstrate that the algorithm generalizes well across planning scenarios while maintaining near-optimal performance when evaluated on the linear component of the objective function. Although this comparison is limited to the linear formulation, it nonetheless provides strong evidence of the GA’s robustness and its capacity to generalise when exact optimisation of the full nonlinear formulation is infeasible. A natural next step for strengthening this validation would be to compare the GA outcomes with the full nonlinear objective function against those from other metaheuristics, thereby testing its robustness more broadly. Beyond validation, further advances in BSS planning could come from the use of multi-objective optimization algorithms specifically designed to approximate Pareto fronts, such as NSGA-II or SPEA2. Unlike weighted-sum approaches, which require multiple runs to explore trade-offs, these algorithms generate sets of non-dominated solutions in a single run, offering a more comprehensive and transparent view of the trade-offs between competing objectives such as demand, social equity, and multimodal integration.

A final consideration concerns the assumptions of the framework. Its effectiveness depends on the availability and quality of open data, which may vary across contexts and affect transferability. Another important point is that urban spatial factors are often correlated, blurring the distinction between separate planning goals. For this reason, preliminary analyses should be conducted to understand these interdependencies and remain aware of the trade-offs that arise when combining different factors into planning objectives.

Taken together, the methodological elements of the framework have direct implications for practice. By flexibly adjusting weights and objectives, planners can use it as a multi-objective scenario-testing tool to explore how different planning priorities affect the spatial distribution of stations, by making trade-offs more explicit. The framework is designed to be transferable because it builds on open datasets such as OSM, census, and transport network data, and a modular structure that allows any spatial factor to be incorporated, enabling its application in other cities with similar data coverage. In addition, although not applied in this study, the framework allows the optimization to focus on specific groups of POIs when planners wish to emphasize particular trip generators, such as hospitals, universities, or business districts. This is made possible by the classification of POIs into categories—such as healthcare, education, culture, or retail—described in the data section. Likewise, proximity to cycling infrastructure can be prioritized by assigning higher weights to the bike-lane variable or by focusing the optimization solely on this factor.

Looking ahead, the framework could be enriched with additional layers of realism to further align with planning practice. One promising direction involves incorporating demand patterns, using data derived from mobility surveys or usage records of an existing BSS. Integrating such information would make it possible to guide the optimization toward areas with demonstrated demand and even to validate alternative scenarios through mobility simulations. In addition, considering the availability of public space and operational constraints such as station capacity would help ensure that proposed sites are not only theoretically optimal but also practically feasible. Political and institutional factors, including budgetary limits linked to station number and size, rebalancing operations, or stakeholder preferences regarding social equity, coverage, or multimodal integration, could likewise be incorporated to make the tool more directly applicable in real-world contexts. Acknowledging these potential extensions underscores the framework’s capacity to evolve into a decision-support system that bridges technical optimization with the complexities of urban governance.

\section*{Code and data availability}

The source code implementing the proposed optimization framework is openly available at \url{https://github.com/Jordigres/bss_station_optimization}. 
All spatial datasets used in this study are publicly accessible from the cited providers, including OpenStreetMap, Open Data Barcelona, INE, and OpenTopoData.

\section*{Funding}
J.G.E is a fellow of Eurecat’s “Vicente López” PhD grant program. J.G.E. is also supported by the Doctoral Training Network of EIT Urban Mobility.

\clearpage
\bibliographystyle{unsrt} 

\bibliography{
    references
}

\clearpage
\appendix
\renewcommand{\thefigure}{\thesection\arabic{figure}}
\renewcommand{\thetable}{\thesection\arabic{table}}

\section{Case of study: Barcelona}
\label{app:barcelona_overview}
\setcounter{figure}{0}
\setcounter{table}{0}

\begin{figure}[hpbt!]
  \centering
  \includegraphics[width=1\textwidth]{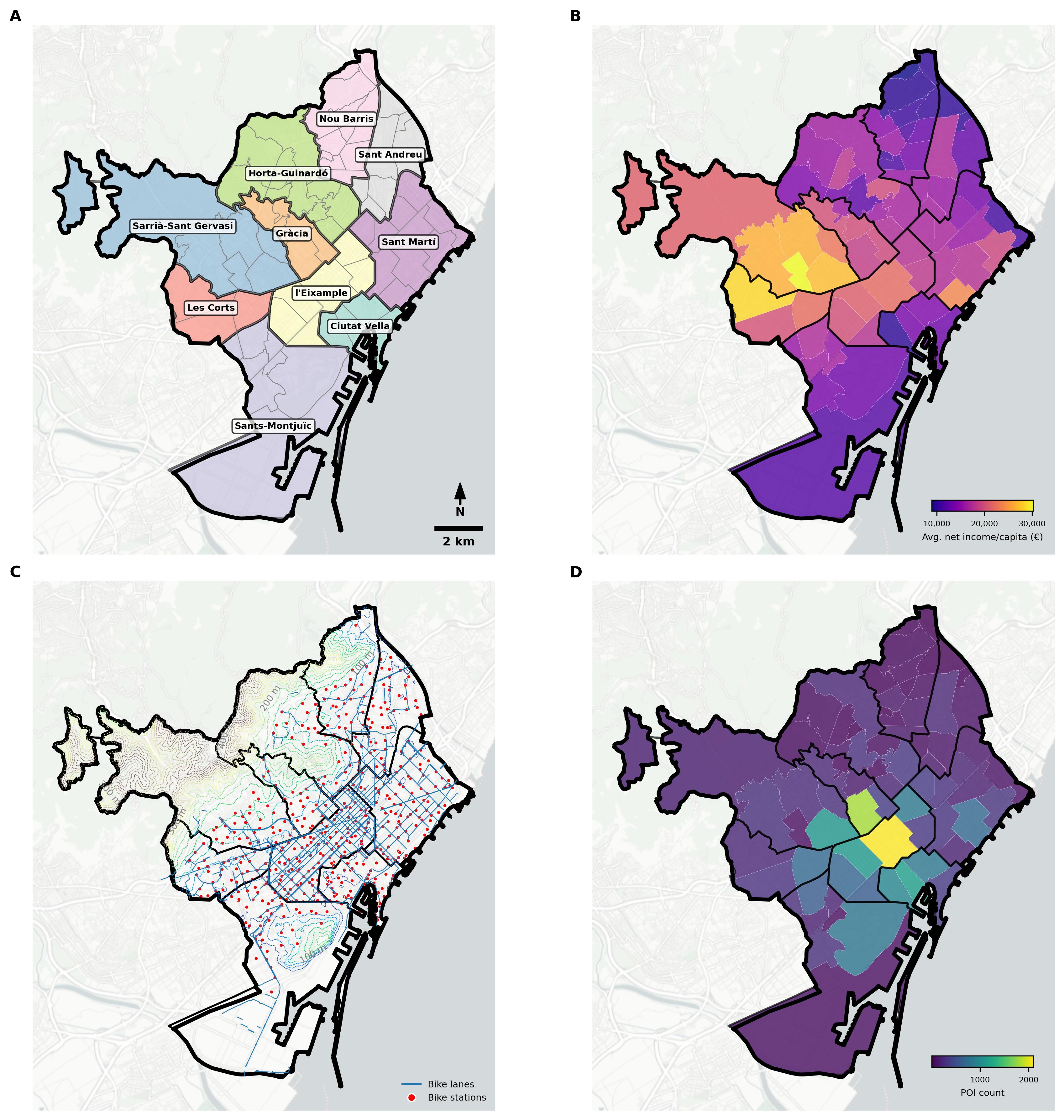}
  \caption{
      \textbf{Overview of Barcelona.} \textbf{(A)} Administrative division of Barcelona: 10 districts (colored polygons) and 73 neighborhoods (gray boundaries). \textbf{(B)} Average net income per capita by neighborhood in 2022, with darker colors indicating lower income levels. \textbf{(C)} City's cycling infrastructure, including the bike lanes and the BSS stations of \textit{Bicing}, overlaid on topographic contours showing elevation variations. \textbf{(D)} Number of POIs per neighborhood, with darker colors indicating less POIs. POIs were obtained from OpenStreetMap.}
  \label{fig:Barcelona_use_case}
\end{figure}

Barcelona, Spain, was selected as the case study to demonstrate the developed methodology due to its diverse urban morphology, pronounced socioeconomic contrasts, and comprehensive open-data resources. Figure~\ref{fig:Barcelona_use_case} summarizes key contextual characteristics, including the city’s administrative division, income distribution, cycling infrastructure, and spatial distribution of points of interest (POIs).

\section{Cleaning the directed bike network}
\label{app:graph_cleaning}
\setcounter{figure}{0}
\setcounter{table}{0} 

\begin{figure}[!htbp]
    \centering
    \includegraphics[width=0.65\linewidth]{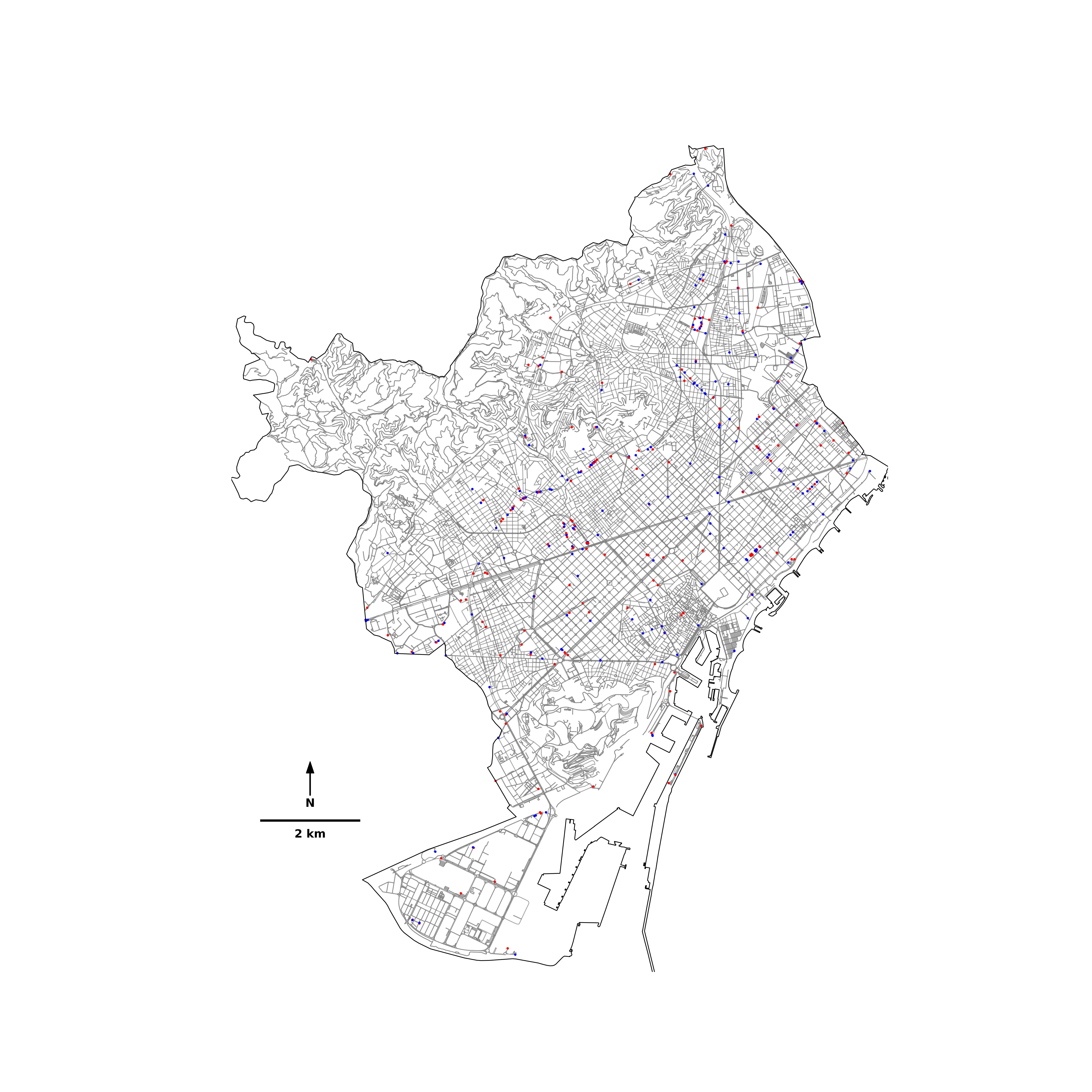}
    \caption{\textbf{Nodes without both incoming and outgoing edges}. Red nodes have only incoming edges and blue nodes have only outgoing edges.}
    \label{fig:bad_nodes}
\end{figure}

\begin{figure}[!htbp]
    \centering
    \includegraphics[width=0.5\linewidth]{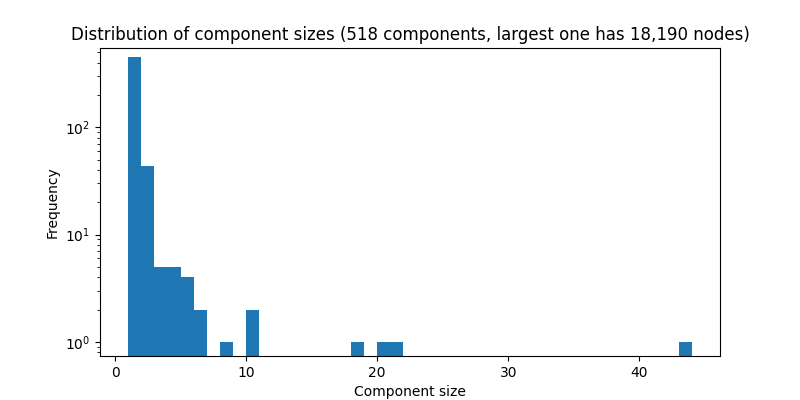}
    \caption{\textbf{Distribution of strongly connected component sizes in the raw directed bike network}. The biggest component of 17,957 nodes is omitted.}
    \label{fig:component_size_distribution}
\end{figure}

To ensure the directed bike‐network graph was fully traversable and without isolated components—thereby avoiding infinite shortest‐path distances—the network was systematically analyzed and modified. Each node was first classified by connectivity into one of four types: no edges, only incoming edges, only outgoing edges, or both. There were 0 nodes with no connections (0.00 \%), 207 with only incoming edges (1.11 \%), and 214 with only outgoing edges (1.14 \%), across 557 strongly connected components (Figures~\ref{fig:bad_nodes}  and \ref{fig:component_size_distribution}). The biggest component included 17,957 nodes and the rest of the 556 was composed of less than 50 nodes, over 80\% of components contained fewer than 10 nodes. Visual inspection of the graph showed that most “one‐way” nodes arose when the network was artificially truncated at the city boundary—cutting off return paths—or when private‐access segments (e.g.\ parking‐lot entrances) created dead ends; only a small fraction were due to OSM tagging errors (Figure~\ref{fig:bad_nodes}).

\begin{figure}[!htbp]
    \centering
    \includegraphics[width=0.8\linewidth]{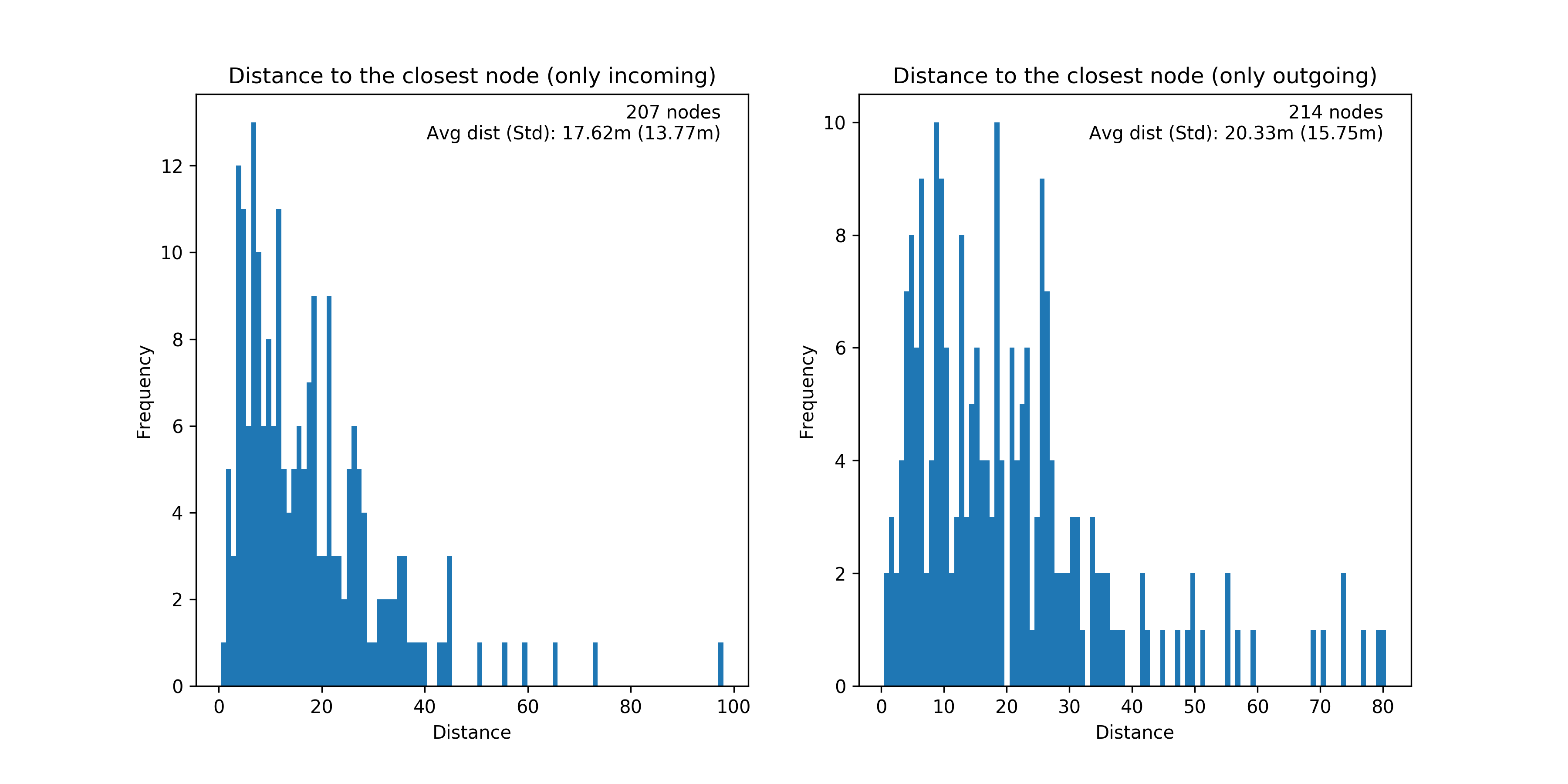}
    \caption{\textbf{Distribution of Euclidean distances between each one‐way node and its nearest neighbor used to restore bidirectional connectivity}. The left panel shows distances for nodes with only incoming edges (new outgoing links), and the right panel for nodes with only outgoing edges (new incoming links). Mean and standard deviation of the new edges' length are indicated on each sub-figure.}
    \label{fig:new_edges_distances}
\end{figure}

Each one‐way node was then reconnected by locating its geographically nearest neighbor and adding the missing link—outgoing if it originally lacked outbound edges, incoming if it lacked inbound edges—to restore bidirectional reachability. After this step, the number of strongly connected components fell from 557 to 122. The added edges exhibited very low Euclidean distances (Figure~\ref{fig:new_edges_distances}): reconnections for nodes missing outgoing links averaged 17.6 m ($\sigma$ = 13.8 m) and, for nodes missing incoming links, 20.3 m ($\sigma$ = 15.8 m), with the vast majority of gaps under 30 m, confirming that network repairs bridge only short spatial discontinuities.

Any remaining disconnected subgraphs were then unified: for each directed edge crossing between components, its reverse was added (preserving only essential attributes). In total, 172 reverse edges were introduced, resulting in a final graph with one strongly connected component.

\section{Data sources}
\label{app:data_sources}

Data for the local factors in Table~\ref{table:local_factors_nodes} were obtained from four open‐data providers: Instituto Nacional de Estadística (INE), Open Data Barcelona (ODB), OpenStreetMaps (OSM), and Open Topo Data (OTD). Below, each source is listed with its relevant variables, geographic coverage, and year of publication.

{ \setlength{\parskip}{0.45\parskip}%
Socioeconomic:
\begin{itemize}
  \item \emph{Population} (INE, 2022): counts of people by sex and five‐year age group, at census‐section level for Spain.
  \item \emph{Income} (INE, 2022): average net income per person, at census‐section level for Spain.
  \item \emph{Education} (ODB, 2024): population aged 16+ by educational attainment and sex, at census‐section level for Barcelona. Levels: less than primary, primary, lower secondary, upper secondary/post‐secondary, tertiary, and not available.
  \item \emph{Nationality} (ODB, 2022): counts by Spanish/EU/other citizenship and sex, at census‐section level for Barcelona.
  \item \emph{Household size} (ODB, 2024): average dwelling size (m²) per census section for Barcelona.
  \item \emph{Vehicle ownership} (ODB, 2024): motorization index (‰) by vehicle category (cars, mopeds, motorcycles, vans, trucks, others) at the census‐section level for Barcelona.
  \item \emph{Unemployment} (ODB, 2024): unemployment rate among population aged 16–64, at neighborhood level for Barcelona.
\end{itemize}

Public Transport:
\begin{itemize}
  \item \emph{Metro, tram, and bus} (OSM, 2025): geographic coordinates for stops and their assigned line names, extracted for the Barcelona area. In the case of the metro the entrances were also obtained. 
\end{itemize}

Built Environment:
\begin{itemize}
  \item \emph{Bike lanes} (OSM, 2025): all OSM ways tagged \texttt{highway=cycleway} in Barcelona.
  \item \emph{Points of Interest} (OSM, 2025): points of interest categorized into healthcare, culture, tourism, recreation, sport, economic and retail, industrial, green, civic, worship, and education, based on a 15-minute city review~\cite{PAPADOPOULOS2023104875}. Additionally, an extra category includes high-traffic locations such as shopping malls, beaches, and large bodies of water. The classification keys and tags are described in the Supplementary material section~\ref{app:pois}.
\end{itemize}

Topography:
\begin{itemize}
  \item \emph{Elevation} (OTD, 2025): Pointwise elevation values retrieved from Open Topo Data API, covering Barcelona’s coordinates.
\end{itemize}
}

\section{OSM POIs}
\label{app:pois}
\setcounter{figure}{0}
\setcounter{table}{0} 

POIs were fetched for the study area using OSMnx’s \texttt{features\_from\_place} function with a broad set of OSM tag keys (amenity, building, craft, healthcare, historic, landuse, leisure, natural, office, shop, sport, tourism, water, waterway). After fetching, each geometry was classified into one or more thematic categories via lookups against predefined tag–value lists (Table~\ref{tab:poi_full_classification}). The specific tag values for each category were chosen by visual inspection of an initial sample to ensure both relevance and completeness.

\begin{scriptsize}
\begin{longtable}{p{0.18\linewidth} p{0.10\linewidth} p{0.65\linewidth}}
  \caption{\textbf{OSM POI classification}.}
  \label{tab:poi_full_classification}\\
  \toprule
  \textbf{Category} & \textbf{Tag key} & \textbf{Values} \\
  \midrule
  \endfirsthead

  \multicolumn{3}{@{}l}{\small\itshape Table \thetable\ continued}\\
  \toprule
  \textbf{Category} & \textbf{Tag key} & \textbf{Values} \\
  \midrule
  \endhead

  \midrule
  \multicolumn{3}{r}{\small\itshape Continued on next page}
  \endfoot

  \bottomrule
  \endlastfoot

  Health \& Care 
    & amenity    & pharmacy, clinic, doctors, hospital, dentist, therapist, nursing\_home, childcare, social\_centre, social\_facility \\
    & building   & hospital \\
    & healthcare & pharmacy, clinic, podiatrist, doctor, dentist, nurse, physiotherapist, laboratory, hospital, alternative;physiotherapist, nutrition\_counselling, therapist, blood\_bank, rehabilitation, dialysis, audiologist, blood\_donation, counselling \\
  \addlinespace

  Culture
    & historic  & memorial, monument, castle, church, archaeological\_site, wayside\_cross, fort, ruins, archaeological\_site;ruins, heritage, aqueduct \\
    & amenity   & library, arts\_centre, exhibition\_centre \\
    & building  & library, museum, chapel, church, cathedral, temple, synagogue \\
    & tourism   & museum, gallery, artwork \\
  \addlinespace

  Tourism
    & tourism & hotel, hostel, guest\_house, motel, museum, theme\_park, artwork, viewpoint, picnic\_site, winery, attraction, gallery, zoo, aquarium, information, chalet \\
    & amenity & attraction, exhibition\_centre, theatre, planetarium \\
    & building & hotel, museum \\
  \addlinespace

  Recreation \& Entertainment
    & amenity  & cinema, theatre, casino, nightclub, bar, pub, restaurant, fast\_food, cafe, ice\_cream, hookah\_lounge, karaoke\_box, toy\_library, food\_court, internet\_cafe \\
    & building & restaurant \\
    & leisure  & amusement\_arcade, escape\_game, bowling\_alley, adult\_gaming\_centre, tanning\_salon, hackerspace, dance, bandstand, marina, esplai, picnic\_table, skill\_game, flight\_simulator, sunbathing, swimming\_area, nature\_reserve \\
    & tourism  & theme\_park, zoo, aquarium, artwork, gallery, museum \\
  \addlinespace

  Sport
    & amenity & gym, track, dojo, sports\_centre, stadium, sports\_hall \\
    & leisure & sports\_centre, swimming\_pool, fitness\_station, bowling\_alley, climbing\_wall, miniature\_golf, horse\_riding \\
    & sport   & table\_tennis, multi, swimming, fitness, yoga, gymnastics, board\_games, climbing, castells, skating, basketball, badminton, taekwondo, martial\_arts, surfing, skate, snowboard, fencing, bodybuilding, boxing, mixed\_martial\_arts, brazilian\_jiu\_jitsu, muay\_thai, kick\_boxing, sailing, skateboard, running, soccer, karate, exercise, chess, orienteering, equestrian, shooting, pilates, padel, calisthenics, billiards, cycling, roller\_skating, paintball, boules, aikido, kayaking, snooker, beachvolleyball, dog\_agility, tennis, bmx, rugby\_league, ice\_skating, field\_hockey, polo, athletics, motor, long\_jump, pole\_vault, archery, racquet, volleyball, futsal, handball, roller\_hockey, cricket, baseball, softball \\
  \addlinespace

  Economic \& Retail
    & amenity & marketplace, atm, bank, bureau\_de\_change, money\_transfer \\
    & office  & insurance, lawyer, estate\_agent, financial, tax\_advisor \\
    & shop    & convenience, supermarket, florist, bakery, butcher, electronics, furniture, clothing, shoes, pharmacy, bookshop, jewelry, pet, hardware, hairdresser, optician, confectionery, gift, kiosk, mobile\_phone, general, \dots \\
  \addlinespace

  Industrial
    & landuse  & industrial, depot, warehouse, quarry \\
    & building & industrial, warehouse, manufacture, factory \\
  \addlinespace

  Green \& Nature
    & landuse & park, forest, meadow, garden, village\_green \\
    & leisure & park, garden, nature\_reserve \\
    & natural & forest, beach, garden, wood, grassland, heath, shrubbery \\
  \addlinespace

  Civic Institutions
    & amenity  & townhall, courthouse, police, prison, fire\_station \\
    & building & government \\
    & office   & government \\
  \addlinespace

  Worship
    & amenity  & place\_of\_worship \\
    & building & church, monastery, synagogue, cathedral, basilica \\
    & historic & church \\
    & landuse  & religious \\
  \addlinespace

  Education
    & amenity  & school, college, university, language\_school, music\_school, driving\_school, beauty\_school, dancing\_school \\
    & building & school, university, college \\
    & office   & educational\_institution \\
  \addlinespace

  High-traffic POIs
    & leisure  & stadium, sports\_centre, sports\_hall, concert\_venue \\
    & tourism  & theme\_park, zoo, aquarium \\
    & natural  & beach, coastline \\
    & landuse  & park \\
    & building & stadium, sports\_hall, concert\_hall \\
    & amenity  & events\_venue, exhibition\_centre, conference\_centre \\
    & shop     & mall, department\_store \\
    & water    & lake, river \\
    & waterway & river \\
\end{longtable}
\end{scriptsize}

\section{Distribution area-level counts across residential buildings}
\label{app:pop_distribution}
\setcounter{figure}{0}
\setcounter{table}{0} 

In some cases, to assign any count‐based attribute more realistically to the network nodes, they were first moved from the area-level aggregates (e.g. census-sections or neighborhoods) to the building scale. Directly allocating section totals by area would scatter counts into non‐residential places (parks, malls, etc.), producing large misallocations. Instead, for each attribute it was (1) extracted and filtered residential building footprints from OSM, and then (2) apportioned each section’s total count across those footprints. This ensures that all attribute counts reside in actual dwellings before they are subsequently aggregated to the bike‐network nodes. 

\subsection{Downloading residential building footprints}

\begin{figure}[!htbp]
  \centering
  \includegraphics[width=0.7\textwidth]{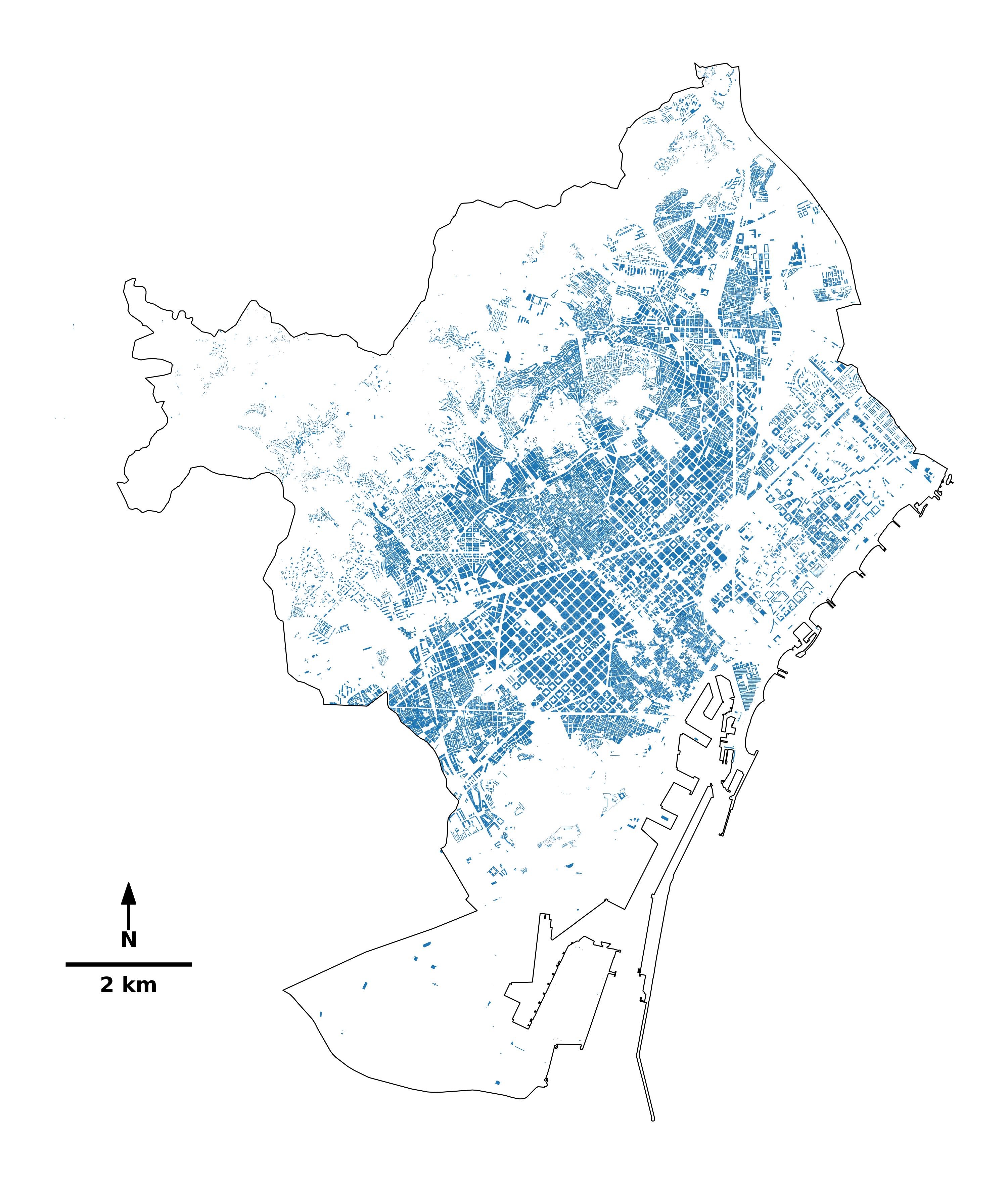}
  \caption{\textbf{Residential building footprints of Barcelona, obtained from OSM}.}
  \label{fig:res_buildings}
\end{figure}

Residential building footprints were first retrieved from OSM via OSMnx’s \texttt{features\_from\_place} using a broad set of tags (\texttt{building}, \texttt{building:condition}, \texttt{amenity}, \texttt{man\_made}, \texttt{office}, \texttt{power} and \texttt{shop}).  The raw footprint layer was then filtered according to predefined tag–value lists (Table~\ref{tab:building_tag_filters}): only features whose \texttt{building} tag matched the residential list were retained, and any footprints matching non-residential building, amenity or tourism exclusion lists were removed.  To resolve remaining ambiguities, footprints were retained if either (a) their \texttt{building} tag differed from “yes” and all \texttt{building:condition}, \texttt{amenity}, \texttt{man\_made}, \texttt{office}, \texttt{power} and \texttt{shop} tags were empty, or (b) their \texttt{building} tag equaled “yes” and, in addition to all those tags being empty, their \texttt{name} was also empty.  The resulting residential footprint layer was saved and inspected visually (Figure~\ref{fig:res_buildings}).  

\begin{scriptsize}

\begin{longtable}{p{0.20\linewidth} p{0.8\linewidth}}
  \caption{\textbf{OSM tags used to filter residential vs.\ non-residential features}}
  \label{tab:building_tag_filters}\\
  \toprule
  \textbf{Category} & \textbf{OSM tag values} \\
  \midrule
  \endfirsthead

  \multicolumn{2}{@{}l}{\small\itshape Table \thetable\ continued}\\
  \toprule
  \textbf{Category} & \textbf{OSM tag values} \\
  \midrule
  \endhead

  \midrule
  \multicolumn{2}{r}{\small\itshape Continued on next page}
  \endfoot

  \bottomrule
  \endlastfoot

  Residential buildings &
    apartments, barracks, bungalow, cabin, detached, annexe, dormitory, farm, ger, hotel, house, houseboat, residential, semidetached\_house, static\_caravan, terrace, tree\_house, trullo, isolated\_dwelling, yes \\

  \addlinespace
  Non-residential buildings &
    commercial, industrial, kiosk, office, retail, supermarket, warehouse, cathedral, chapel, church, temple, kingdom\_hall, monastery, mosque, presbytery, shrine, synagogue, religious, cloister, bakehouse, civic, college, fire\_station, government, hospital, kindergarten, public, school, toilets, train\_station, transportation, university, museum, annex, administrative, library, shelter, sports, market, barn, conservatory, cowshed, farm\_auxiliary, greenhouse, slurry\_tank, stable, sty, grandstand, pavilion, riding\_hall, sports\_hall, sports\_centre, stadium, allotment\_hall, boathouse, hangar, hut, shed, car\_garage, garage, garages, parking, digester, service, tech\_cab, transformer\_tower, water\_tower, storage\_tank, silo, manufacture, air\_shaft, beach\_hut, bunker, castle, construction, container, guardhouse, military, outbuilding, pagoda, quonset\_hut, roof, ruins, tent, tower, windmill \\

\addlinespace
  Non-residential tourism &
    aquarium, attraction, gallery, hostel, information, motel, theme\_park, zoo, winery 
\\
  \addlinespace
  Non-residential landuse &
    construction, industrial, farmland, farmyard, orchard, greenhouse\_horticulture, plant\_nursery \\
  \addlinespace
  Non-residential amenities &
    bicycle\_parking, motorcycle\_parking, parking\_space, parking\_entrance, bicycle\_rental, shelter, research\_institute, smoking\_area, table, toilets, waste\_disposal, recycling, garden, coworking\_space, dog\_toilet, drinking\_water, bbq, public\_building, atm, beauty\_school, bench, bicycle\_rental;left\_luggage, charging\_station, childcare, clock, compressed\_air, dance\_school, disused, dojo, dressing\_room, fixme, flight\_attendant, grit\_bin, grocery, hookah\_lounge, karaoke\_box, kick-scooter\_parking, letter\_box, loading\_dock, locker, lounger, luggage\_locker, office, parcel\_locker, pastries, photo\_booth, piano, post\_box, relay\_box, sailing\_school, scooter\_rental, shower, signs, stock\_exchange, surf\_school, swingers\_club, tap, telephone, therapist, union, vacuum\_cleaner, vending\_machine, warehouse, waste\_basket, water\_point, watering\_place, wifi, ticket\_validator, Geldwechselstube, lavoir, meeting\_point, parking\_exit, retirement\_home, bar, bear\_box, biergarten, cafe, canteen, fast\_food, food\_court, ice\_cream, pub, restaurant, market, college, dancing\_school, driver\_training, driving\_school, kindergarten, language\_school, library, toy\_library, music\_school, prep\_school, school, ski\_school, training, university, bicycle\_repair\_station, boat\_rental, boat\_sharing, boat\_storage, bus\_station, car\_rental, car\_sharing, car\_wash, ferry\_terminal, fuel, motorcycle\_rental, parking, taxi, traffic\_park, vehicle\_inspection, weighbridge, payment\_terminal, bank, bureau\_de\_change, money\_transfer, payment\_centre, baby\_hatch, clinic, dentist, doctors, hospital, nursing\_home, pharmacy, social\_facility, veterinary, arts\_centre, brothel, casino, cinema, community\_centre, conference\_centre, events\_venue, exhibition\_centre, fountain, gambling, love\_hotel, music\_venue, nightclub, planetarium, public\_bookcase, social\_centre, stage, stripclub, studio, swingerclub, theatre, exhibition\_hall, courthouse, fire\_station, police, post\_depot, post\_office, prison, ranger\_station, townhall, animal\_boarding, animal\_breeding, animal\_shelter, animal\_training, baking\_oven, crematorium, dive\_centre, funeral\_hall, grave\_yard, hunting\_stand, internet\_cafe, kitchen, kneipp\_water\_cure, marketplace, monastery, mortuary, place\_of\_mourning, place\_of\_worship, public\_bath, refugee\_site, security\_control
 \\
\end{longtable}
\end{scriptsize}

\subsection{Distributing counts}

Count‐based variables were redistributed from their original spatial aggregations to residential building footprints via an area‐share method. Buildings were first spatially joined to the polygon in which they lie. For each polygon with total count \(C\) and containing \(n\) buildings of areas \(a_1,\dots,a_n\), each building \(i\) was assigned  
\[
C_i = C \times \frac{a_i}{\sum_{j=1}^n a_j}\,.
\]  
Point‐geometry buildings were buffered by 25 m to ensure intersection with their parent polygon. The result is a building‐level dataframe in which each residential footprint carries its share of the count. These building‐level counts are subsequently aggregated to network nodes (e.g.\ by summation within a node buffer) to derive node‐level attributes that more accurately reflect the spatial distribution of the original data.

\section{Normalization procedure}
\label{app:normalization}
\setcounter{figure}{0}
\setcounter{table}{0} 

Prior to computing utility scores, all local‐factor variables were rescaled to a common \([0,1]\) range. Selecting an appropriate normalization technique for each variable is essential to prevent any single factor from disproportionately influencing the composite utility. While the ultimate normalization choices should reflect the specific BSS planning objectives and the local data distributions, general guidelines are presented in this study. Multiple techniques were considered, each chosen according to the variable’s bounds, distribution, and interpretability:

\begin{itemize}
  \item  Min–max scaling: linearly maps each value between its observed minimum and maximum.  Ideal for variables with known finite bounds or approximately uniform distributions without extreme outliers, as it preserves the original range and interpretative endpoints.

  \item Z‐score rescaling: standardizes to zero mean and unit variance, then applies min–max mapping.  Suited to roughly symmetric, bell-shaped distributions that differ only in scale, ensuring comparability without distortion by differing variances.

  \item Robust scaling: centering on the median and dividing by the inter-quartile range, with extreme values clipped to the \([0,1]\) interval. Reserved for heavy-tailed variables with extreme outliers, it limits the influence of rare extremes on the normalized range.      

  \item Log‐transform (\(\log(1+x)\)): compresses right-skewed, strictly positive variables before scaling. Note that any zeros were shifted by \(+1\) to satisfy the positivity requirement. This transformation is particularly effective on variables where a small number of very large values would otherwise dominate the normalized range.

  \item Box–Cox transformation: power‐based variance stabilization for strongly skewed distributions, followed by rescaling. As with the log, all inputs must be strictly positive; zeros were replaced with a small constant (\(\epsilon=10^{-6}\)) prior to transformation. Ideal when a simple log underperforms in reducing skewness.

\end{itemize}

Additionally, for variables where lower raw values indicate more favorable conditions (e.g.\ travel cost, slope penalties), the normalized score was inverted via $v_{inv} = 1 - v_{norm}$,
so that higher values consistently correspond to more desirable outcomes across all factors.

\begin{figure}[!htbp]
    \centering
    \includegraphics[width=0.9\linewidth]{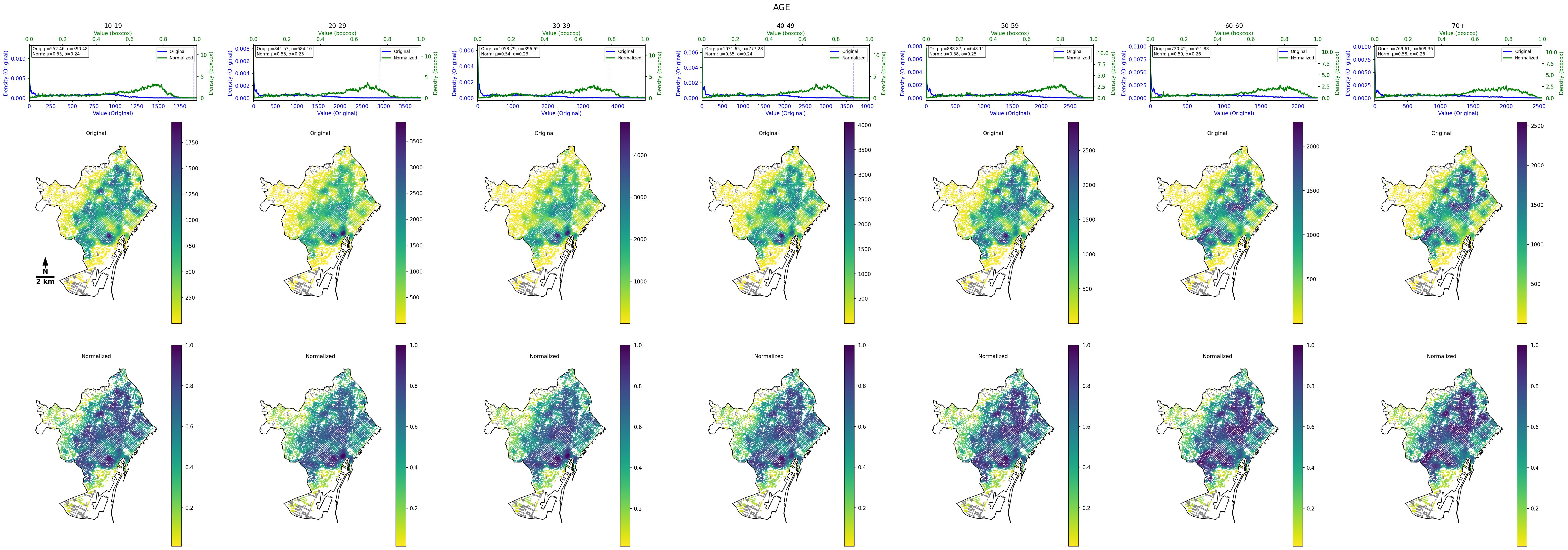}
   \caption{\textbf{Age‐group distributions and spatial patterns before and after normalization}. Columns show cohorts 10–19, 20–29, 30–39, 40–49, 50–59, 60–69, and 70+. \emph{Top row:} kernel density estimates of the raw counts (blue) and the normalized values (green), with vertical dashed lines indicating the lower and upper outlier thresholds for the normalized data (first and third quartiles ± 1.5 × IQR).  
    \emph{Middle row:} raw values mapped at each network node (zero‐value nodes in gray).  
    \emph{Bottom row:} normalized values mapped at the same nodes (zero‐value nodes in gray).}

    \label{fig:age_normalization}
\end{figure}

\begin{figure}[!htbp]
    \centering
    \includegraphics[width=0.9\linewidth]{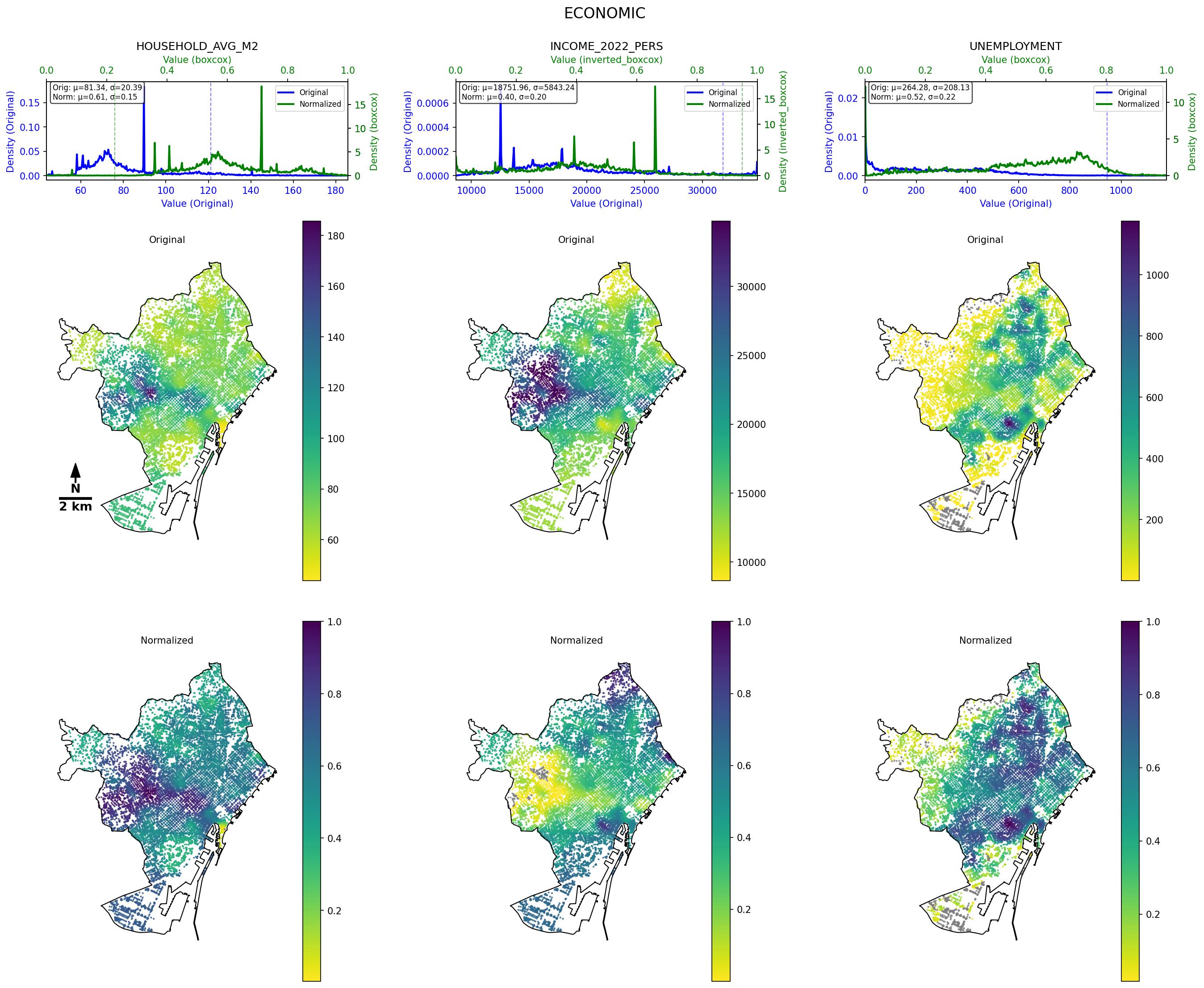}
   \caption{\textbf{Distributions and spatial patterns before and after normalization of the average household size in $m^{2}$, average income per person in 2022 and unemployment percentage}. \emph{Top row:} kernel density estimates of the raw counts (blue) and the normalized values (green), with vertical dashed lines indicating the lower and upper outlier thresholds for the normalized data (first and third quartiles ± 1.5 × IQR).  
    \emph{Middle row:} raw values mapped at each network node (zero‐value nodes in gray).  
    \emph{Bottom row:} normalized values mapped at the same nodes (zero‐value nodes in gray).}

    \label{fig:economic_normalization}
\end{figure}

\begin{figure}[!htbp]
    \centering
    \includegraphics[width=0.9\linewidth]{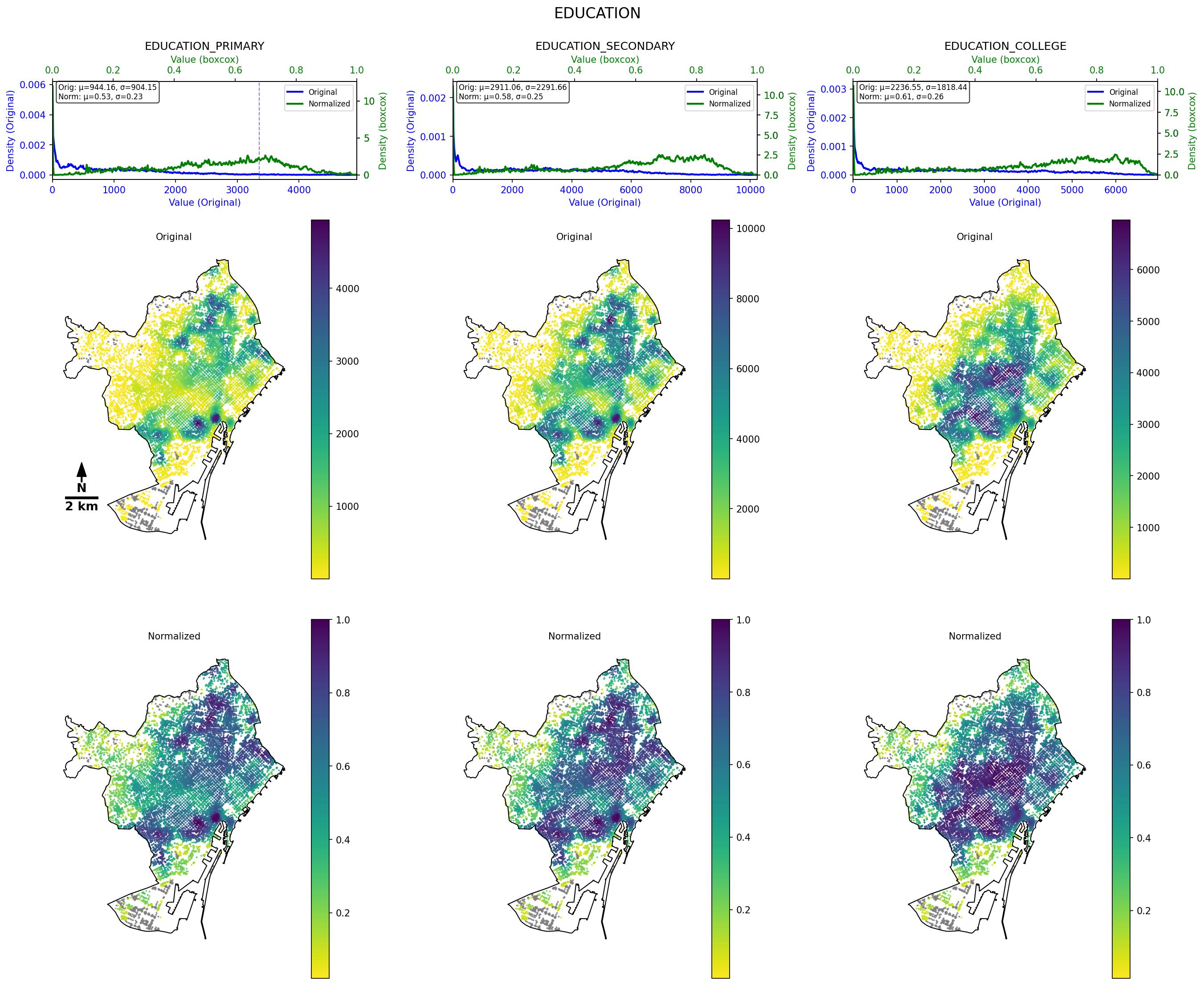}
   \caption{\textbf{Distributions and spatial patterns before and after normalization of the number of people with up to primary education, with up to secondary education and with college education}. \emph{Top row:} kernel density estimates of the raw counts (blue) and the normalized values (green), with vertical dashed lines indicating the lower and upper outlier thresholds for the normalized data (first and third quartiles ± 1.5 × IQR).  
    \emph{Middle row:} raw values mapped at each network node (zero‐value nodes in gray).  
    \emph{Bottom row:} normalized values mapped at the same nodes (zero‐value nodes in gray).}

    \label{fig:education_normalization}
\end{figure}

\begin{figure}[!htbp]
    \centering
    \includegraphics[width=0.9\linewidth]{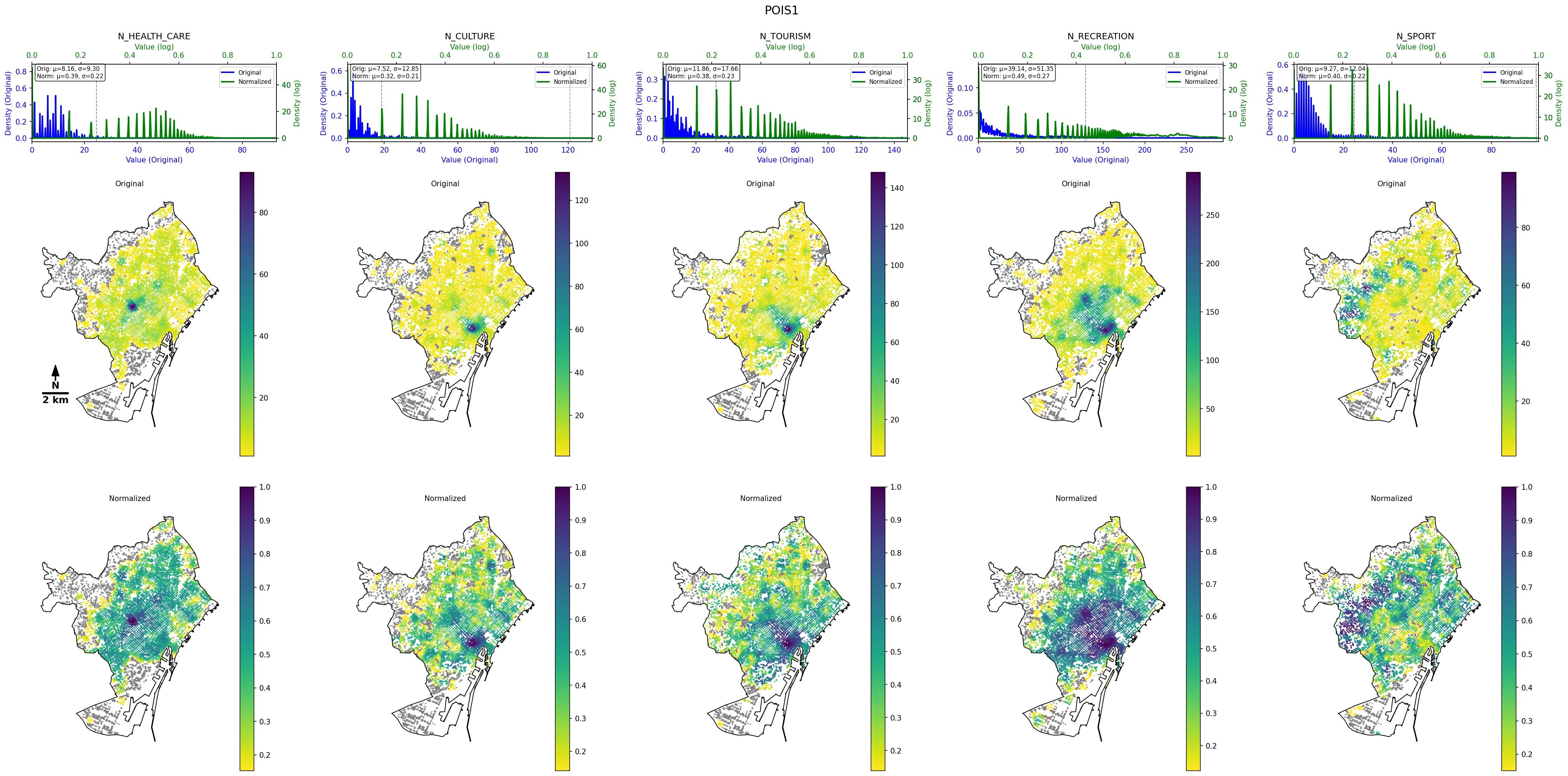}
    \caption{\textbf{Distributions and spatial patterns before and after normalization of category‐specific POI counts for health \& care, culture, tourism, recreation, sport locations}.  
    \emph{Top row:} kernel density estimates of the raw counts (blue) and the normalized values (green), with vertical dashed lines indicating the lower and upper outlier thresholds for the normalized data (first and third quartiles ± 1.5 × IQR).  
    \emph{Middle row:} raw values mapped at each network node (zero‐value nodes in gray).  
    \emph{Bottom row:} normalized values mapped at the same nodes (zero‐value nodes in gray).}
    \label{fig:pois1_normalization}
\end{figure}

\begin{figure}[!htbp]
    \centering
    \includegraphics[width=0.9\linewidth]{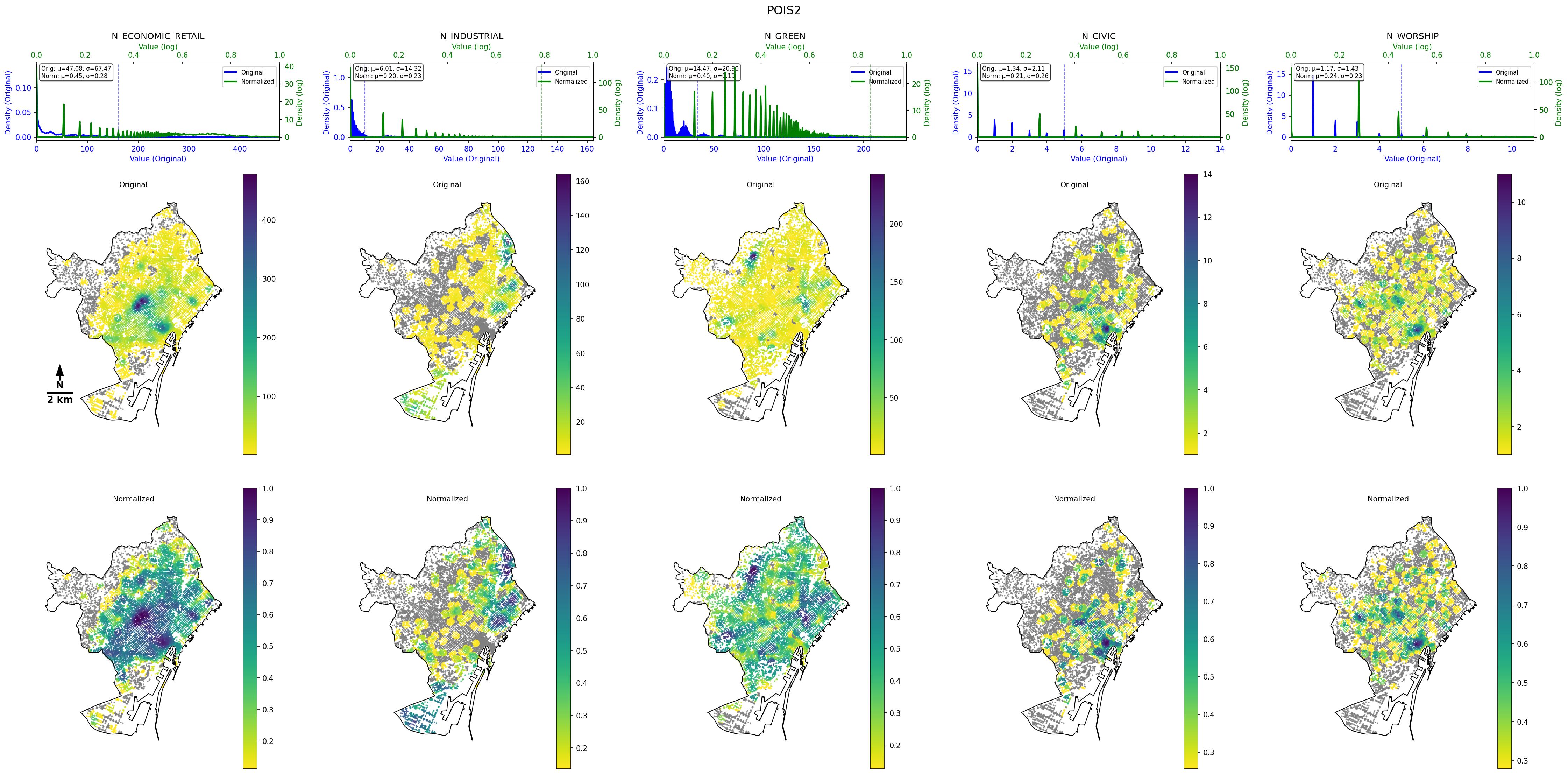}
    \caption{\textbf{Distributions and spatial patterns before and after normalization of category‐specific POI counts for economic/retail, industrial, green, civic and worship locations}.  
    \emph{Top row:} kernel density estimates of the raw counts (blue) and the normalized values (green), with vertical dashed lines indicating the lower and upper outlier thresholds for the normalized data (first and third quartiles ± 1.5 × IQR).  
    \emph{Middle row:} raw values mapped at each network node (zero‐value nodes in gray).  
    \emph{Bottom row:} normalized values mapped at the same nodes (zero‐value nodes in gray).}
    \label{fig:pois2_normalization}
\end{figure}

\begin{figure}[!htbp]
    \centering
    \includegraphics[width=0.9\linewidth]{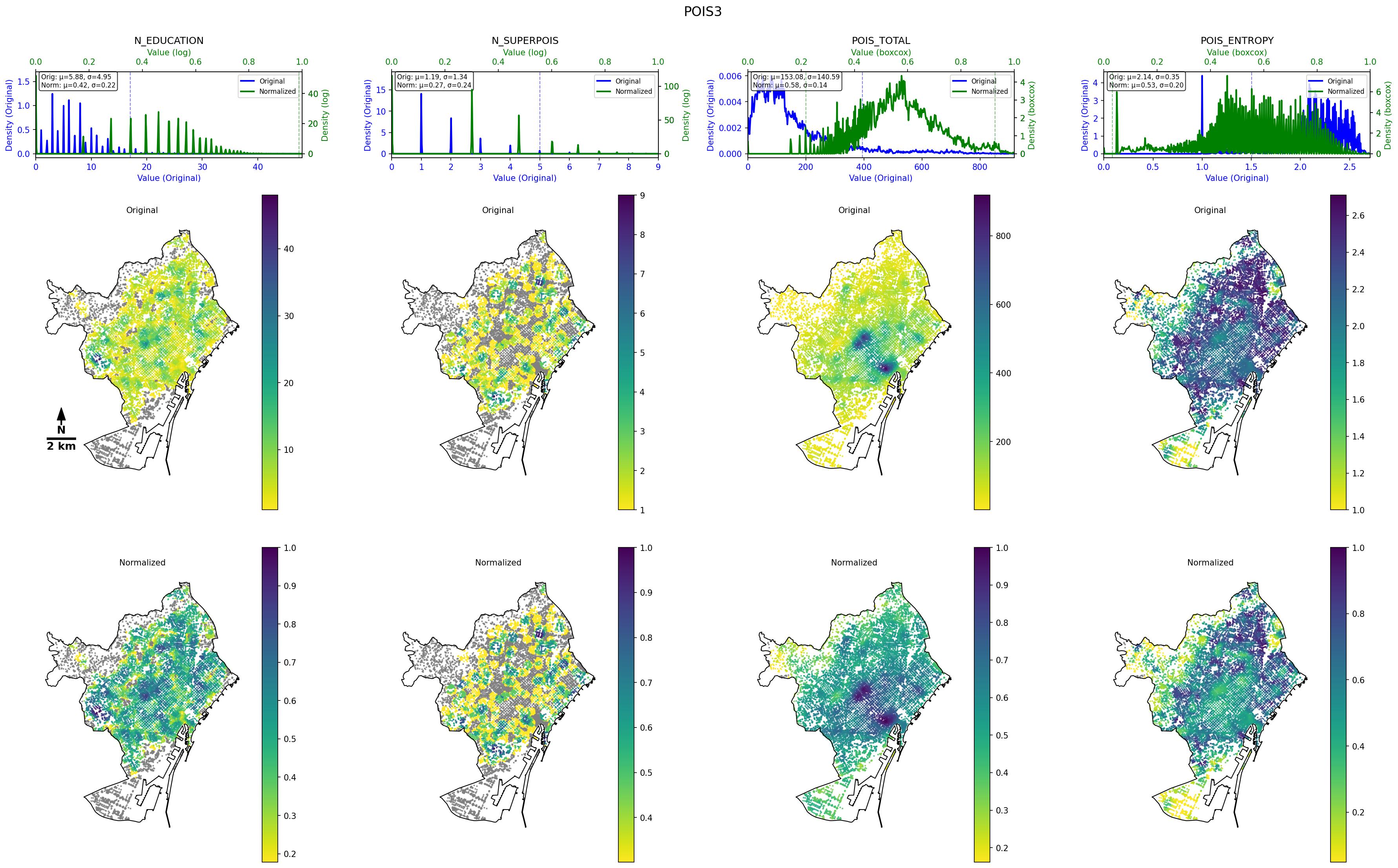}
    \caption{\textbf{Distributions and spatial patterns before and after normalization of category‐specific POI counts for for education, high-traffic “super” POIs, the total POI count, and POI diversity (entropy)}.  
    \emph{Top row:} kernel density estimates of the raw counts (blue) and the normalized values (green), with vertical dashed lines indicating the lower and upper outlier thresholds for the normalized data (first and third quartiles ± 1.5 × IQR).  
    \emph{Middle row:} raw values mapped at each network node (zero‐value nodes in gray).  
    \emph{Bottom row:} normalized values mapped at the same nodes (zero‐value nodes in gray).}
    \label{fig:pois3_normalization}
\end{figure}

\begin{figure}[!htbp]
    \centering
    \includegraphics[width=0.9\linewidth]{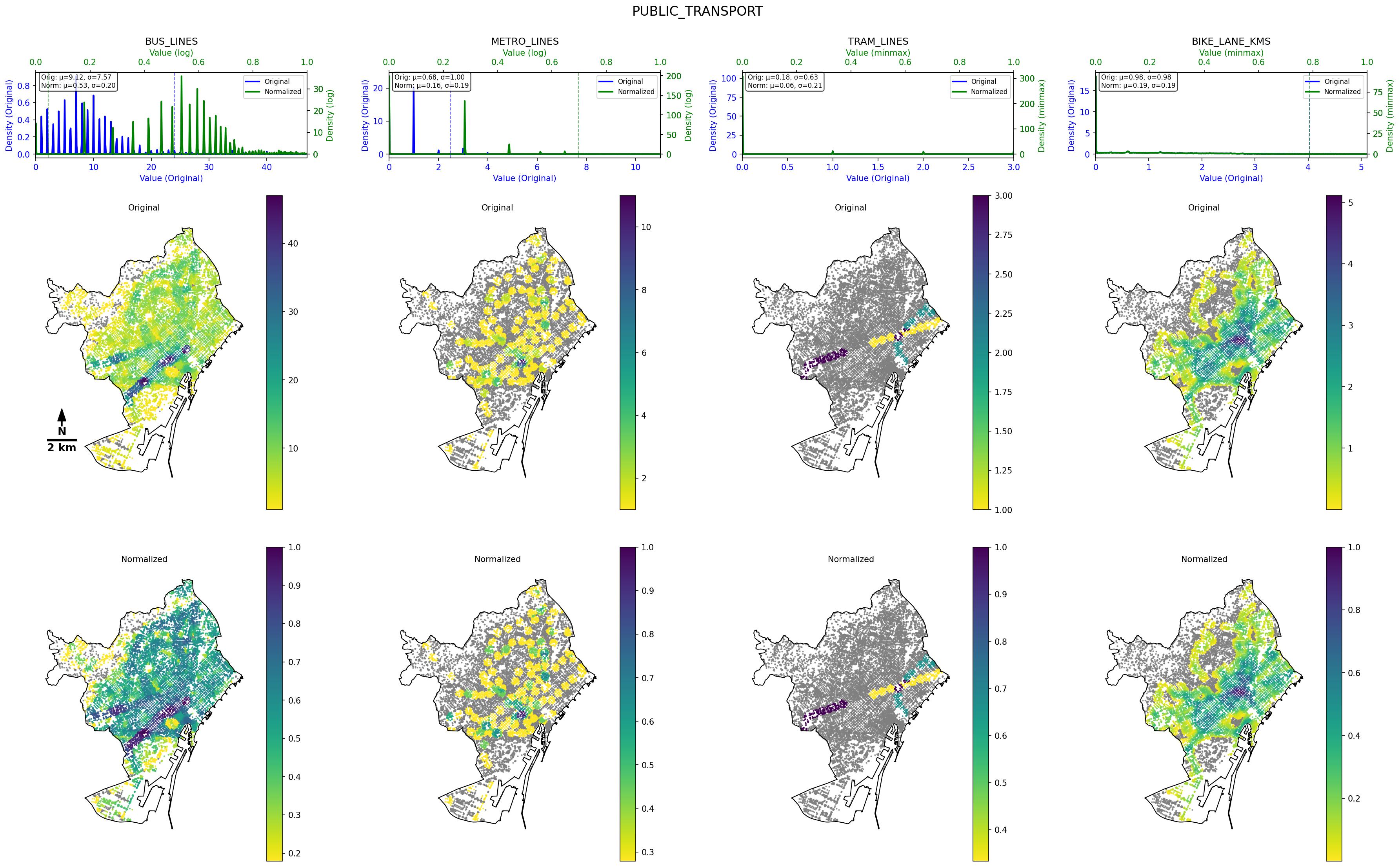}
    \caption{\textbf{Distributions and spatial patterns before and after normalization of the public transport variables}. \emph{Top row:} kernel density estimates of the raw counts (blue) and the normalized values (green), with vertical dashed lines indicating the lower and upper outlier thresholds for the normalized data (first and third quartiles ± 1.5 × IQR).  
    \emph{Middle row:} raw values mapped at each network node (zero‐value nodes in gray).  
    \emph{Bottom row:} normalized values mapped at the same nodes (zero‐value nodes in gray).}
    \label{fig:pt_normalization}
\end{figure}

\begin{figure}[!htbp]
    \centering
    \includegraphics[width=0.9\linewidth]{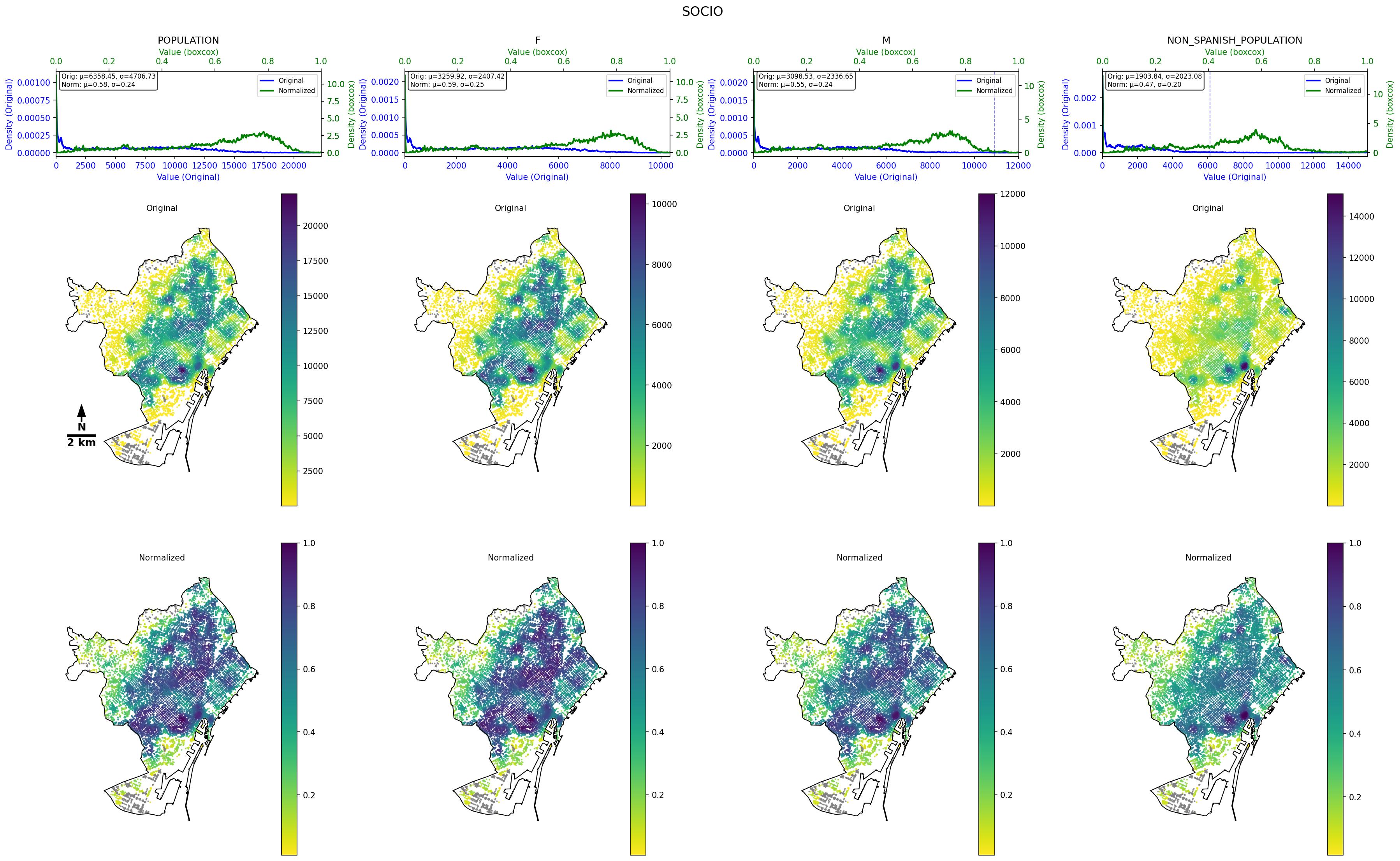}
    \caption{\textbf{Distributions and spatial patterns before and after normalization of the sociodemographic variables: total population, female population, male population, and non-Spanish population}. \emph{Top row:} kernel density estimates of the raw counts (blue) and the normalized values (green), with vertical dashed lines indicating the lower and upper outlier thresholds for the normalized data (first and third quartiles ± 1.5 × IQR).  
    \emph{Middle row:} raw values mapped at each network node (zero‐value nodes in gray).  
    \emph{Bottom row:} normalized values mapped at the same nodes (zero‐value nodes in gray).}
    \label{fig:socio_normalization}
\end{figure}

\begin{figure}[!htbp]
    \centering
    \includegraphics[width=0.9\linewidth]{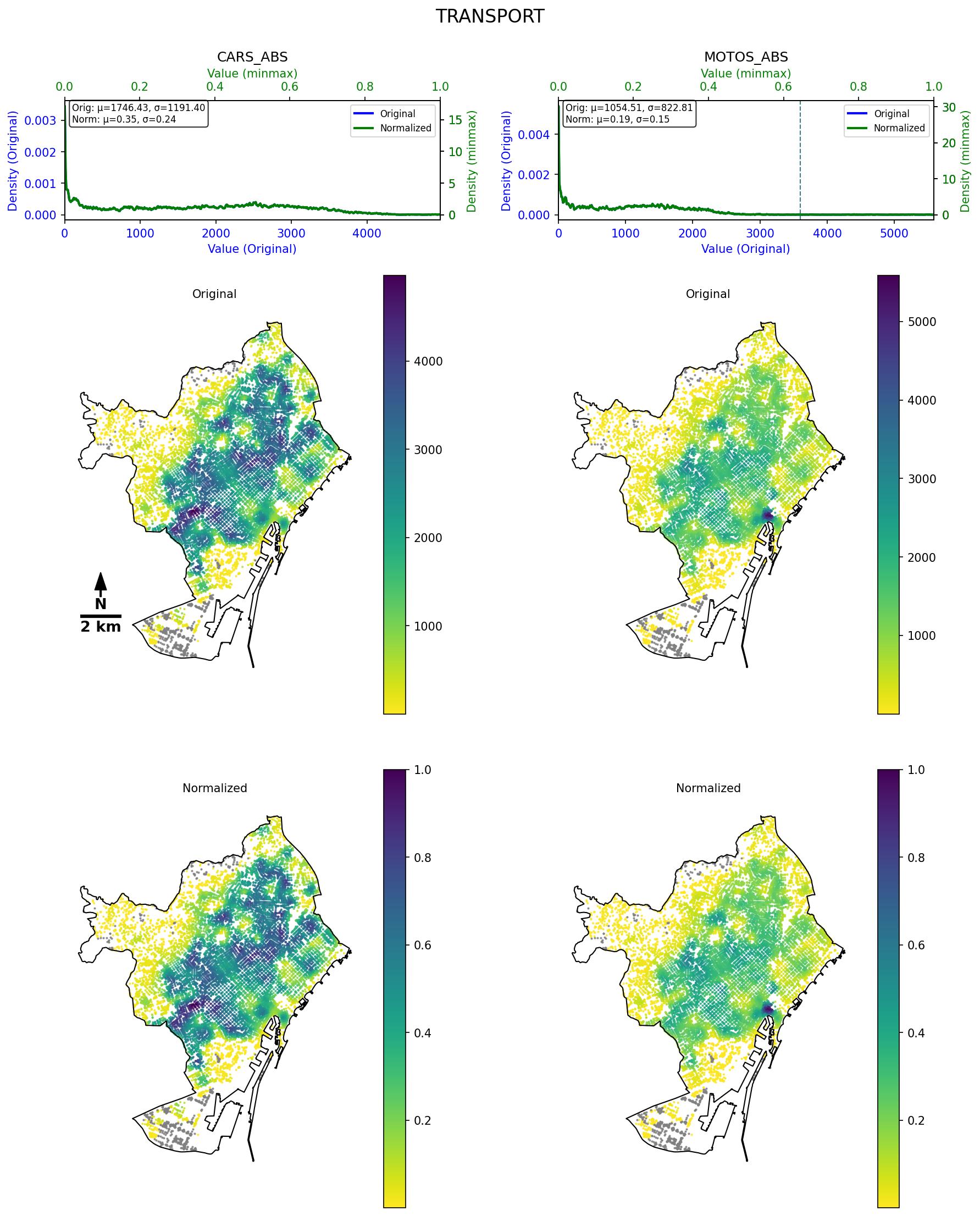}
    \caption{\textbf{Distributions and spatial patterns before and after normalization of the car and motos count}. \emph{Top row:} kernel density estimates of the raw counts (blue) and the normalized values (green), with vertical dashed lines indicating the lower and upper outlier thresholds for the normalized data (first and third quartiles ± 1.5 × IQR).  
    \emph{Middle row:} raw values mapped at each network node (zero‐value nodes in gray).  
    \emph{Bottom row:} normalized values mapped at the same nodes (zero‐value nodes in gray).}
    \label{fig:transport_normalization}
\end{figure}

Each selected transformation was followed by a final min–max rescaling to ensure every variable lies in \([0,1]\). Table~\ref{tab:normalization_map} lists the normalization method applied to each variable.  Detailed diagnostic figures, which display both the statistical and spatial effects of each transformation, are provided below (Figures \ref{fig:age_normalization}, \ref{fig:economic_normalization}, \ref{fig:education_normalization}, \ref{fig:pois1_normalization}, \ref{fig:pois2_normalization}, \ref{fig:pois3_normalization}, \ref{fig:pt_normalization}, \ref{fig:socio_normalization}, \ref{fig:transport_normalization})

\begin{table}[!htbp]
  \centering
  \caption{\textbf{Normalization method assigned to each variable}}
  \label{tab:normalization_map}
  \footnotesize
  \renewcommand{\arraystretch}{1.2}
  \begin{tabular}{p{0.20\linewidth} p{0.75\linewidth}}
    \toprule
    \textbf{Method} & \textbf{Variables} \\
    \midrule
    
    Min–max scaling & \texttt{tram\_lines}, \texttt{bike\_lane\_kms}, \texttt{cars\_ownership}, \texttt{motos\_ownership} \\
    
    Log–transform & \texttt{bus\_lines}, \texttt{metro\_lines}, \texttt{n\_health\_care}, \texttt{n\_culture}, \texttt{n\_tourism}, \texttt{n\_recreation}, \texttt{n\_sport}, \texttt{n\_economic\_retail}, \texttt{n\_industrial}, \texttt{n\_green}, \texttt{n\_civic}, \texttt{n\_worship}, \texttt{n\_education}, \texttt{n\_superpois} \\
    
    Box–Cox transformation & \texttt{age\_10-19}, \texttt{age\_20-29}, \texttt{age\_30-39}, \texttt{age\_40-49}, \texttt{age\_50-59}, \texttt{age\_60-69}, \texttt{age\_70+}, \texttt{household\_avg\_m2}, \texttt{unemployment}, \texttt{education\_primary}, \texttt{education\_secondary}, \texttt{education\_college}, \texttt{pois\_count}, \texttt{pois\_entropy}, \texttt{population}, \texttt{female}, \texttt{male}, \texttt{non\_spanish\_population}\\
    
    Inverted Box–Cox & \texttt{income}\\
    \bottomrule
  \end{tabular}
\end{table}

\section{Results}

\subsection{Spatial distributions of local factors for scenarios S1–S3}
\label{app:scenarios} 
\begin{figure}[!htbp]
    \centering
    \includegraphics[width=0.65\linewidth]{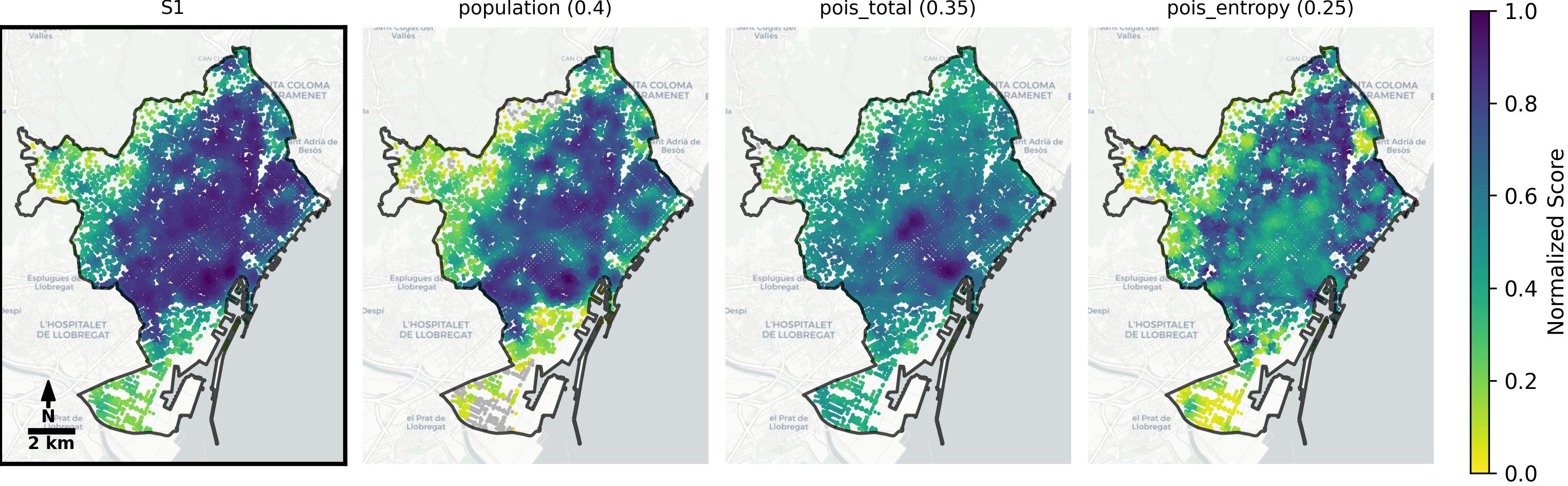}
    \caption{\textbf{Spatial distributions for Scenario S1.} The leftmost panel shows each node’s normalized utility score, computed as a weighted sum of the local factors. In the remaining panels, the spatial pattern of each individual factor is displayed, with the factor’s assigned weight indicated in that panel’s title.}
    \label{fig:spatial_distribution_s1}
\end{figure}

\begin{figure}[!htbp]
    \centering
    \includegraphics[width=0.65\linewidth]{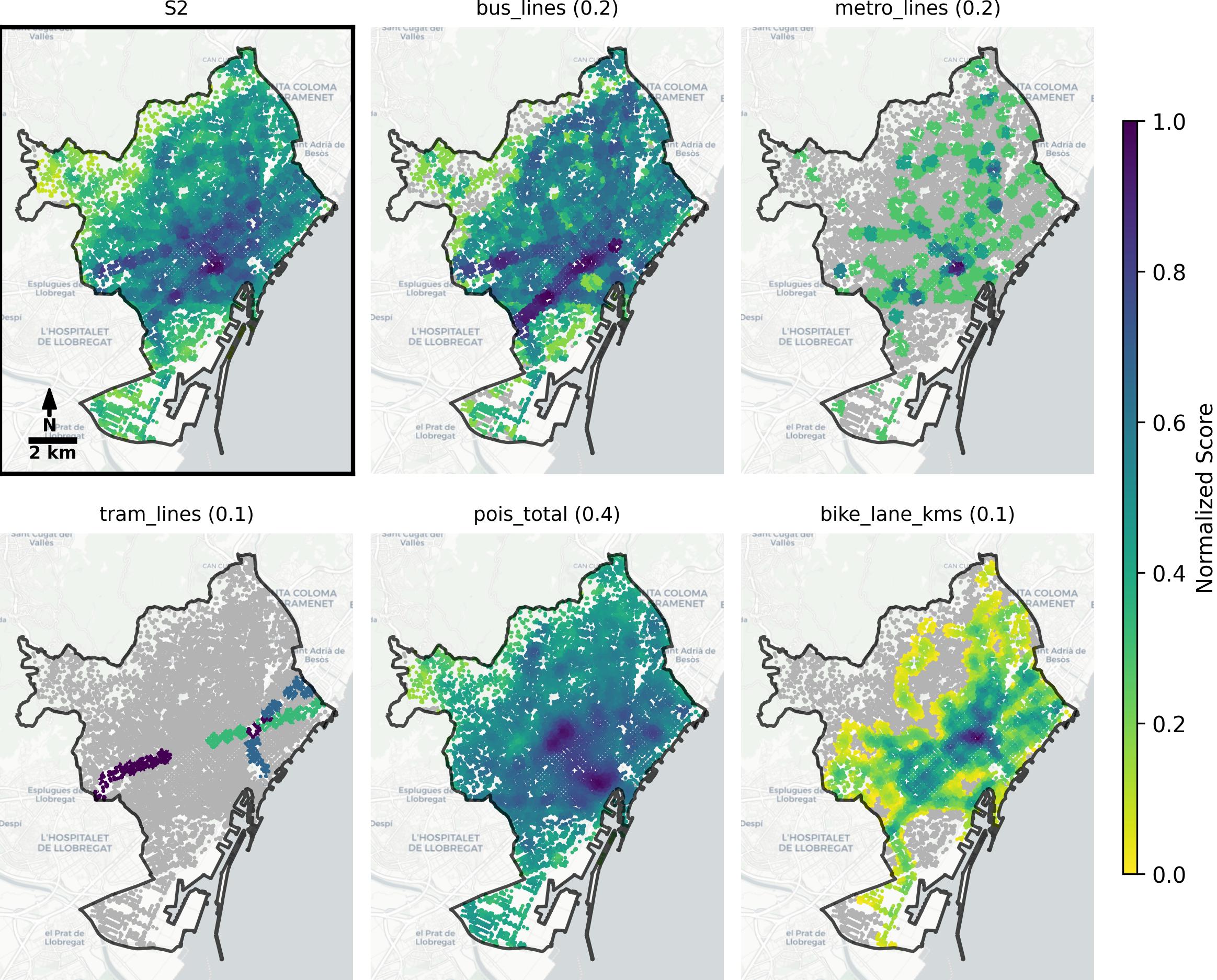}
    \caption{\textbf{Spatial distributions for Scenario S2.} The leftmost panel shows each node’s normalized utility score, computed as a weighted sum of the local factors. In the remaining panels, the spatial pattern of each individual factor is displayed, with the factor’s assigned weight indicated in that panel’s title.}
    \label{fig:spatial_distribution_s2}
\end{figure}

\begin{figure}[!htbp]
    \centering
    \includegraphics[width=0.65\linewidth]{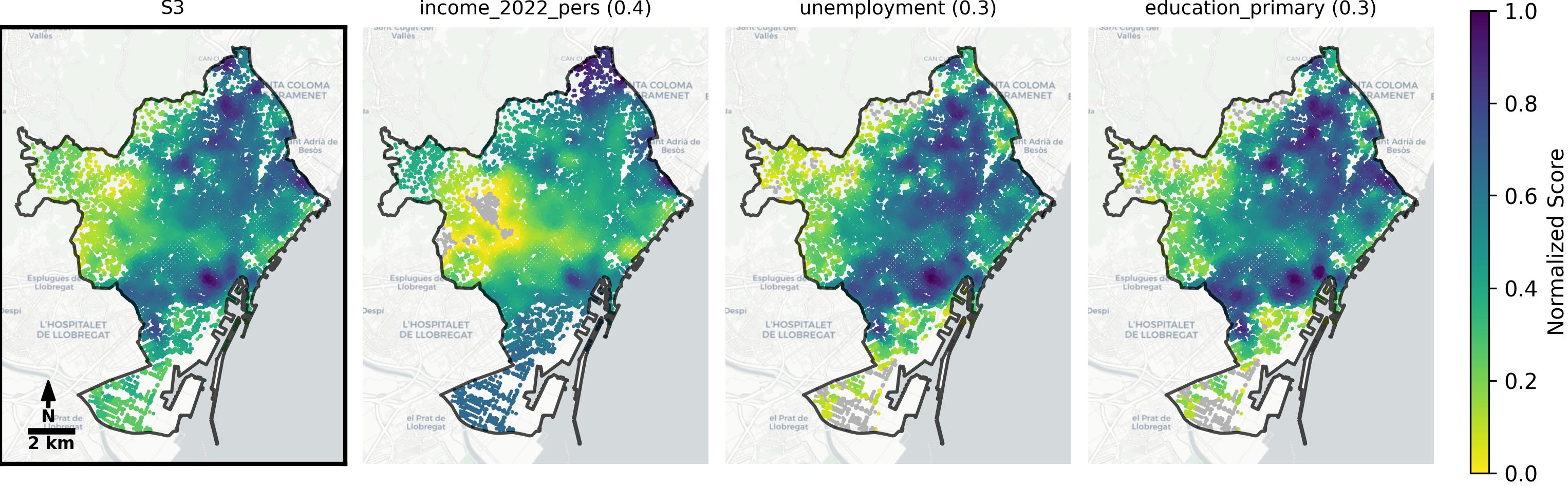}
    \caption{\textbf{Spatial distributions for Scenario S3.} The leftmost panel shows each node’s normalized utility score, computed as a weighted sum of the local factors. In the remaining panels, the spatial pattern of each individual factor is displayed, with the factor’s assigned weight indicated in that panel’s title.}
    \label{fig:spatial_distribution_s3}
\end{figure}

\clearpage

\subsection{Network-based station optimization for scenarios S2 and S3}
\label{app:network_scenarios} 

 \begin{figure}[!htbp]
     \centering
     \includegraphics[width=1\linewidth]{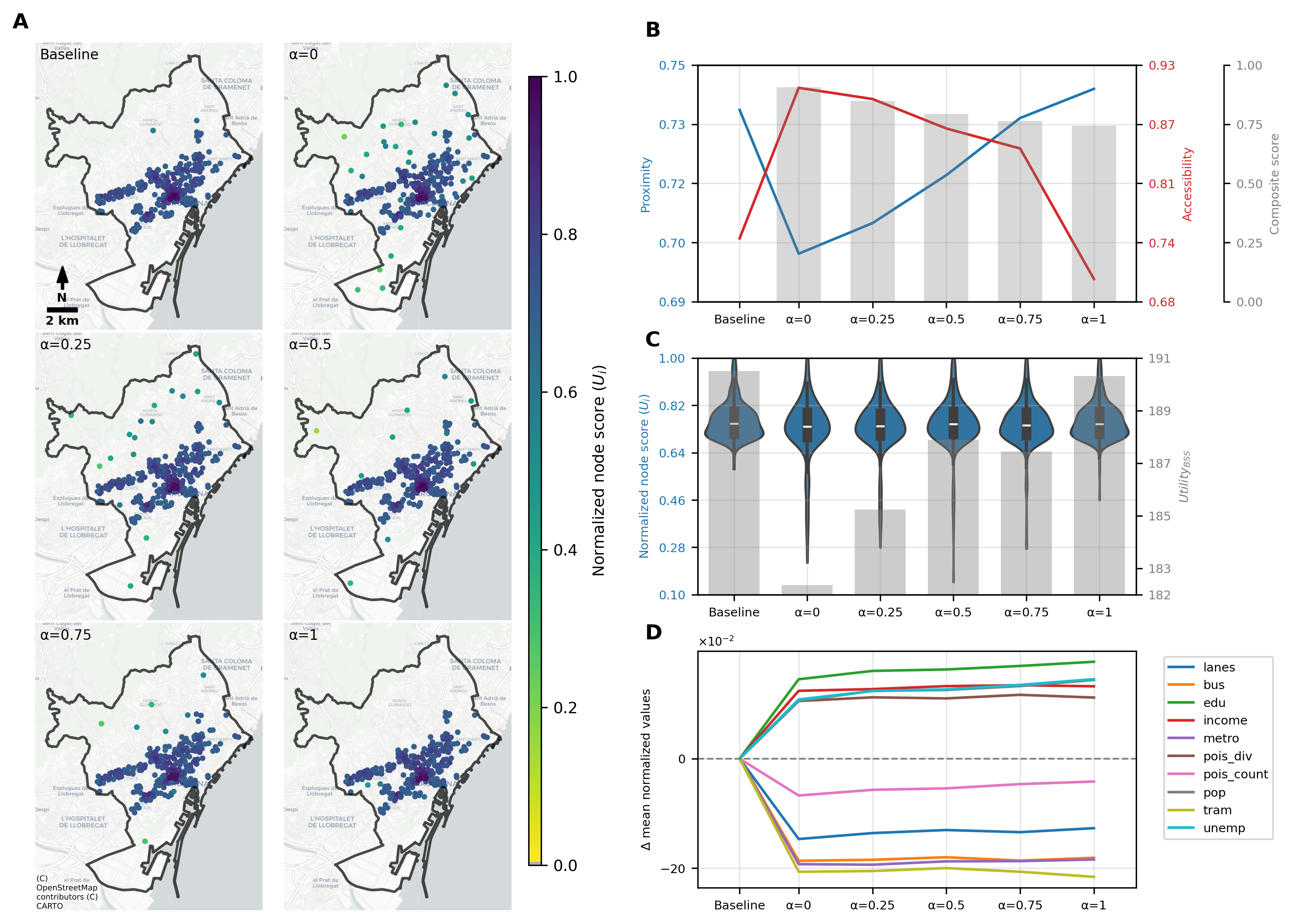}
     \caption{\textbf{Impact of network‐metrics trade-off ($\alpha$) on scenario S2.}  All panels use the same S2 input weights to illustrate how modifying $\alpha$ shifts the BSS network and the node utilities $U_i$. \textbf{(A)} Maps of selected stations under six conditions: a baseline optimization without network metrics, and optimizations with $\alpha$ = 0, 0.25, 0.50, 0.75, and 1. \textbf{(B)} displays proximity score $S_{pro}$ and accessibility score $S_{acc}$ versus $\alpha$. \textbf{(C)} BSS utility $U_{BSS}$ alongside box-plots of individual station utilities for each $\alpha$. \textbf{(D)} Change in mean local factor values of the selected stations relative to the no network metrics baseline.}
     \label{fig:alpha_screening_s2}
 \end{figure}

 \begin{figure}[!htbp]
     \centering
     \includegraphics[width=1\linewidth]{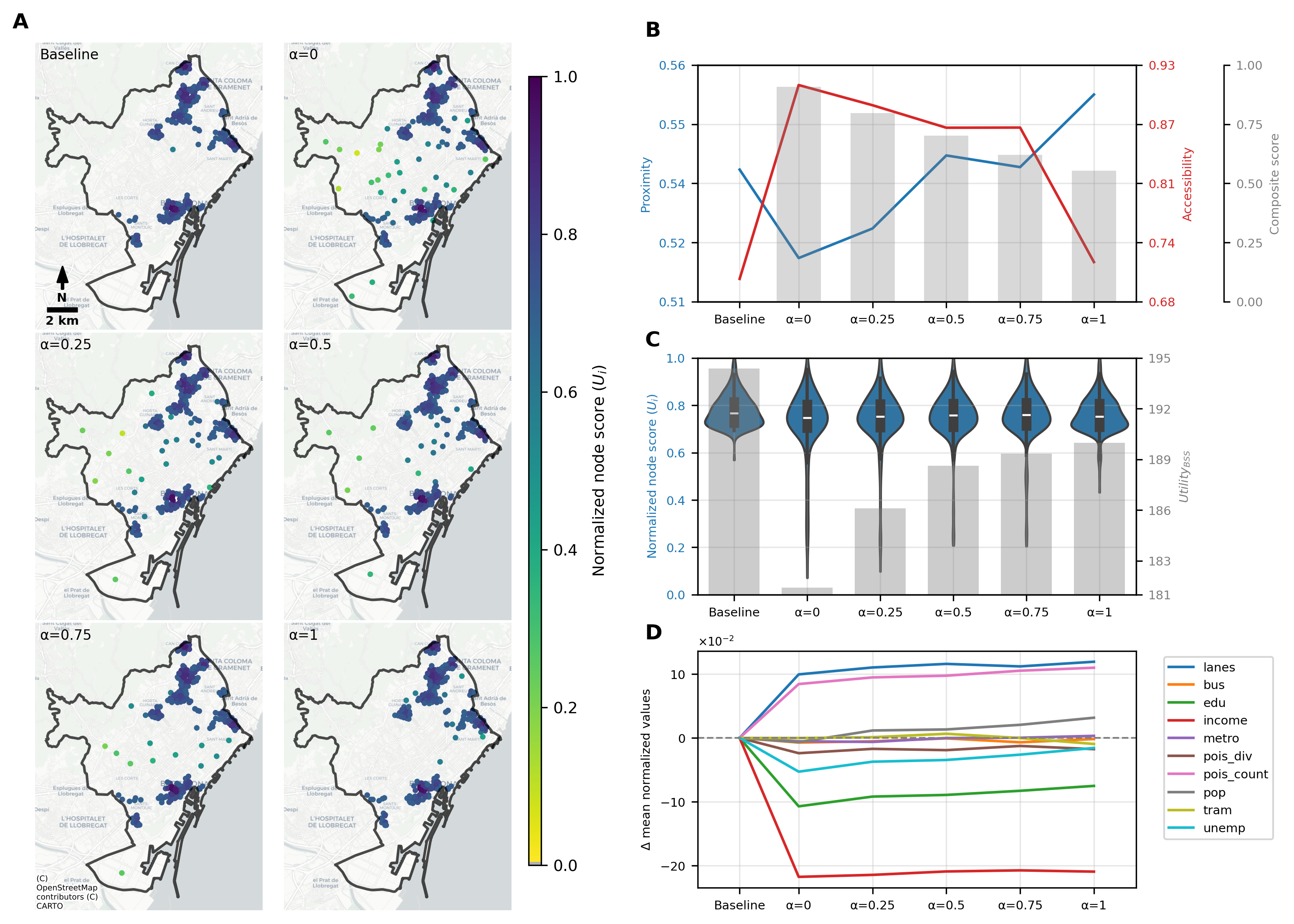}
     \caption{\textbf{Impact of network‐metrics trade-off ($\alpha$) on scenario S3.}  All panels use the same S3 input weights to illustrate how modifying $\alpha$ shifts the BSS network and the node utilities $U_i$. \textbf{(A)} Maps of selected stations under six conditions: a baseline optimization without network metrics, and optimizations with $\alpha$ = 0, 0.25, 0.50, 0.75, and 1. \textbf{(B)} displays proximity score $S_{pro}$ and accessibility score $S_{acc}$ versus $\alpha$. \textbf{(C)} BSS utility $U_{BSS}$ alongside box-plots of individual station utilities for each $\alpha$. \textbf{(D)} Change in mean local factor values of the selected stations relative to the no network metrics baseline.}
     \label{fig:alpha_screening_s3}
 \end{figure}

\clearpage

\subsection{Altitude-adjusted optimizations}
\label{app:results_altitude}

\begin{figure}[!htbp]
  \centering
  \includegraphics[width=1\linewidth]{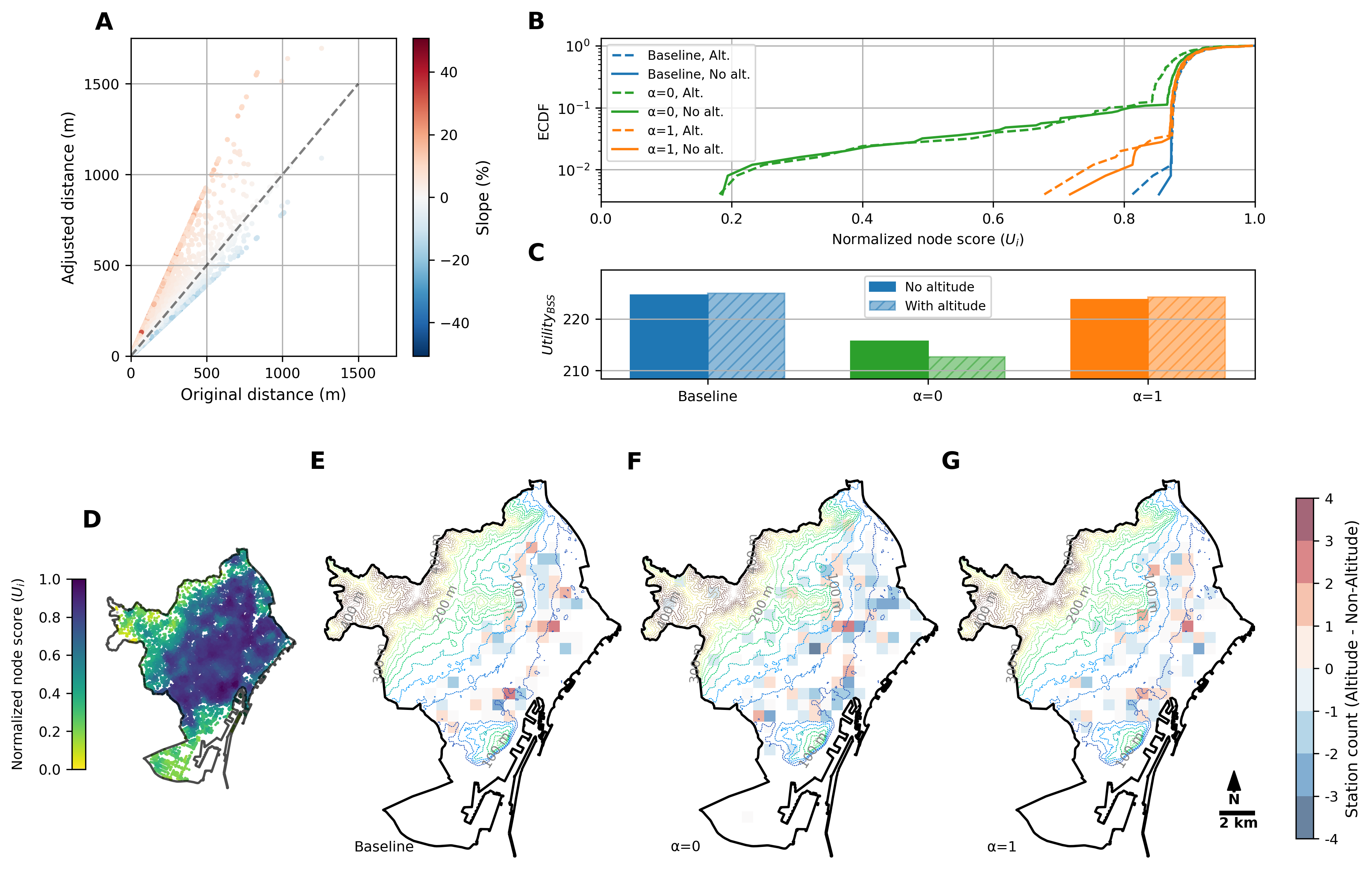}    \caption{
\textbf{Impact of altitude-adjusted distances on station selection.} \textbf{(A)} Original edge distances vs. altitude-adjusted distances, coloured by slope of the between origin and destination. \textbf{(B)} ECDF of normalized node scores for different $\alpha$ values under the S1 scenario, with and without altitude adjustments; the baseline corresponds to the optimization without network metrics for S1. \textbf{(C)} Composite utility $U^*_{BSS}$ for the resulting station sets, showing a slight trade-off when elevation is considered. \textbf{(D)} Spatial distribution of normalized node scores in the study area. \textbf{(E–G)} Differences in selected station locations (grid cell counts) between configurations with and without altitude adjustment under the \textbf{(E)} S1 scenario, and its network-aware versions using \textbf{(F)}~$\alpha=0$ and \textbf{(G)}~$\alpha=1$. Grids cells are 500 m by 500 m. Positive values indicate more stations were selected in a given cell when altitude was considered. Background contour lines represent elevation levels.}
  \label{fig:altitude_s1}
\end{figure}

\begin{figure}[!htbp]
    \centering
    \includegraphics[width=1\linewidth]{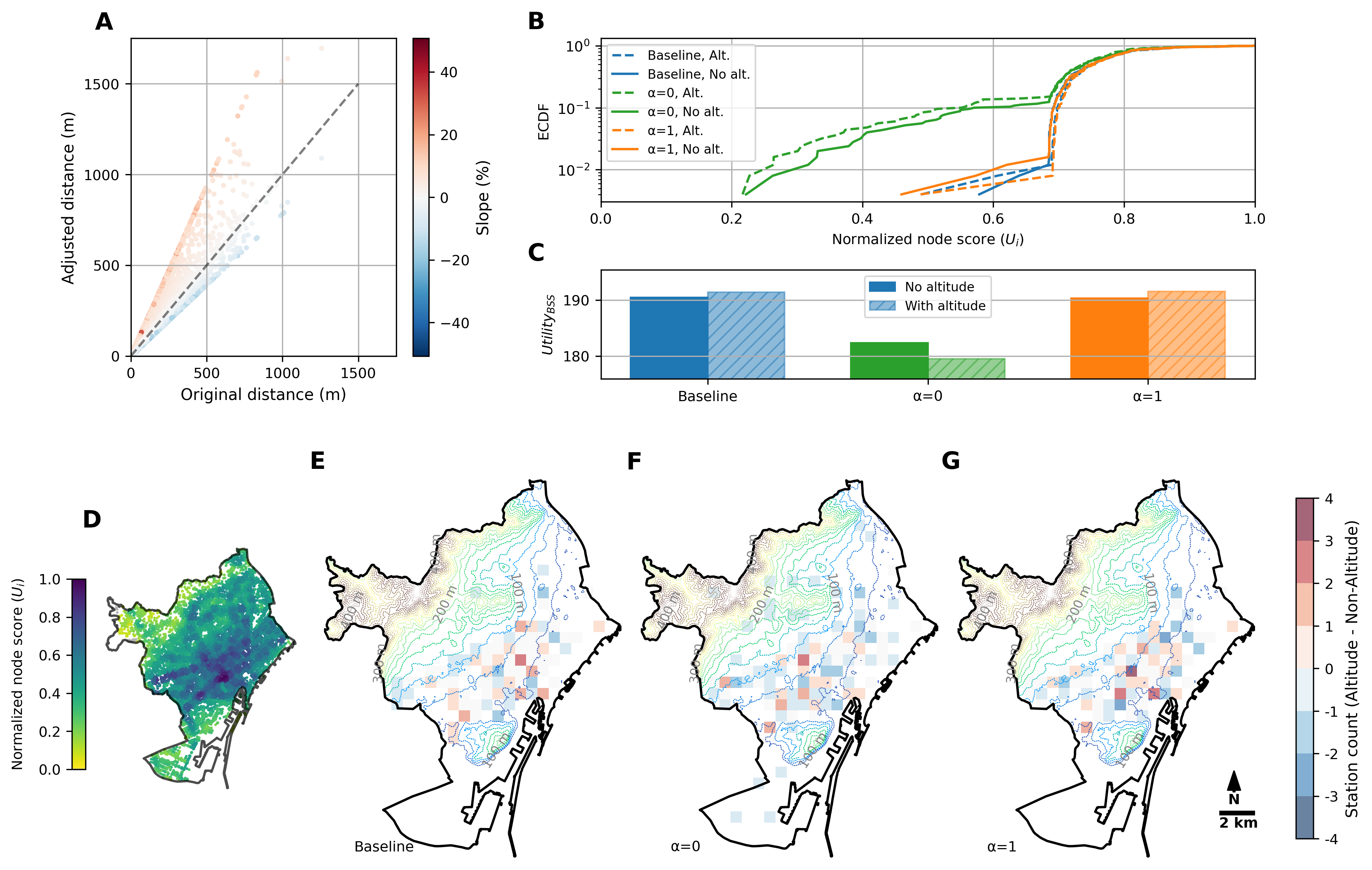}    \caption{
\textbf{Impact of altitude-adjusted distances on station selection.} \textbf{(A)} Original edge distances vs. altitude-adjusted distances, coloured by slope of the between origin and destination. \textbf{(B)} ECDF of normalized node scores for different $\alpha$ values under the S2 scenario, with and without altitude adjustments; the baseline corresponds to the optimization without network metrics for S2. \textbf{(C)} Composite utility $U^*_{BSS}$ for the resulting station sets, showing a slight trade-off when elevation is considered. \textbf{(D)} Spatial distribution of normalized node scores in the study area. \textbf{(E–G)} Differences in selected station locations (grid cell counts) between configurations with and without altitude adjustment under the \textbf{(E)} S2 scenario, and its network-aware versions using \textbf{(F)}~$\alpha=0$ and \textbf{(G)}~$\alpha=1$. Grids cells are 500 m by 500 m. Positive values indicate more stations were selected in a given cell when altitude was considered. Background contour lines represent elevation levels.}
    \label{fig:altitude_s2}
\end{figure}

\subsection{BSS expansion}
\label{app:results_bss_expansion}

\begin{figure}[!htbp]
    \centering
    \includegraphics[width=0.5\linewidth]{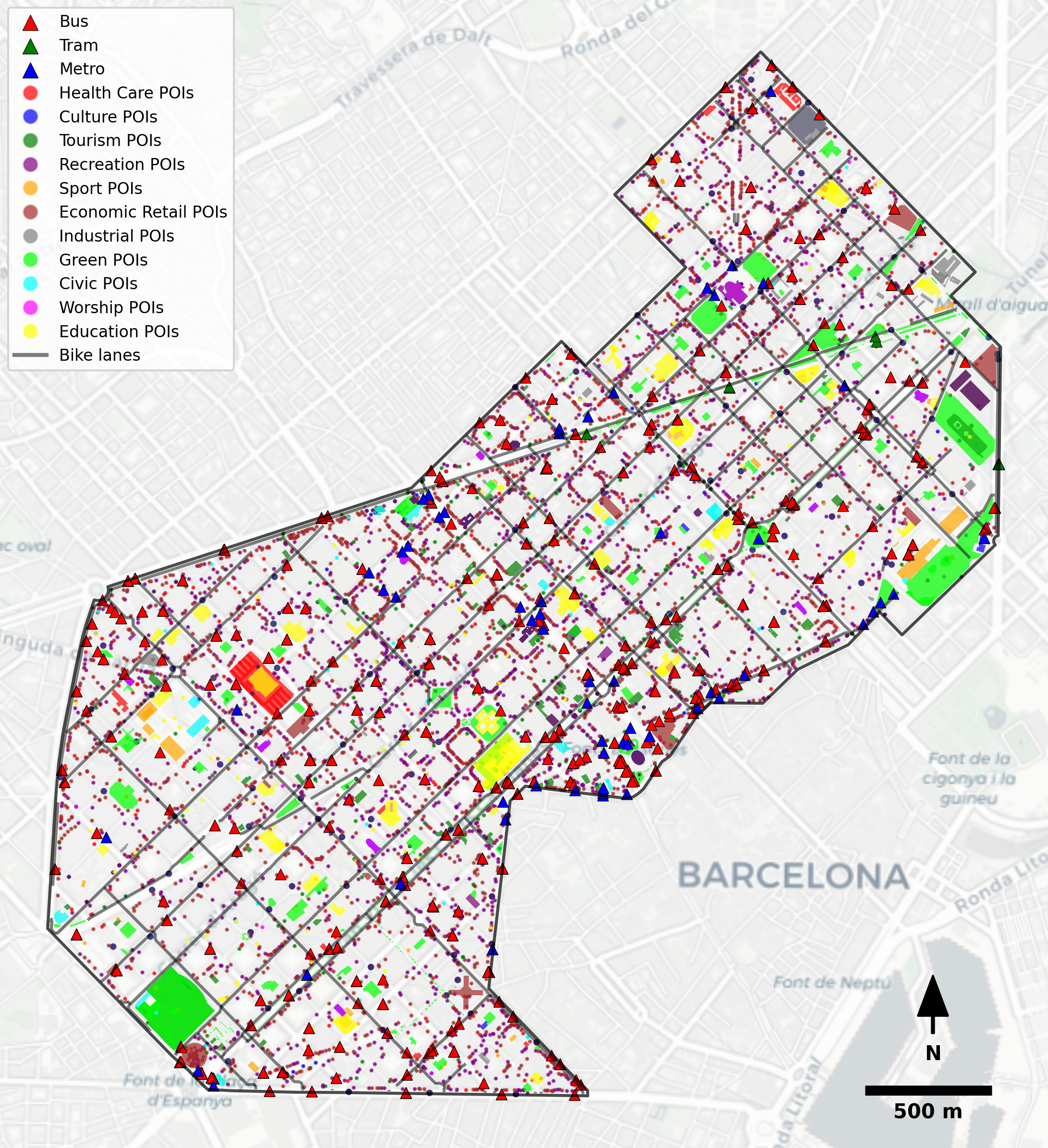}
    \caption{\textbf{Spatial distribution of public transport stations and POIs across the Eixample district in Barcelona.} Public transport nodes include bus and tram stops, and metro entrances.}

    \label{fig:pois_and_pt_eixample}
\end{figure}

\end{document}